\documentclass[english]{emulateapj}
\usepackage[T1]{fontenc}
\usepackage{amstext}
\usepackage{amssymb}
\usepackage{graphicx}

\makeatletter

\DeclareRobustCommand{\greektext}{%
  \fontencoding{LGR}\selectfont\def\encodingdefault{LGR}}
\DeclareRobustCommand{\textgreek}[1]{\leavevmode{\greektext #1}}
\DeclareFontEncoding{LGR}{}{}
\DeclareTextSymbol{\~}{LGR}{126}

\usepackage[hyperfootnotes=false]{hyperref}
\hypersetup{colorlinks=true,citecolor=blue}
\shorttitle{OBSERVATION OF THE FUV CONTINUUM BACKGROUND}
\shortauthors{SEON ET AL.}

\makeatother

\usepackage{babel}
\begin{document}

\title{Observation of The Far-Ultraviolet Continuum Background with SPEAR/FIMS}

\author{Kwang-Il Seon\altaffilmark{1,2}, Jerry Edelstein\altaffilmark{3},
Eric Korpela\altaffilmark{3}, Adolf Witt\altaffilmark{4}, Kyoung-Wook
Min\altaffilmark{5}, Wonyong Han\altaffilmark{1,2}, Jongho Shinn\altaffilmark{1},
Il-Joong Kim\altaffilmark{1}, Jae-Woo Park\altaffilmark{6}}

\altaffiltext{1}{Korea Astronomy and Space Science Institute, Daejeon, Republic of Korea, 305-348; email: kiseon@kasi.re.kr}
\altaffiltext{2}{University of Science and Technology, Daejeon, Republic of Korea, 305-350}
\altaffiltext{3}{Space Sciences Laboratory, University of California, Berkeley, CA, USA 94720}
\altaffiltext{4}{Ritter Astrophysical Research Center, University of Toledo, Toledo, OH, USA 43606}
\altaffiltext{5}{Korea Advanced Institude of Science and Technology, Daejeon, Republic of Korea 305-701}
\altaffiltext{6}{Korea Intellectual Property Office, Daejeon, Republic of Korea 302-701} 
\begin{abstract}
We present the general properties of the far-ultraviolet (FUV; 1370--1710
\AA) continuum background over most of the sky, obtained with the
\textit{Spectroscopy of Plasma Evolution from Astrophysical Radiation}
instrument (SPEAR, also known as FIMS), flown aboard the STSAT-1 satellite
mission. We find that the diffuse FUV continuum intensity is well
correlated with $N_{{\rm HI}}$, 100 $\mu$m, and H$\alpha$ intensities
but anti-correlated with soft X-ray intensity. The correlation of
the diffuse background with the direct stellar flux is weaker than
the correlation with other parameters. The continuum spectra are relatively
flat. However, a weak softening of the FUV spectra toward some sight
lines, mostly at high Galactic latitudes, is found not only in direct-stellar
but also in diffuse background spectra. The diffuse background is
relatively softer than the direct stellar spectrum. We also find that
the diffuse FUV background averaged over the sky has a bit softer
spectrum compared to direct stellar radiation. A map of the ratio
of 1370--1520 \AA\ to 1560--1710 \AA\ band intensity shows that
the sky is divided into roughly two parts. However, this map shows
a lot of patchy structures on small scales. The spatial variation
of the hardness ratio seems to be largely determined by the longitudinal
distribution of OB-type stars in the Galactic plane. A correlation
of the hardness ratio with the FUV intensity at high intensities is
found but an anti-correlation at low intensities. We also find evidence
that the FUV intensity distribution is log-normal in nature.
\end{abstract}


\section{Introduction}

Measurement of the interstellar radiation field (ISRF), comprised
of the direct-stellar and diffuse background radiation fields, is
of considerable interest because the ISRF, in particular at the far-ultraviolet
(FUV) waveband, controls the physics and chemistry of the interstellar
medium (ISM). The FUV ISRF is often expressed by a simple approximation
referred to as the Habing radiation field \citep{Harbing68} and was
measured for the first time by \citet{Henry77b} with two Geiger counters
on board an Aerobee rocket. \citet{Draine78} obtained another simple
analytic expression that appeared to be in good agreement with all
the previous results at that time, and it has been referred to as
the Draine's ``standard'' UV background. \citet{GON80} computed a
theoretical model and tabulated the average UV ISRF scaled to fit
the radiation field observed by the S2/68 telescope aboard the TD-1
satellite. \citet{Mathis83} redetermined the ISRF by fitting the
spectrum derived by \citet{GON80} to power-law and scaling up by
15\% to bring it into better agreement with the measurement of \citet{Henry80}
and extended the calculation to cover wavelengths from 0.09 $\mu$m
to 1000 $\mu$m.

The majority of the diffuse FUV background radiation or the diffuse
galactic light is generally believed to be of Galactic origin, starlight
scattered off by interstellar dust. Observations of the diffuse background
then give us information on the scattering properties of the dust
grains. However, early attempts to measure the diffuse FUV background
were motivated by the hope that a large fraction of the radiation
in this waveband would be extragalactic (see reviews by \citealp{PJ1980},
\citealp{BOW1991}, and \citealp{Henry91}). There had been serious
disagreements among observers regarding the spatial and spectral variabilities
of the FUV background radiation. Discrepancies among early measurements
of the diffuse FUV radiation have lead to different interpretations
as to its dominant contributor, either extragalactic light or galactic
light scattered off dust grains. A correlation between the FUV background
and Galactic neutral hydrogen column density ($N_{{\rm HI}}$) was
presented for the first time by \citet{PMB1980}, thereby implying
that most of the FUV background is Galactic in origin, consistent
with scattering of the integrated FUV radiation field by interstellar
dust. Substantial evidence supporting the conclusion has been reported
later \citep{MJoubert80,Joubert83,Jakobsen84,Onaka91}.

Some studies, however, have suggested that the correlation with Galactic
latitude is more significant than that with Galactic \ion{H}{1} column
density \citep{Weller83,FCF89}, or that the FUV background intensity
shows only a weak correlation with Galactic observables \citep{Murthy89,Murthy90}.
However, \citet{Perault91} confirmed the correlation with Galactic
\ion{H}{1} column density after subtraction of the inverse sin$|b|$
dependence from the data obtained with the D2B-Aura satellite. \citet{HBM91}
found that their data taken with the Berkeley UVX spectrometer was
three times better in dependence with Galactic \ion{H}{1} column
density than with Galactic latitude. \citet{Jakobsen87}, \citet{Perault91},
and \citet{Sasseen96} showed also that a correlation exists between
the diffuse infrared background intensity at 100 $\mu$m as measured
by the Infrared Astronomical Satellite (IRAS) and the diffuse background
intensity in the FUV wavelengths. \citet{Sasseen95} demonstrated
a similarity between the spatial power spectra of the IRAS 100 $\mu$m
cirrus images and the diffuse FUV background images obtained with
the Far Ultraviolet Space Telescope (FAUST). In the scale of a single
cloud, \citet{Haikala1995} discovered a Galactic cirrus cloud, G251.2+73.3,
near the north Galactic pole in FUV waveband and obtained a good correlation
between the FUV intensity and IRAS 100 $\mu$m surface brightness.
The existence of these properties thus provides strong observational
support for the dust-scattered origin of most, if not all, of the
FUV background.

Different contributors to the diffuse FUV background have been suggested
as well. \citet{DW1980} proposed fluorescent H$_{2}$ emission as
a significant contributor to the observed FUV background in addition
to scattering by interstellar dust grains. \citet{Jakobsen82} showed
that the reported correlations could be explained by a combination
of starlight scattered by interstellar dust and the H$_{2}$ fluorescence.
Later, the H$_{2}$ fluorescence emission lines from the diffuse ISM
were observed with the Berkeley UVX Shuttle Spectrometer \citep{MartinHB90}
and then by SPEAR \citep{Lee06,Lee08,Ryu06}. Those measurements make
it clear that the molecular hydrogen emissions are not the dominant
contributor of the diffuse FUV background. \citet{Jakobsen81} suggested
that emission lines from a hot ($\gtrsim10^{5}$ K) Galactic coronal
gas would contribute to the diffuse FUV background, although this
source would not be of sufficient intensity to influence the measurements
of diffuse FUV continuum background. \citet{Deharveng82} estimated
the possible contribution of a warm ($\sim10^{4}$ K) ionized medium
(WIM) to the FUV background and attributed a FUV continuum intensity
of $\sim5-70$ photons cm$^{-2}$ s$^{-1}$ sr$^{-1}$ \AA$^{-1}$
(continuum unit, hereafter CU) to the two-photon continuum emission
from the WIM (see also \citealt{HBM91}). \citet{Reynolds92} showed
that the two-photon emission from the WIM accounts for approximately
20\% of the diffuse FUV intensity near 1600 \AA.

Disagreements on the spectral shape of the diffuse FUV continuum background
are also noticeable. \citet{Henry78} measured the diffuse FUV background
at the north and south Galactic pole regions, using a spectrometer
flown on \emph{Apollo} 17, and reported a spectrum that is flat between
1300 and 1525 \AA\ and declining in intensity toward 1625 \AA. The
observations by \citet{Hua79} and \citet{Zvereva82} supported the
decrease of intensity with increasing wavelength at high latitudes,
using a spectrometer on board the Soviet spacecraft ``Prognoz-6''.
They also reported the flat spectrum in the intermediate latitudes
and a rapid increase of the intensity with wavelength near and inside
the Milky Way. However, \citet{Anderson79} reported some evidence
for a sharp rise in intensity longward of 1680 \AA\ with instruments
flown on an \emph{Aries} sounding rocket. The data observed by \citet{Henry80}
also showed a rise longward of about 1500 \AA. \citet{Murthy89}
found a spectrally flat background over the wavelength band from 1200
to 1700 \AA\ with the Johns Hopkins UVX experiment on board the Space
Shuttle. The spectrum obtained at a high Galactic latitude with the
Berkeley UVX FUV spectrometer is relatively flat between 1400 and
1850 \AA\ \citep{MartinB90}.

Such discrepancies would be better understood with spectrophotometric
observations over large area of the sky. However, only a few studies
have mapped a substantial fraction of sky in the FUV waveband. \citet{GON80}
presented the spatial distribution of the FUV stellar radiation at
four passbands centered at 1565 \AA, 1965 \AA, 2365 \AA, and 2740
\AA\ from observations made with the S2/68 sky-survey telescope.
The first coarse map of the diffuse FUV radiation was presented by
\citet{Lequeux90} and later updated by \citet{Perault91} using the
data obtained with the ELZ spectrophotometer on board the D2B-Aura
satellite \citep[see also,][]{MJoubert78,MJoubert80,Joubert83}. The
most recent FUV band map is from the Narrowband Ultraviolet Imaging
Experiment for Wide-Field Surveys (NUVIEWS; \citealt{SCH2001}), which
mapped one-quarter of the sky in FUV radiation ($\lambda=1740$ \AA,
$\Delta\lambda=100$ \AA\ FWHM) with $\sim5'-10'$ imaging resolution,
including features such as the Upper Scorpius region.

Spectroscopy of Plasma Evolution from Astrophysical Radiation instruments
(SPEAR, also known as Far Ultraviolet Imaging Spectrograph or FIMS)
provided the first large-area spectral sky survey of diffuse FUV radiation
\citep{Edelstein06a,Edelstein06b}. The mission observed $\sim80\%$
of the sky and conducted deep pointed observations toward selected
targets, such as supernova remnants \citep{Seonetal06,Nishikida06,Shinn06,Shinn07,Kim07,Kim2010,Kim2010b},
superbubbles \citep{Kregenow06,Park07,Ryu08}, photoionized \ion{H}{2}
regions \citep{Park2010}, and H$_{2}$ molecular clouds \citep{Lee06,Lee08,Ryu06,Park2009}.
The emission lines from multiply-ionized carbon, oxygen, silicon,
and aluminum toward the north ecliptic and Galactic pole regions observed
with SPEAR/FIMS were presented in detail by \citet{Korpela2006} and
\citet{Welsh2007}.

In this paper, we describe some general results on the diffuse FUV
(1370--1710 \AA) continuum background observed with SPEAR/FIMS. Emission
line sky-maps obtained with SPEAR/FIMS is described in \citet{Kregenow2007}
and Edelstein et al. (2011, in preparation). This paper is organized
as follows. In \S2 we describe the observation and data reduction.
The FUV intensity maps, comparison of the diffuse FUV background with
a model of the stellar radiation field are shown in \S3. Spectral
shapes of the diffuse FUV background is described in \S4. We compare
the diffuse FUV continuum map with maps of $N_{{\rm HI}}$, H$\alpha$,
100 $\mu$m, and soft X-ray (1/4 keV) in \S5. A discussion is presented
in \S6. Finally, we conclude with a summary of results in \S7.

\section{Observation and data reduction}

\subsection{Observation}

SPEAR/FIMS%
\footnote{http://spear.ssl.berkeley.edu/.%
} is a dual-channel FUV imaging spectrograph (``Short'' wavelength
channel {[}S-band{]} 900--1150 \AA, ``Long'' wavelength channel {[}L-band{]}
1350--1750 \AA; $\lambda/\Delta\lambda\sim550$) with a large field
of view (S-band, $4^{\circ}\times4'.6$; L-band, $7^{\circ}.4\times4'.3$),
designed to observe diffuse FUV emission lines. SPEAR is the primary
payload on the first Korean science and technology satellite, \emph{STSAT-1},
and was launched into a $\sim700$ km Sun-synchronous orbit on 2003
September 27. The SPEAR/FIMS mission, the instruments, their on-orbit
performance, and the basic processing of the instrument data are described
in detail by \citet{Edelstein06a,Edelstein06b}. The spectrographs
consist of the gratings, order baffling, the detector, a shutter-slit
unit, a mirror unit with field baffling, and a deployable dust cover.
The shutter can be selected to admit 0\%, 1\%, 10\%, or 100\% of the
available light for safe and photometric observations of bright sources.
The instrument provides for the first time accurate determination
of the FUV background spectral map covering a substantial fraction
of sky. The first Galactic map of diffuse FUV (L-band) background
measured by SPEAR/FIMS was presented in \citet{Edelstein06a}. Because
S-band lacks the Ly$\alpha$-rejection filter that L-band has, the
major continuum source for S-band is the instrumental background \citep{Edelstein06b,Kregenow2007}.
In addition, S-band has lower sensitivity than L-band, as well as
a higher non-astrophysical background, including a strong Ly$\beta$
(1026 \AA) airglow line. Therefore, in the present study, we analyze
only the L-band data. Given the wavelength dependence of the dust
albedo and the fact that the composite stellar spectrum, which is
dominated by OB-type stars, peaks near L-band, we expect that the
astrophysical continuum in S-band is no higher than in L-band.

SPEAR/FIMS observed the sky during the all-sky survey observation
mode by scanning the entire sky along the short axis of the slit,
i.e., along the $\sim5'$ field direction of the $\sim8^{\circ}\times5'$
FOV. In each orbit, the FOV scanned $\sim180^{\circ}$, from an ecliptic
pole to the opposite ecliptic pole, during the eclipse time ($\sim$
25 min), and the pole-to-pole scanning drifts $360^{\circ}$ along
the ecliptic equator for one year because of the properties of a sun-synchronous
orbit. The FUV sky maps for L-band were created by binning photon
and exposure events using the \emph{HEALPix} tessellation scheme \citep{Gorski05}
with $\sim1^{\circ}$ pixels (corresponding to the resolution parameter
$N_{{\rm side}}=64$), unless otherwise specified, and L-band wavelength
bins of 1.0 \AA. Only the data from 1370 to 1710 \AA, excluding
the intense \ion{O}{1} airglow line at 1356 \AA, were used for the
present analysis. The data were selected for attitude-knowledge quality
by using times when the derived attitude error is $\le30'$ and contain
$4.1\times10^{7}$ and $2.0\times10^{7}$ photons before and after
removal of locally bright stars, respectively. The diffuse background
map%
\footnote{The maps presented in the present study can be downloaded from the
following URL: http://kiseon.kasi.re.kr/FIMS.html.%
} was obtained by eliminating photon and exposure events in the locally
intense pixels (i.e., stars) from the starting sky map with $\sim14'$
pixels ($N_{{\rm side}}=256$). Each orbital scan was made into an
image and convolved with an edge detection kernel shaped like the
stellar point spread function. Any region with counts 3$\sigma$ above
the local background was considered a star. All photon events within
a $3'$ radius of the star centroid were flagged in the data. This
strategy effectively removed most faint and all bright stars (Edelstein
et al. 2011, in preparation). The remaining photon and exposure events
were binned into a map with $\sim1^{\circ}$ pixels ($N_{{\rm side}}=64)$.

\begin{figure*}[t]
\begin{centering}
\includegraphics[clip,scale=0.5,angle=90]{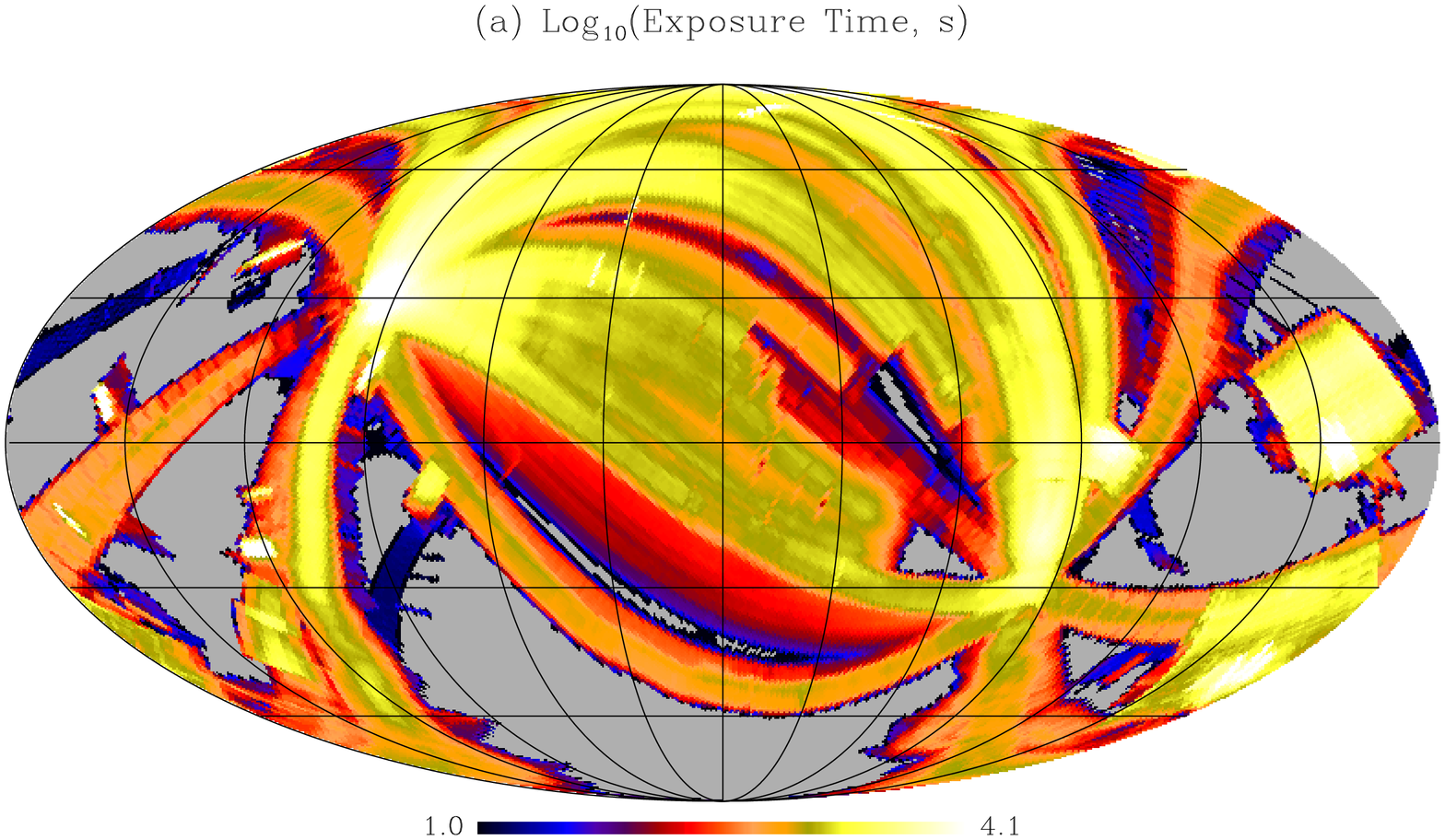}
\par\end{centering}

\begin{centering}
\bigskip{}
\includegraphics[clip,scale=0.5,angle=90]{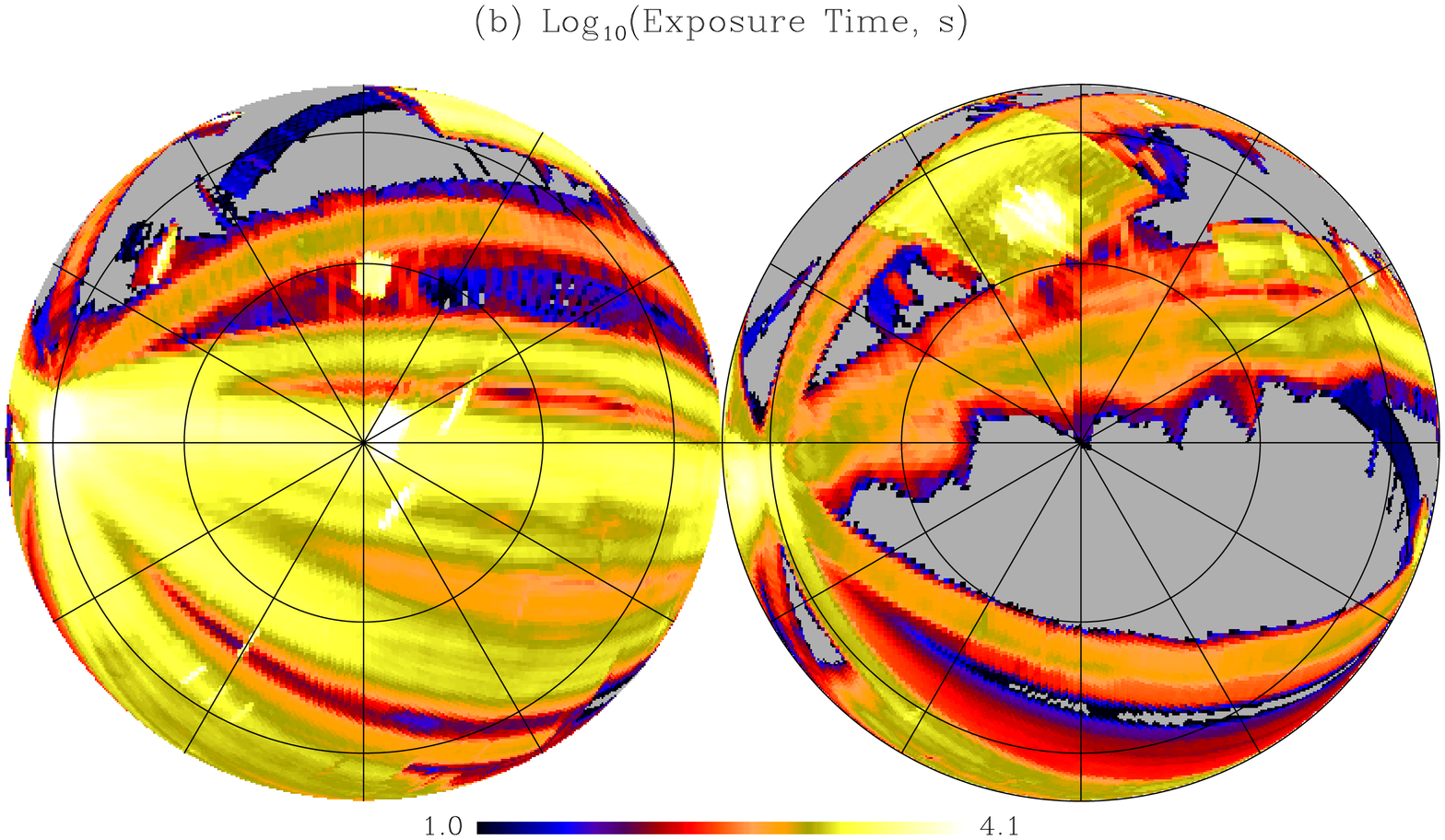}
\par\end{centering}

\begin{centering}
\medskip{}

\par\end{centering}

\caption{\label{exposure_map}(a) Mollweide and (b) orthographic projections
of sky exposure for the SPEAR/FIMS L-band observations. The map was
made using the \emph{HEALPix} scheme with $\sim1^{\circ}$ pixels
($N_{{\rm side}}$ = 64). The intensity scales are logarithmic across
the color bars. Exposure time observed with both 100\% and 10\% shutter
apertures are combined. (a) Galactic coordinates centered at $(l,b)=(0^{\circ},0^{\circ})$
with the longitude increasing toward the left are shown with latitude
and longitude lines on a $30^{\circ}$ grid. (b) The left and right
sides of orthographic projections are centered at the northern and
southern Galactic polar caps, respectively, and their longitude increases
clockwise and counter-clockwise, respectively. In both projections,
$l=0^{\circ}$ is at the six o'clock position.}
\end{figure*}

We repaired orbit number mismatches that were found after the publication
of \citet{Edelstein06a,Edelstein06b}, and included about 100 more
orbits other than those used in \citet{Edelstein06a}. Sky exposure
was derived by generating time-marked exposure events and mapping
the events to a sky position as detailed in \citet{Edelstein06a}.
The corresponding sky exposure map for the L-band with $\sim1^{\circ}$
pixels ($N_{{\rm side}}=64$), shown in Figure \ref{exposure_map},
covers $\sim80$\% of the sky and includes features such as deep
exposures ($>10$ ks) toward calibration and pointed study fields.
About 15\% of these observations were taken using the 10\% shutter
position. The exposures taken with the 10\% shutter aperture were
scaled by 0.1, as determined in \S\ref{sub:Instrumental-Background},
and added to the map taken with the 100\% shutter position. In the
figure, the gray color represents locations with zero exposure.

In L-band, there are several spectral features, such as atomic emission
lines (\ion{Si}{4} $\lambda$1398, \ion{Si}{2}{*} $\lambda$1532,
\ion{C}{4} $\lambda\lambda$1548, 1551, and \ion{Al}{2} $\lambda$1671),
quasi-band structures of H$_{2}$ fluorescence lines, and two-photon
continuum emission. Since the physical processes that give rise to
the emission lines are different than dominant continuum scattering,
the variation of the emission lines may be independent from the dust-scattering
processes that provide the bulk of the signal. If this emission is
included when estimating the ``continuum'' intensity, the ability
to compare the true FUV continuum background and other ISM tracers
is weakened, as is any discussion of the statistical properties of
the FUV continuum. To obtain the true intensity of the FUV continuum,
we may have to model all possible emission lines and continuum using
spectral fits, as in \citet{Korpela2006}. However, the signal-to-noise
ratio of the data with $\sim1^{\circ}$ pixels is typically only $\sim10$
(as described in \S\ref{sub:Statistics}), which is too small to
provide reliable fits. To obtain reliable fits, large pixel sizes
of $\sim2^{\circ}-7.5^{\circ}$ ($N_{{\rm side}}=32-8)$ were required,
especially at high Galactic latitudes, depending on locations in the
sky (see also Edelstein et al. 2011, in preparation). In this paper,
instead of using spectral fits and variable pixel sizes, we excluded
the wavelength regions of the strongest emission lines (\ion{Si}{2}{*},
\ion{C}{4}, and \ion{Al}{2}, as shown in Figures \ref{spec_vs_lon}
and \ref{spec_vs_lat}) and averaged the data in the wavelength regions
of 1310--1520 \AA, 1560--1660 \AA, and 1680--1710 \AA. Unless otherwise
specified, the FUV intensities in this paper were obtained in this
way.

We report how the emission lines contribute to the total intensity
of L-band. Edelstein et al. (2011, in preparation) showed that, on
average, all atomic emission lines and H$_{2}$ fluorescence lines
contribute about 5.1\% and 2.9\% (respectively) to the intensity of
SPEAR/FIMS L-band. Our independent modeling using spectral fits indicated
that the maximum contribution of emission lines is about 10\%. Notably,
the present results are not significantly different from results estimated
using the whole L-band, which implies that the emission lines do not
play a significant role in analyzing the general properties of the
FUV continuum background. We therefore use the terms ``diffuse FUV
background'' and ``FUV continuum background'' interchangeably. The
contribution of two-photon emission to the diffuse FUV continuum background
is discussed in \S\ref{sub:two-photon}.

\subsection{Instrumental Background\label{sub:Instrumental-Background}}

The SPEAR/FIMS data are affected by instrumental (or detector) background
due to cosmic rays, radioactive decay within the detector, and thermal
charged particles entering the instrument. These background sources
are relatively constant in time and uniform across the face of the
detectors, implying that the instrumental background is spectrally
flat intrinsically \citep{Korpela2006,Kregenow2007}.

Since the instrumental background is independent of the size of the
shutter aperture while the astrophysical signal is not, the background
can be determined by comparing the data observed with both 100\% and
10\% shutter apertures toward the same sky. We thus determine the
instrumental background rate assuming that the intensities measured
with 100\% and 10\% shutter apertures are $I_{100\%}=I_{{\rm sky}}+I_{{\rm det}}$
and $I_{10\%}=aI_{{\rm sky}}+I_{{\rm det}}=aI_{{\rm 100\%}}+(1-a)I_{{\rm det}}$,
respectively. Here, $a$, $I_{{\rm sky}}$, and $I_{{\rm det}}$ denote
the scale factor between two shutter apertures, the sky intensity
measured with 100\% shutter aperture, and the detector background,
respectively.

\begin{figure}[t]
\begin{centering}
\includegraphics[scale=0.55]{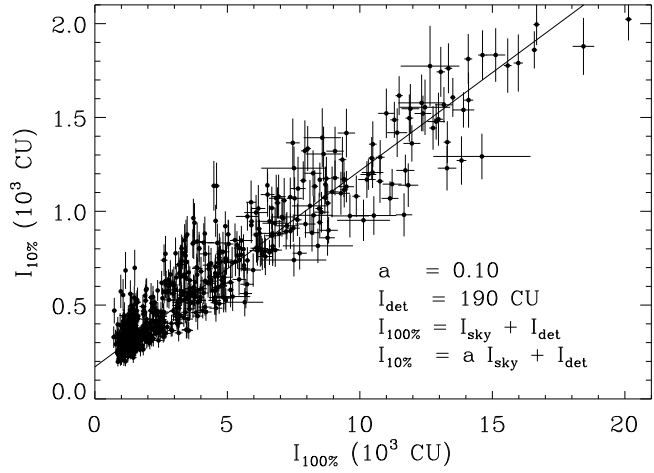}
\par\end{centering}

\caption{\label{compare_100_10}Pixel by pixel comparison of the 100\% and
10\% shutter aperture data. By fitting the data with a linear function,
the scale factor for the 10\% shutter aperture data and detector background
rate are estimated. The comparison is made with the $N_{{\rm side}}=64$
maps. Here, CU represents the continuum unit (photons cm$^{-2}$ s$^{-1}$
\AA$^{-1}$ sr$^{-1}$ CU).}
\end{figure}

We first estimate the detector background intensity averaged over
the whole L-band instead of applying the method to every spectral
bin, and then calculate the instrumental background spectrum in terms
of CU. We note that each pixel of a \emph{HEALpix} map is hierarchically
subdivided into many smaller pixels as the resolution parameter $N_{{\rm side}}$
increases. Some of the subdivided pixels from a non-zero-exposure
pixel at $N_{{\rm side}}=64$ can have zero exposures, especially
when the pixel is located at the boundary of each orbital scan. Hence,
when comparing the 100\% and 10\% maps at $N_{{\rm side}}=64$, two
coinciding pixels may not have the same sky coverage at higher resolutions.
This discrepancy can produce a large variation when comparing 100\%
and 10\% maps with $N_{{\rm side}}=64$. Thus, we only used pixels
($N_{{\rm side}}=64$) with 100\% sky coverage at the level of $N_{{\rm side}}=256$,
i.e., those pixels whose subdivided pixels at the level of $N_{{\rm side}}=256$
all have non-zero exposures. As shown in Figure \ref{compare_100_10},
over the whole L-band, the averaged intensity of instrumental background
that best fits the measurements is $\sim190\pm10$ CU. Similar results
that are consistent within $\sim10$\% were obtained using the higher
or lower resolution parameters. In the fit, we fully adopted errors
in both axes and found the best-fit linear straight line between the
two intensities \citep{Press92}. Because the noise-property of the
SPEAR/FIMS data is dominated by photon-noise, we used only photon-noise
in the present study. The estimated instrumental background was also
justified by an arbitrary increase or decrease of the background intensity
and through visual inspection of the combined map. Increasing or decreasing
$I_{{\rm det}}$ by more than $\sim$ 50 CU caused the observed sky
regions with the 10\% aperture to appear unrealistically bright or
faint in the combined map. The scale factor between the two shutter
apertures is $\sim0.10\pm0.01$ as originally designed.

\begin{figure}[t]
\begin{centering}
\includegraphics[scale=0.55]{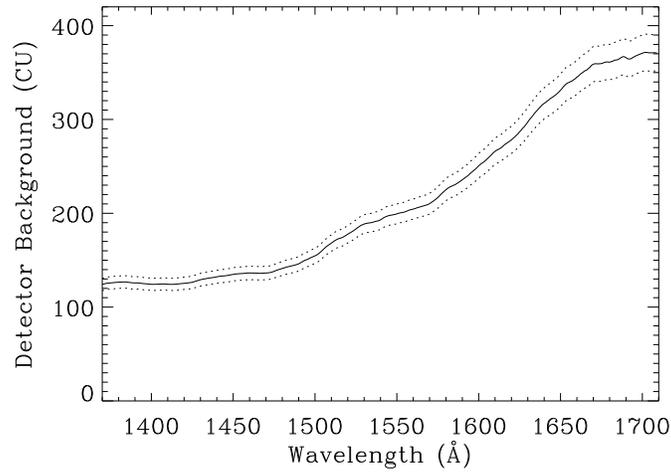} 
\par\end{centering}

\caption{\label{detector_spec}Instrumental background spectrum. The spectrum
is essentially the inverse of the effective area curves, as shown
in \citet{Edelstein06b}. An $1\sigma$ error range for the average
of detector background value is also shown with dotted lines.}
\end{figure}

Denoting the instrumental background rate by $R_{{\rm det}}$ counts
s$^{-1}$ \AA$^{-1}$ in the detector space, the instrumental background
spectrum $I_{{\rm det}}(\lambda_{i})$ in units of CU in an $i^{{\rm th}}$
spectral bin is given by

\begin{equation}
I_{{\rm det}}(\lambda_{i})=\frac{R_{{\rm det}}\Delta T\Delta\lambda}{A(\lambda_{i})\Delta\Omega\Delta T\Delta\lambda}=\frac{R_{{\rm det}}}{A(\lambda_{i})\Delta\Omega},\label{eq:1}
\end{equation}
where $A(\lambda_{i})$, $\Delta\Omega$, $\Delta\lambda$, and $\Delta T$
denote the effective area at the wavelength $\lambda_{i}$, solid
angle, wavelength binsize, and exposure time, respectively. Applying
the maximum likelihood method to Poisson data, the mean instrumental
background intensity over the L-band is given by

\begin{equation}
I_{{\rm det}}=\frac{\sum_{i}R_{{\rm det}}\Delta T\Delta\lambda}{\sum_{i}A(\lambda_{i})\Delta\Omega\Delta T\Delta\lambda}=\frac{R_{{\rm det}}N_{{\rm bin}}}{\sum_{i}A(\lambda_{i})\Delta\Omega},\label{eq:2}
\end{equation}
 where $N_{{\rm bin}}$ is the number of wavelength bins. We should
note here that $I_{{\rm det}}\neq\sum_{i}I_{{\rm det}}(\lambda_{i})/N_{{\rm bin}}$.
Thus the instrumental background spectrum is obtained by

\begin{equation}
I_{{\rm det}}(\lambda_{i})=I_{{\rm det}}\frac{\sum_{j}A(\lambda_{j})}{A(\lambda_{i})N_{{\rm bin}}},\label{eq:3}
\end{equation}
and is shown in Figure \ref{detector_spec}. In the figure, $1\sigma$
error limits are also shown in dotted lines. The spectrum of the instrumental
background is inversely proportional to the effective area.

Using the solid angle of L-band $\Delta\Omega$ of $\sim1.6\times10^{-4}$
sr and average effective area $\Sigma_{j}A(\lambda_{j})/N_{{\rm bin}}$
of $\sim0.19$ cm$^{-2}$, we obtain the detector background rate
$R_{{\rm det}}=I_{{\rm det}}\Sigma_{j}A(\lambda_{j})\Delta\Omega/N_{{\rm bin}}\sim0.006$
counts s$^{-1}$ \AA$^{-1}$. The instrumental background was independently
estimated to be 0.02--0.04 counts s$^{-1}$\AA$^{-1}$ by summing
many shutter-closed dark exposures of 42 ks from 250 orbits \citep{Edelstein06b}.
\citet{Lee06} adopted the same method and found a bit weaker detector
dark background of about 0.01 counts s$^{-1}$\AA$^{-1}$. We note
that the instrumental background rate estimated in the present study
is even smaller than the value obtained by \citet{Lee06}. The discrepancy
is attributable to the light leakage in shutter-closed dark exposures.

\subsection{Basic Statistics\label{sub:Statistics}}

\begin{figure}[t]
\begin{centering}
\includegraphics[scale=0.65]{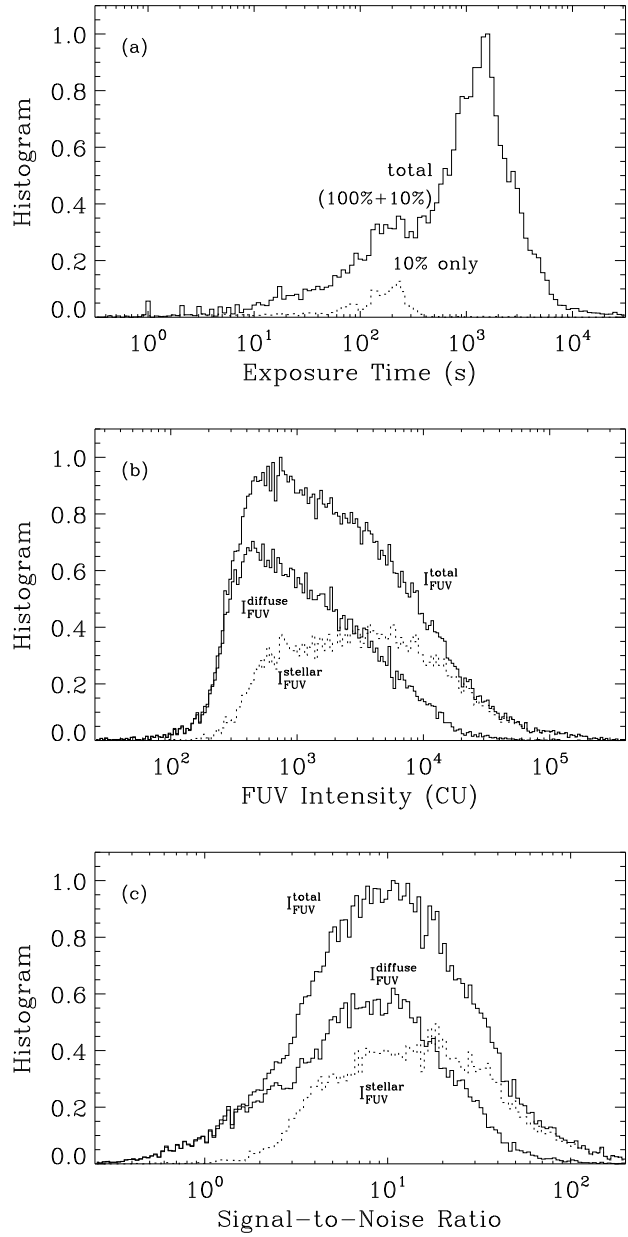}
\par\end{centering}

\caption{\label{histograms}Histograms of (a) exposure time (s), (b) the FUV
intensity (photons s$^{-1}$ cm$^{-1}$ sr$^{-1}$ \AA$^{-1}$, CU)
over the observed sky, and (c) signal-to-noise ratio with bins scaled
logarithmically. The histograms are made with about $1^{\circ}$ pixels
($N_{{\rm side}}=64$). A background of 190 CU was removed from the
data to take into account the instrumental dark background.}
\end{figure}

The histogram of the sky exposure is shown in Figure \ref{histograms}(a)
on a logarithmic scale. The dotted line in the figure shows the exposure
distribution for the data observed with the 10\% shutter aperture
only. Figure \ref{histograms}(b) shows the intensity distributions
of the observed total FUV intensity ($I_{{\rm FUV}}^{{\rm total}}$)
including both diffuse ($I_{{\rm FUV}}^{{\rm diffuse}}$) and direct
stellar ($I_{{\rm FUV}}^{{\rm stellar}}$) intensities, and the diffuse
FUV background only. The FUV background map has median and mean intensities
of $\sim1250$ and of $\sim3170$ CU, respectively. The reason for
the difference between the median and mean values are obvious from
the asymmetry of the intensity histogram. The difference in the median
value from \citet{Edelstein06a} is mainly due to subtraction of the
instrumental background and inclusion of more orbits. The total FUV
intensity map has median and mean intensities of $\sim1610$ and $\sim7080$
CU, respectively. The distribution of signal-to-noise ratio is also
shown in Figure \ref{histograms}(c). For $\sim1^{\circ}$ ($N_{{\rm side}}=64)$
pixels, about 84\% and 87\% of pixels have significance of $3\sigma$
or higher from the diffuse FUV sky and the total sky, respectively.
Mean and median signal-to-noise ratios of $I_{{\rm FUV}}^{{\rm diffuse}}$
are $\sim13$ and $\sim10$, respectively, and the ratios of $I_{{\rm FUV}}^{{\rm total}}$
are $\sim17$ and $\sim10$, respectively. The statistical properties
of the FUV intensities estimated over the whole L-band including the
strong emission line regions are not significantly different.

\section{FUV Intensity}

\subsection{Morphology}

\begin{figure*}[t]
\begin{centering}
\includegraphics[angle=90,scale=0.5]{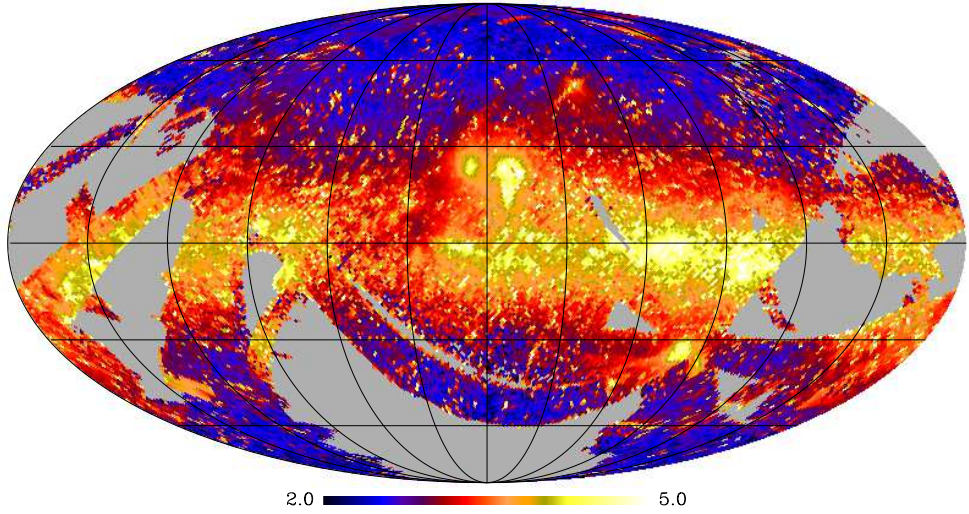}
\par\end{centering}

\bigskip{}

\begin{centering}
\includegraphics[scale=0.5,angle=90]{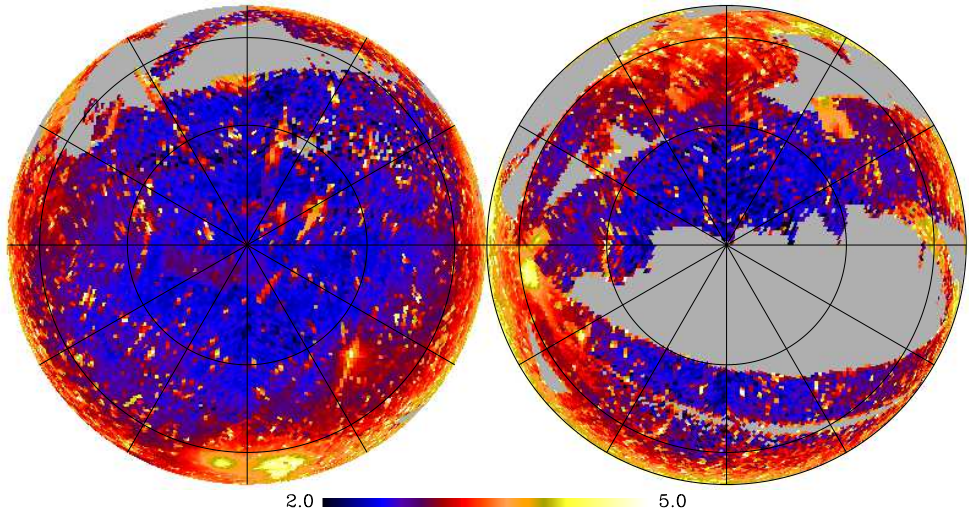}
\par\end{centering}

\begin{centering}
\medskip{}

\par\end{centering}

\caption{\label{isrf_map}(a) Mollweide and (b) orthographic projections of
the combined total (diffuse + direct stellar) FUV map observed with
SPEAR/FIMS, after removal of the instrumental background. The left
and right sides of orthographic projections are centered at the northern
and southern Galactic polar caps, respectively, and their longitude
increases clockwise and counter-clockwise, respectively. In both projections,
$l=0^{\circ}$ is at the six o'clock position. The intensity scales
are logarithmic across the color bars.}
\end{figure*}

The maps of the total intensity ($I_{{\rm FUV}}^{{\rm total}}$) and
the diffuse background intensity ($I_{{\rm FUV}}^{{\rm diffuse}}$)
observed with SPEAR/FIMS are shown in Figure \ref{isrf_map} and \ref{diffuse_map}
together with the signal-to-noise ratio maps. As observed by \citet{Henry77b}
and \citet{GON80}, the sky is strongly anisotropic in FUV. Not only
is there a strong variation with Galactic latitude, but also with
longitude. For example, much more FUV radiation is observed between
longitude $180^{\circ}$ and $360^{\circ}$ than between $0^{\circ}$
and $180^{\circ}$ in the Galactic plane. Gould's belt, inclined at
$\sim18^{\circ}$ to the Galactic plane with the direction of tilt
toward Orion, shows up well in Figure \ref{isrf_map} (\citealp{Stothers74};
\citealp{Elias06} and references therein).

Obviously, the diffuse FUV background seems to follow direct starlight
in general. However, it will be shown in \S3.3 and \S5 that $I_{{\rm FUV}}^{{\rm diffuse}}$
correlates better with dust, traced by 100 $\mu$m emission, and \ion{H}{1}
column density rather than with the direct stellar intensity. The
largest diffuse intensity is toward the Galactic plane and other regions
where bright early-type stars coexist with suitable columns of interstellar
dust, such as the obvious feature of the Sco-Cen association and the
Vela region. We note that the sky is brightest at the Vela region
in both the total and diffuse background intensities.

We confirm the existence of a significant depression in both the diffuse
and total FUV maps at high latitudes above less intense segments of
the Galactic plane from $l=20^{\circ}$ to $l=60^{\circ}$, as noted
in \citet{SCH2001} and \citet{GON80}. \citet{GON80} showed that
this direction corresponds to one of the most heavily reddened directions
in the sky by comparing the FUV map with the reddening data of \citet{Nandy78}.
We also confirmed the strong reddening toward this direction with
the more recent results given by \citet{Arenou92}, \citet{Chen98},
and \citet{Joshi05}. The dark region in $15^{\circ}<l<40^{\circ}$
and $-6^{\circ}<b<+20^{\circ}$ is in fact a molecular cloud referred
to as Aquila Rift, which emits strong CO line emission and has a maximum
extinction of $A_{{\rm V}}\sim3$ \citep{Dame1985,Straizys2003}.

As can be seen in the orthographic projections of Figure \ref{diffuse_map}(b),
the map reveals lots of detectable structures everywhere in the sky.
A full examination of the rich detail in the diffuse FUV map observed
by SPEAR/FIMS is beyond the scope of this paper. However, it would
be worthwhile to note that a large structure centered at $(l,b)\sim(316^{\circ},51^{\circ})$
is a feature produced by dust-scattered off starlight from a B1III/IV
+ B2V spectroscopic binary $\alpha$ Vir with a distance of 87 pc
\citep{Park2010}. An \ion{H}{2} region surrounding the star has
been detected by \citet{Reynolds85}.

\subsection{Dependencies on Galactic Coordinates}

Figures \ref{intensity_vs_lon} and \ref{intensity_vs_lat} examine
the distribution of $I_{{\rm FUV}}^{{\rm diffuse}}$ as a function
of Galactic coordinates. The top panel of each figure plots a two-dimensional
histogram in the space of intensity versus coordinates, while the
bottom panel shows the median and the standard deviation of intensities
from the median value within each of coordinate bins. The sharp cutoff
at low intensities in Figure \ref{histograms}(b) and the shape of
the latitude profile of the diffuse FUV emission in Figure \ref{intensity_vs_lat}
resemble the corresponding features shown in the diffuse H$\alpha$
emission (Figures 10 and 11, respectively, in \citealt{Haffner03}).
These similarities between the diffuse FUV and H$\alpha$ emissions
may be explained to some extent by a simple plane-parallel model.
In the idealized model, the cosecant effect is removed by plotting
$I_{{\rm FUV}}^{{\rm diffuse}}\sin|b|$ versus $\sin|b|$. This is
examined in Figure \ref{isinb_vs_sinb}. The top panel again plots
a two-dimensional histogram and the bottom panel shows the median
and standard deviation, with the trend for the northern Galactic hemisphere
split from that of the southern hemisphere. It should be noted, however,
that the plane-parallel model turns out to be only approximate in
describing the diffuse FUV background, as is also the case for the
\ion{H}{1} gas and the WIM \citep{Dickey90,Haffner03}. The horizontal
line in Figure \ref{isinb_vs_sinb} delineates the overall median
value of $I_{{\rm FUV}}^{{\rm diffuse}}\sin|b|=525.4$ CU. The decrease
in $I_{{\rm FUV}}^{{\rm diffuse}}\sin|b|$ at $\sin|b|<0.05$ (i.e.,
$|b|<3^{\circ}$) is attributed to a strong dust extinction in the
Galactic disk.

The diffuse FUV intensity decreases from $\sin|b|=0.2$ through 1.0
in both Galactic hemispheres. Similar effects have been observed in
the H$\alpha$ and \ion{H}{1} distributions, and the departure from
the $\csc|b|$ law was attributed at least in some extent to the presence
of the Local Bubble \citep{Cox87}, a localized region of very low-density
gas around the Sun, as pointed out by \citet{Dickey90} for \ion{H}{1}.
Since the Local Bubble is believed to be elongated perpendicular to
the Galactic plane \citep{Lallement03}, a larger fraction of column
density of the ISM is carved out at high latitudes, yielding the apparent
departure from the $\csc|b|$ law. However, the main reason for the
departure in the diffuse FUV background is due to the larger contribution
of dust-scattered FUV radiation at low latitude, as shown for the
H$\alpha$ radiation by \citet{Wood99}.

\subsection{Comparison with Stellar Radiation Fields}

\citet{Henry77} showed that simple addition of the predicted UV light
from all of the stars listed in common star catalogs provides an estimate
of the expected ISRF in the UV band which is in good agreement with
the more complex models \citep[see also, ][]{Henry77b}. Thus, we
made a model of direct starlight using the \emph{Hipparcos} catalog
\citep{Perryman97,vanLeeuwen2007} for stellar location, distance,
spectral type, and brightness. The \emph{Hipparcos} catalog has a
limiting magnitude of $V\sim12.4$ and was chosen for the calculation
of three-dimensional locations of stars. The model is basically the
same as in \citet{Henry93}, \citet{Henry02}, and \citet{Sujatha2004}.

Based on the spectral type of each star, we derived a temperature
and effective gravity using the tables from \citet{Straizys81} and
calculated a spectral energy distribution for each star with a new
grid of Kurucz models \citep{Castelli03}, rather than using the original
grid of Kurucz models. The intrinsic FUV luminosity of each star was
obtained by scaling the model flux with the factor derived from the
formula, as in \citet{Sujatha2004}:
\begin{equation}
\frac{4\pi R^{2}}{4\pi d^{2}}F_{V}e^{-\tau_{V}}=(f_{V})_{0}10^{-V/2.5},\label{eq:scaling_r2}
\end{equation}
where $R^{2}$ is the scaling factor corresponding to the stellar
radius, $d$ the distance from the Sun, $(f_{V})_{0}=3.631\times10^{-9}$
ergs cm$^{-2}$ sec$^{-1}$ $\textrm{\AA}^{-1}$ the absolute visual
magnitude corresponding to zero magnitude \citep{BCP1998}, $F_{V}$
the model flux (ergs cm$^{-2}$ sec$^{-1}$ $\textrm{\AA}^{-1}$)
at $V$ band filter, $V$ the observed Johnson magnitude, and $\tau_{V}$
the optical depth at $V$ band filter which can be estimated from
$\tau_{V}=A_{V}/1.0863$. Here, the absolute visual magnitude is calculated
from the standard relation $A_{V}=R_{V}E(B-V)$ and $R_{V}=3.1$,
and the color excess $E(B-V)$ obtained from the observed $B-V$ and
the theoretical $B-V$ as derived from the Kurucz model spectrum.
In many cases multiple or complex spectral types were associated with
a single star; in such cases, we used the first listed spectral type.
We then integrated the stellar flux attenuated by the dust distribution
along given lines of sight. The resulting stellar fluxes in every
pixel are summed up to yield the ``stellar equivalent diffuse intensity
(SEDI)'' as defined by \citet{HBM91}: 
\begin{equation}
I_{{\rm SEDI}}^{{\rm Hipparcos}}=\sum_{i=1}^{{\rm all\, stars}}\frac{L_{i}}{4\pi d_{i}^{2}}\frac{1}{\Omega_{{\rm Healpix}}}e^{-\tau_{i}},\label{eq:sedi_eq}
\end{equation}
where $L_{i}$ is the luminosity of each star at the 1370--1710 \AA\
band (excluding the strongest emission line regions) in photons s$^{-1}$
\AA$^{-1}$, $d_{i}$ is the distance to the star in centimeters,
$\tau_{i}$ is the optical depth to each star at the waveband, and
$\Omega_{{\rm Healpix}}=4\pi/N_{{\rm pix}}$ is the solid angle (steradians)
of a pixel. The optical depth at the FUV wavelength is calculated
by $\tau_{{\rm FUV}}=2.86(\sigma_{{\rm FUV}}/\sigma_{V})E(B-V)$ where
$\sigma_{{\rm FUV}}$ and $\sigma_{V}$ are the dust extinction cross-sections
at the FUV wavelength and the V band filter for the average Milky
Way dust with $R_{V}=3.1$ as from \citet{WD2001} and \citet{DRA2003}.
We used the resolution parameter $N_{{\rm side}}=32$ corresponding
to an angular resolution of $\sim1.8^{\circ}$ and the number of pixels
$N_{{\rm pix}}=12,288$ in order to increase the number of pixels
including stars.

We also calculated the SEDI using the TD-1 stellar catalog \citep{Thompson1978}.
The TD-1 satellite performed the first UV all-sky survey and measured
the absolute UV fluxes of point sources down to the tenth visual magnitude
for unreddened early B stars. The TD-1 catalog presents the stellar
fluxes in four bands (each 310--330 \AA\ wide, centered at 1565,
1965, 2365, and 2740 \AA) for 31,215 stars. The SEDI of the TD-1
stars is defined by $I_{{\rm SEDI}}^{{\rm TD-1}}=\Sigma_{i}F_{i}/\Omega_{{\rm Healpix}}$
using the stellar fluxes $F_{i}$ at the 1565 \AA\ band. \citet{SCH2001}
also used the TD-1 star catalog to model the diffuse FUV background,
and they found that stars with 10 times higher flux than the TD-1
catalog cutoff contribute about 21 CU to the FUV background. Therefore,
the TD-1 catalog is complete enough for the analysis of the diffuse
FUV background.

First of all, the TD-1 SEDI is compared with the \emph{Hipparcos}
SEDI and the direct stellar intensity obtained from SPEAR/FIMS data
in Figures \ref{compare_stellar}(a) and (b), respectively. The \emph{Hipparcos}
SEDI predicts the TD-1 SEDI very well in the statistical sense. However,
the \emph{Hipparcos} SEDI is a bit higher than the TD-1 SEDI at intensities
lower than $\sim100$ CU, implying that the stars in the TD-1 catalog
are not complete down to the intensity level that was estimated with
the \emph{Hipparcos} catalog, although the incompleteness is not so
significant. Larger scatters in Figure \ref{compare_stellar}(b) than
those shown in Figure \ref{compare_stellar}(a) are mainly due to
the coarse algorithm used to identify stars in the SPEAR/FIMS data
analysis. However, the correlation between the TD-1 SEDI and the SPEAR/FIMS
$I_{{\rm FUV}}^{{\rm stellar}}$ is good enough to move into the following
arguments.

We note in Figures \ref{compare_stellar}(c) and (d) that $I_{{\rm FUV}}^{{\rm diffuse}}$
shows more or less a general correlation with the \emph{Hipparcos}
and TD-1 SEDIs. However, as noted by \citet{HBM91}, as long as one
avoids the brightest early-type stars in the sky by more than a few
degrees, there should not be a strong correlation between line-of-sight
starlight and line-of-sight dust illumination. In fact, we found that
the correlation with the direct stellar intensity is rather weaker
than that with the infrared 100 $\mu$m emission and $N_{{\rm HI}}$,
as discussed in \S5. It is also obvious that the $I_{{\rm FUV}}^{{\rm total}}$
is brighter than both SEDIs, as in Figures \ref{compare_stellar}(e)
and (f), especially at relatively low intensities $I_{{\rm FUV}}^{{\rm total}}\lesssim10,000$
CU. The deviation of $I_{{\rm FUV}}^{{\rm total}}$ from the \emph{Hipparcos}
and TD-1 SEDIs increases as the SEDIs decrease. This trend is a direct
result of the intensity distribution of $I_{{\rm FUV}}^{{\rm diffuse}}$
as can be seen in Figures \ref{compare_stellar}(c) and (d), in which
$I_{{\rm FUV}}^{{\rm diffuse}}$ shows no strong correlation with
the SEDIs. More specifically, within each ``pixel'' ($\sim1.8^{\circ}$),
there is an excess of diffuse FUV, which is above the estimate from
the stellar fluxes that are directly detected and/or modeled within
that pixel and does not correlate with the direct stellar flux in
the pixel. For low-intensity pixels, the excess of diffuse FUV would
mainly be due to dust-scattered light from stars that are not located
along the line of sight of the pixels. The conclusion is discussed
much more clearly in \S4.2, which attempts to link a harder spectrum
at low intensities to this excess. In fact, luminous UV stars produce
extended dust-scattered halos that are far greater than the volume
sample of the pixels here \citep[e.g., ][]{MurthyHenry2011}. We thus
conclude that simply adding stellar flux underestimates ISRF, especially
at the low intensity of $I_{{\rm ISRF}}\lesssim10,000$ CU, against
the results of \citet{Henry77} and \citet{Henry77b}. Three-dimensional
radiative transfer models may help to assert what portion of the ISRF
is the result of the dust-scattered FUV background.

It would be worthwhile to note that the contribution of unresolved
faint stars to our diffuse FUV background map is negligible. The diffuse
background or the total intensity would be correlated at low intensity
pixels with the model SEDI if this was the case. However, it is obvious
in Figure \ref{compare_stellar}(c) and (d) that this is not the case.
The difference of more than an order between $I_{{\rm SEDI}}$ and
the mean value of $I_{{\rm FUV}}^{{\rm total}}$ at $I_{{\rm SEDI}}\sim50$
CU in Figures \ref{compare_stellar}(e) and (f) is unlikely due to
the contribution from unresolved stars. We thus conclude that residual
starlight is not a significant contributor to the diffuse FUV background.
The contribution from faint stars seen in Figure \ref{compare_stellar}(a)
is much smaller than the difference between $I_{{\rm SEDI}}$ and
$I_{{\rm FUV}}^{{\rm total}}$ in Figures \ref{compare_stellar}(e)
and (f). Many studies \citep[i.e., ][]{HBM91,Henry02} have estimated
the contribution of unresolved faint stars with several methods in
their analysis of the FUV background data, and they all have concluded
that the contribution is not significant. Additional support for this
conclusion will be provided in \S4 from an argument based on hardness
ratio. We also note that the large difference between the diffuse
FUV intensity and the model SEDIs cannot be due to H$_{2}$ fluorescence
lines and two-photon continuum emission, which may account for $\sim2.9$\%
(Edelstein et al. 2011, in preparation) and $\sim4-9$\% (see \S\ref{sub:two-photon})
of the diffuse intensity, respectively.

Zodiacal light, sunlight scattered by interplanetary dust, is also
negligible at these wavelengths since the solar G0 spectrum falls
steeply at the FUV wavelengths \citep{Tennyson88}, and we have confirmed
that our data show no systematic trend with ecliptic latitude. Scattering
of diffuse Ly$\alpha$ radiation (geocoronal/interplanetary) as determined
from laboratory measurements is also negligible. The insignificance
of unresolved faint stars, zodiacal light, and Ly$\alpha$ scattering
is also justified from the agreement of the so-called ``isotropic
component'' estimated in \S5 with the previous measurements.

\section{FUV Spectrum}

\subsection{Spectral Variations}

The direct stellar and diffuse background intensities should show
strong anisotropy not only in intensity but also in spectral shape.
In Figures \ref{spec_vs_lon} and \ref{spec_vs_lat}, the average
spectra of the diffuse background, direct starlight, and total radiation
in various Galactic longitude and latitude ranges, respectively, are
shown. The direct stellar spectra were obtained by subtracting the
diffuse background spectra from the total spectra in each coordinate
range. We note that the FUV intensities are strongest in $240^{\circ}<l<270^{\circ}$
and in $-10^{\circ}<b<0^{\circ}$, where the Vela region is located,
as was seen in Figure \ref{diffuse_map}. The Vela region is the brightest
not only in emission lines due to the Vela supernova remnant but also
in the FUV continuum \citep[cf. ][]{Nishikida06}.

The average background intensity is about the same as the direct stellar
intensity in all coordinate ranges shown in the figures, and the total
intensity is approximately two times higher than the direct stellar
intensity. Ignoring the latitude range $b<-10^{\circ}$, where exposures
and signal-to-noise ratios are small, the intensity ratio of the diffuse
background to the direct starlight decreases with Galactic latitude,
and then increases at the highest latitude range of $70^{\circ}<b<90^{\circ}$.
The trend is easily understandable, at least qualitatively, by adopting
the results from a simplified spherical dust model with an embedded
source. The scattered to absorbed stellar flux ratio in the model
increases monotonically with optical depth and can easily be of the
order of unity or higher for optical depth of $\tau=1-2$ \citep{Witt1982}.
The optical depth of $\tau\sim1.26$ at 1550 \AA\ is estimated at
$N_{{\rm HI}}=1\times10^{21}$ cm$^{-2}$ using the standard gas-to-dust
ratio \citep{BOH1978} and the Milky Way dust extinction cross-section
\citep{WD2001,DRA2003}. Therefore, the approximate equality between
the diffuse and direct stellar intensities at the Galactic plane is
obtained. The decrease of the diffuse to direct stellar intensity
ratio with Galactic latitude is also explained since the optical depth
decreases with latitude. However, at the highest Galactic latitude
range $70^{\circ}<b<90^{\circ}$, the ``isotropic component'' of $\sim300$
CU discussed in \S6 contributes significantly to the diffuse background
in addition to the scattered radiation of Galactic starlight in the
latitude range, and thus causes the ratio to increase.

A weak rise in the diffuse spectra longward of about 1550 \AA\ is
shown in most of the latitudes in Figure \ref{spec_vs_lat}, although
the FUV continuum spectra are flat in general. However, the dependence
of the rise on the Galactic latitude is not clear. The rise is also
noticeable at $120^{\circ}<l<180^{\circ}$ in Figure \ref{spec_vs_lon},
where most of observations were performed at high latitudes as seen
in Figure \ref{exposure_map}. It should be noted as well that the
softer the diffuse spectrum is, the softer the direct stellar spectrum
also is. A much sharper rise longward of $\sim1550$ \AA\ than that
found here was obtained by \citet{Henry80}, in which they concluded
that the rise was due to very large numbers of faint stars contributing
to the ISRF. Later, \citet{Henry02} reanalyzed the data and attributed
the rise to an actual rise in the interstellar grain albedo. However,
these features may not be due to a single cause. Firstly, a stronger
extinction at a shorter wavelength leads to the reddening or softening
not only of direct stellar, but also of diffuse spectra, unless the
dust albedo decreases with wavelength as will be shown in \S6 (Figure
\ref{aeffect}). Secondly, if there is no significant contribution
from bright early-type stars near a given sight line, faint late-type
stars should contribute, at least in part, to the spectra in addition
to the dust-scattered light from distant early-type stars.

It should also, more importantly, be noted that the diffuse background
spectrum is always softer than the direct stellar spectrum. The rise
of the dust albedo with wavelength is in fact closely related to the
relative softness of the scattered to the direct stellar spectra,
as will be discussed in \S6. We also note that both the direct stellar
and diffuse spectra are harder, or bluer, in the Galactic longitude
range $240^{\circ}<l<270^{\circ}$, where the Vela region dominates
the radiation field, than other longitude ranges, implying the existence
of lots of hot stars and/or less extinction in the region. A relatively
sharp rise longward of about 1690 \AA\ at $|b|>50^{\circ}$ is also
found mostly in the diffuse background spectrum. However, the origin
of this rise is not clear at this time.

We now compare the spatially-averaged total (direct stellar + diffuse)
FUV spectrum observed with SPEAR/FIMS with the previous models and
observations in Figure \ref{isrf_compare}. It can be easily noted
that all previous results are higher, up to a factor of more than
3 for the Draine's model, than the SPEAR/FIMS data. In Figure \ref{spec_vs_lat},
the observed total spectrum is stronger only at $-10^{\circ}<b<0^{\circ}$
than the Draine's ``standard'' radiation field, while the intensities
in other latitude ranges are in general weaker. The relative weakness
of the observed data compared to the model ISRFs indicates that the
previous works are mostly biased toward bright regions in Galactic
plane. In Figure \ref{isrf_compare}, we also plotted the average
spectra of the diffuse background and the direct starlight. The figure
shows once again that the direct starlight and the diffuse background
contributes approximately the same amounts to the total spectrum on
average, implying underestimation of the ISRF when only direct starlights
are added.

The spectral shapes of the model ISRFs calculated by \citet{GON80}
and the Habing field are more or less similar to the SPEAR/FIMS data,
while the Draine's ``standard'' and Mathis background radiation fields
are a bit harder than that of SPEAR/FIMS. The observed spectrum of
\citet{Henry80} is softer than the SPEAR/FIMS data. We also note
that the diffuse background spectrum averaged over all the sky is
a bit softer than the stellar spectrum, and that discrete spectral
features such as atomic emission lines are mainly shown in the diffuse
spectrum only, as already shown in Figures \ref{spec_vs_lon} and
\ref{spec_vs_lat}. The relative softness or redness of the diffuse
background compared to the direct stellar spectrum is once again attributed
to rise of the dust-scattering albedo at a longer wavelength, as will
be discussed in \S6.

\subsection{Hardness Ratio}

In order to further investigate spectral variations of the diffuse
background, we plotted a map of the hardness ratio defined as the
ratio of average intensities at $\lambda\lambda1370-1520$ \AA\ and
at $\lambda\lambda1560-1710$ \AA\ (excluding $\lambda\lambda1660-1680$
\AA\ to avoid the \ion{Al}{2} emission line) in Figure \ref{hardness_map}.
In the figure, we selected a color scheme to represent the region
with relatively blue or hard spectra as blue colored. The spectral
hardness map shows highly patched structures and no clear trend along
with Galactic latitudes. Instead, we note that the sky can be largely
divided into two blocks mainly along Galactic longitudes. In fact,
we smoothed the hardness ratio map using a gaussian kernel with full
width at half maximum of $\sim3.5^{\circ}$ and saturated the color
range to clarify the division of the sky based on the hardness ratio.
A bit harder radiation is observed between longitude $180^{\circ}$
and $360^{\circ}$, where stronger radiation is found in Figure \ref{diffuse_map},
than between $0^{\circ}$ and $180^{\circ}$. Interestingly, we found
that the hardness ratio of the diffuse background follows in general
the spatial distribution of OB-type stars in the Galactic plane. In
Figure \ref{td1_stars}, we plotted maps for numbers of OB-type stars
and of A-type stars using the TD-1 stellar catalog. We note from the
figure that the Galactic plane can be divided into three regions:
$240^{\circ}<l<360^{\circ}$ where OB-type stars are predominantly
found, $30^{\circ}<l<90^{\circ}$ and $180^{\circ}<l<240^{\circ}$
where A-type stars are dominant, and $90^{\circ}<l<150^{\circ}$ where
both OB- and A-types are rare. In the Galactic longitudes where OB-type
stars are dominant, the hardness ratio is generally higher than the
other regions.

We also plotted the ratio versus Galactic longitude and latitude in
Figures \ref{hardness_vs_lon} and \ref{hardness_vs_lat}, respectively.
In Figure \ref{hardness_vs_lon}, the median value of the hardness
ratio at each longitude bin is generally lower in $0^{\circ}<l<180^{\circ}$
than in $180^{\circ}<l<360^{\circ}$. On the other hand, Figure \ref{hardness_vs_lat}
shows no clear trend along Galactic latitude. In order to confirm
the close relation between the hardness ratio and the number of OB-type
stars, we calculated the average hardness ratio and the number of
OB-type stars within each of the $3^{\circ}$ longitude bins, and
plotted the average hardness ratio versus the number of OB-type stars
in Figure \ref{hardness_obtype}. The figure shows a clear correlation
between the average hardness ratio and the number of OB-type stars.
We also note a dip at $120^{\circ}<l<180^{\circ}$ in Figure \ref{hardness_vs_lon}.
The dip does not seem to be an artifact caused by a relatively low
exposure at low latitudes in this longitude range. In the light of
the argument described in the previous paragraph, the softness in
the longitude region would be a direct consequence of no bright stars
in the range.

Figure \ref{hardness_vs_intensity} shows the hardness ratio versus
the diffuse FUV intensity. Rises of the hardness ratio at high and
low intensities are noted in the figure. To understand the trend,
we calculated the hardness ratio versus the direct stellar intensity
using the previous stellar model; the result is shown in Figure \ref{hardness_model}.
In the calculation of the hardness ratio, we incorporated the wavelength
dependence of dust extinction as well. We note that the model ratios
predict the observed values nicely at high intensities ($I_{{\rm FUV}}^{{\rm diffuse}}>2000$
CU), of which most of the radiation is from low Galactic latitudes.
Therefore, we attribute the rise of the hardness ratio at high intensities
($I_{{\rm FUV}}^{{\rm diffuse}}>2000$ CU) mostly to the spectral
hardness of stars that are located at low Galactic latitudes.

However, the slight rise of the hardness ratio at low intensities
($I_{{\rm FUV}}^{{\rm diffuse}}<2000$ CU) cannot be explained by
stellar types at high Galactic latitudes. One possible explanation
for this rise is the contribution of dust-scattered starlight from
earlier stars rather distant than nearby stars that are located along
the same line of sight for each pixel. This is supported by Figures
\ref{compare_stellar}(c) and (d), which show clear excesses of the
diffuse background above the nearby direct stellar intensity. In fact,
our preliminary results of the Monte Carlo simulation, which will
be presented in detail elsewhere, show that B-type stars are significant
sources of the diffuse FUV background even at the high Galactic latitudes
where most stars are A-types. \citet{Henry02} also noted that B-type
stars contribute the largest amount to the ISRF in his simplified
model. We thus conclude that both the rise of the hardness ratio and
the excess of the diffuse background over the direct stellar intensity
at low intensities are largely due to the contribution of B-type starlight
scattered off dust grains. However, at the lowest intensities of less
than a few hundred CU, the ``isotropic component'' would contribute
significantly. A detailed study on the relative significance between
the two components may need extensive radiative transfer models that
are beyond the scope of this paper.

\section{Comparison with various galactic quantities}

We now examine correlations of the observed FUV background with various
galactic quantities, not only with the \ion{H}{1} column density
and 100 $\mu$m emission but also with the H$\alpha$ and soft X-ray
(1/4 keV) emissions. Although extensive correlation studies between
the diffuse FUV background, the \ion{H}{1} column density, and 100
$\mu$m emission have been performed, no detailed correlation study
with other waveband observations has been done. The all-sky map of
\ion{H}{1} column density are from the Leiden/Argentine/Bonn (LAB)
Survey \citep{Kalberla2005}. The 100 $\mu$m and H$\alpha$ emission
maps are from \citet{Schlegel98} and \citet{Finkbeiner03}, respectively.
The soft X-ray (1/4 keV) map is from the \emph{ROSAT }All-Sky Survey
(RASS) map \citep{Snowden97}. The \ion{H}{1} data interpolated onto
a HEALPix projection were obtained from the Legacy Archive for Microwave
Background Data Analysis (LAMBDA).

Correlations between the diffuse FUV intensity and other quantities
are shown in Figure \ref{correlation} in log-log scales. Two-dimensional
histogram images and contours are shown in the figure instead of the
usual pixel-by-pixel comparison to clarify their correlation. Correlation
coefficients estimated in logarithmic scale are also shown in Figures
\ref{correlation}. It is also worthwhile to note that the correlation
coefficient between the diffuse FUV background and the direct stellar
radiation is $\sim0.46$, much less than those for $N_{{\rm HI}}$
and $I_{{\rm 100\mu m}}$. In Figure \ref{correlation_linear}, we
examine the correlations in linear-linear scales, especially at high
Galactic latitudes ($b>25^{\circ}$). The general similarity in the
latitude dependence of the data sets (the \ion{H}{1} column density,
100 $\mu$m, H$\alpha$, and FUV emissions) may dominate their correlations.
To reduce the latitude dependence, we therefore calculated the average
intensities multiplied by $\sin|b|$ within each of the $\Delta\sin|b|=0.02$
latitude intervals. The correlations between the quantities multiplied
by $\sin|b|$ are shown in Figure \ref{correlation_sinb}, in which
only the data sets observed at high Galactic latitudes ($|b|\ge30^{\circ}$)
were examined, in order to minimize the dust extinction effect at
low Galactic latitudes.$ $ The numbers outside and inside the parentheses
in Figure \ref{correlation_sinb} are correlation coefficients calculated
in linear and logarithmic scales, respectively. Here, we note that
Figure \ref{correlation_sinb} shows strong correlations between emission
tracers, implying that the correlations between the FUV background
and other emission tracers are not simply due to the latitude dependence
of the Galactic observables.

\subsection{Correlation with $N_{{\rm HI}}$ and 100$\mu$m emission}

Correlation plots of the diffuse FUV background with \ion{H}{1} column
density and 100 $\mu$m intensity are shown in Figures \ref{correlation}(a)
and (b), respectively. There is large scatter in the correlation plots.
We note that the diffuse FUV continuum intensity saturates at about
$N_{{\rm HI}}\sim10^{21}$ cm$^{-2}$, which was found for the first
time by \citet{HBM91}. The same effect of saturation is also shown
in comparison with the 100 $\mu$m intensity. This saturation at a
high intensity can be attributed to strong dust absorption of the
FUV radiation at low Galactic latitudes. Flattening at a low intensity,
which has been often attributed to extragalactic background, is also
seen in Figures \ref{correlation}(a) and (b).

As is obvious in the figures, the diffuse FUV intensity for a given
$N_{{\rm HI}}$ or 100 $\mu$m brightness varies more than by a factor
of 2--3, up to an order of magnitude, with directions in the sky.
\citet{WIT1997} argued that most of the observed variation is a result
of the spatially, severely anisotropic, ISRF. Since the diffuse FUV
background depends not only on the interstellar dust but also on the
in-situ stellar radiation field, the usual pixel-by-pixel comparison
might be affected by the strong anisotropic stellar radiation field.
We thus averaged the FUV data corresponding to a given \ion{H}{1}
column density or a given 100 $\mu$m brightness and compared the
average FUV intensity with $N_{{\rm HI}}$ and 100 $\mu$m brightness.
In this way, the effect of the spatially varying stellar radiation
field may be averaged, and only the effect due to \ion{H}{1} or dust
column density can be investigated. The average FUV intensity versus
$N_{{\rm HI}}$ and $I_{{\rm 100\mu m}}$ are shown in Figure \ref{correlation_linear},
together with the best regression lines. In the figure, the data used
in the fit are denoted by filled circles, and the data not used in
the fit by hollow circles. In the fit, we used only the data for $b>25^{\circ}$.
The best-fit regression lines are given by 
\begin{equation}
I_{{\rm FUV}}^{{\rm diffuse}}=(1.49\pm0.07)\times\frac{N_{{\rm HI}}}{10^{18}{\rm cm}^{-2}}+(271.2\pm26.7)\,\,{\rm CU,}\label{eq:fuv_nhi}
\end{equation}
and
\begin{equation}
I_{{\rm FUV}}^{{\rm diffuse}}=(158.3\pm11.7)\times\frac{I_{{\rm 100\mu m}}}{{\rm MJy/sr}}+(243.1\pm44.4)\,\,{\rm CU}\label{eq:fuv_ir100}
\end{equation}
for \ion{H}{1} column density and 100 $\mu$m emission, respectively.

\citet{PMB1980} reported variations in the slopes and brightness
axis intercepts corresponding to the various scans. As noted by \citet{Jakobsen84},
there is no unique, canonical relationship between background intensity
and a hydrogen column density that is obeyed everywhere in the sky.
Principally, this result is not surprising, since the detailed viewing
geometry and the illuminating UV stellar radiation field are highly
anisotropic, and several of the physical parameters, such as dust
properties, are expected to vary from region to region in the ISM.
However, it is obvious, as shown in Figure \ref{correlation}(a) and
(b), that there is a general correlation in the global scale between
the diffuse FUV background and both the neutral hydrogen column density
$N_{{\rm HI}}$ and IR 100 $\mu$m emission. The linear correlation
has been measured by many different investigators, and the results
published before 1991 were summarized in \citet{BOW1991} as $I_{{\rm FUV}}^{{\rm diffuse}}\simeq(0.3-2.5)\times(N_{{\rm HI}}/10^{18}{\rm cm}^{-2})+300$
CU, which is consistent with our result.

The flaring and large scatters shown in these linear-linear correlation
plots are in general due to large variations of the diffuse FUV intensity
as noted in the log-log correlation plots of Figure \ref{correlation}.
Note that the larger scatters are found at the higher mean intensities,
the property that will be discussed in more detail in \S6. We also
note a straight-line correlation for $I_{{\rm 100}\mu{\rm m}}<8$
MJy/sr in Figure \ref{correlation_linear}(b). \citet{Witt2008} found
that there was also a linear correlation between the optical surface
brightness of high-latitude cirrus clouds and their 100 $\mu$m intensity
at least up to $\sim$ 8 MJy/sr. This is the range in which these
objects remain optically thin and single scattering by dust grains
is dominant. The systematic deviation from a straight-line correlation
for $I_{{\rm 100}\mu{\rm m}}>8$ MJy/sr in Figure \ref{correlation_linear}(b)
would be due to the fact that at the higher $I_{{\rm 100}\mu{\rm m}}$
values we are entering lower galactic latitudes where the effects
of the strongly forward-directed scattering phase function of dust
produces strongly enhanced FUV intensities compared to those produced
by the scattering at mostly larger scattering angles at higher latitudes,
where the 100 $\mu$m intensities are generally low.

\subsection{Correlation with the Diffuse H$\alpha$ Emission}

In linear scale, the best regression line between $I_{{\rm FUV}}^{{\rm diffuse}}$
and $I_{{\rm H}\alpha}$ is given by 
\begin{equation}
I_{{\rm FUV}}^{{\rm diffuse}}=(456.0\pm23.5)\times\frac{I_{{\rm H}\alpha}}{{\rm R}}+(309.2\pm31.4)\,\,{\rm CU,}\label{eq:fuv_halpha}
\end{equation}
where R denotes Rayleigh (1 R $=10^{6}/4\pi$ photons cm$^{-2}$ s$^{-1}$
sr$^{-1}$). The average $I_{{\rm FUV}}^{{\rm diffuse}}$ versus $I_{{\rm H\alpha}}$
and best-fit line are shown in Figure \ref{correlation_linear}(d).
Only the data for $b>25^{\circ}$ were used in the fit.

We note a very nice, unexpected, correlation between the FUV background
and the diffuse H$\alpha$ emission, some of which is known to be
emitted from the WIM. As can be seen in Figures \ref{correlation}
and \ref{correlation_linear}, the low intensity behavior in the correlation
of $I_{{\rm FUV}}^{{\rm diffuse}}$ with $I_{{\rm H}\alpha}$ is slightly
different from in the correlation with either $N_{{\rm HI}}$ or $I_{{\rm 100}\mu{\rm m}}$.
Figure \ref{correlation} shows that the correlation between $I_{{\rm FUV}}^{{\rm diffuse}}$
and $I_{{\rm H}\alpha}$ is overall straight down to the lowest intensities,
while other relations become flat or curved a bit upward. Moreover,
in Figure \ref{correlation_linear}, at the low intensity of $I_{{\rm FUV}}^{{\rm diffuse}}$
($\sim400-500$ CU), the diffuse FUV intensity seems to decouple from
the 100 $\mu$m intensity (and a bit from the \ion{H}{1} column density),
while the correlation between $I_{{\rm FUV}}^{{\rm diffuse}}$ and
$I_{{\rm H}\alpha}$ is still retained. These results seem to indicate
that the FUV background has a stronger correlation with the H$\alpha$
emission, at least qualitatively, than with the other emission tracers,
although the correlation coefficient between $I_{{\rm FUV}}^{{\rm diffuse}}$
and $I_{{\rm H}\alpha}$ is not significantly higher than the other
correlation coefficients (Figure \ref{correlation}).

Figure \ref{correlation_sinb} shows the results after the removal
of inverse-$\sin|b|$ dependence due to the plane-parallel medium,
offering clearer evidence that the correlation between $I_{{\rm FUV}}^{{\rm diffuse}}$
and $I_{{\rm H}\alpha}$ is stronger than other correlations. Since
the far-IR emission from interstellar dust correlates well with neutral
hydrogen emission \citep[e.g., ][]{Boulanger1988}, the diffuse FUV
continuum correlates well with both $I_{{\rm 100}\mu{\rm m}}$ and
$N_{{\rm HI}}$ at the same significance, as indicated by the correlation
coefficients in Figure \ref{correlation_sinb}(a) and (b). If the
correlation between the FUV and H$\alpha$ backgrounds is simply due
to the co-existence of dust and H$\alpha$-emitting gas, the correlation
of $I_{{\rm FUV}}^{{\rm diffuse}}$ and $I_{{\rm H}\alpha}$ would
have the same significance as with either $I_{{\rm 100}\mu{\rm m}}$
or $N_{{\rm HI}}$. On the other hand, the correlation with the H$\alpha$
background would be enhanced when the physical origin of a large portion
of the H$\alpha$ background is analogous to the FUV background. We
may then conjecture that the better correlation of the FUV with H$\alpha$
intensities is attributed to the similarity of the physical origins
of the diffuse FUV and H$\alpha$ emissions. The fact that the correlation
between $I_{{\rm FUV}}^{{\rm diffuse}}$ and $I_{{\rm H}\alpha}$
appears to exist not only on a large scale but also in smaller-cloud
scales supports the idea wherein their physical origins are at least
in part related. We discuss this connection briefly in \S6.1 and
in more detail in a paper in preparation.

We here consider several possible explanations for the relationship
between the FUV continuum and H$\alpha$ emissions, assuming that
the physical origins of the backgrounds are indeed closely related
with each other. First of all, a large fraction of the correlation
between $I_{{\rm FUV}}^{{\rm diffuse}}$ and $I_{{\rm H}\alpha}$
may be caused by the two-photon continuum emission in FUV wavelengths,
which originates from the WIM. However, the contribution of two-photon
emission to the FUV continuum ($I_{2\gamma}/I_{{\rm H}\alpha}\approx4-9$\%,
as found in \S\ref{sub:two-photon}) is not large enough to explain
the correlation relation Eq. (\ref{eq:fuv_halpha}).

\citet{Wood99} estimated that the scattered H$\alpha$ intensity
at high Galactic latitudes is $\sim5-20$\% of the total H$\alpha$
intensity. Recently, \citet{Mattila07} and \citet{Lehtinen2010}
found that in some dust clouds, the H$\alpha$ radiation is mostly
due to scattering by dust grains that are illuminated by a H$\alpha$-emitting
source off the line of sight. \citet{Witt2010} showed that a substantial
fraction of the diffuse high-latitude H$\alpha$ background is caused
by interstellar dust scattering of H$\alpha$ photons that originate
elsewhere in the Galaxy. In particular, they found many extended regions
where the fraction of the scattered H$\alpha$ intensity is of an
order of 50\% or higher. \citet{Witt2010} and \citet{Dong2011} also
found that the most likely average fraction of scattered H\textgreek{a}
intensity is about 20\% of the total H\textgreek{a} intensity. Thus,
the scattered component of the H$\alpha$ emission may strengthen
the correlation between the diffuse H$\alpha$ and FUV intensities,
especially at high latitudes.

It should be noted here that the differences between the data in Figure
\ref{correlation_linear}(b) and Figure \ref{correlation_linear}(d)
accord with the increasing significance of the dust-scattered H$\alpha$
emission with latitude. At low latitudes the diffuse H$\alpha$ emission
is mostly from ionized gas and not related to dust-scattering, while
the FUV background is from dust-scattering. Therefore, at low latitudes
and high values of the diffuse FUV intensity, a much wider range of
variability in the H$\alpha$ intensity is found for a given value
of the diffuse FUV intensity (Figure \ref{correlation_linear}(d))
while the diffuse FUV intensity still correlates with 100 $\mu$m
emission showing an enhancement due to the effects of strong forward-scattering
(see \S5.1). On the other hand, at high latitudes and low values
of the diffuse FUV intensity, much of the structure in the faint H$\alpha$
background may be due to scattering. Thus, we are seeing a linear
correlation with the diffuse FUV intensity there. In other words,
the H$\alpha$ contribution from dust-scattering becomes increasingly
important at higher latitudes. Such a strong linear correlation seems
to imply that a significant portion of H$\alpha$ at high latitudes
is due to dust-scattering. A more detail comparison of the diffuse
FUV and H$\alpha$ maps in both large and small scales is needed to
address how large portion of the diffuse H$\alpha$ emission is due
to dust-scattering.

Finally, there is one possibility beyond the scattered H$\alpha$
that is worth considering. The ISM may have low-density paths and
voids that allow for ionizing photons from midplane OB stars to reach
and ionize gas many kiloparsecs above the Galactic plane \citep{Wood2010}.
Pathways that provide lower than average densities to high latitudes
could enhance both scattered FUV and gas ionized by the Lyman continuum
from the same OB stars that are much closer to the Galactic plane.
Combined with the scattered H$\alpha$ emission, the presence of such
pathways could provide a stronger correlated link between FUV and
H$\alpha$. In this regard, we note that if ionizing photons could
use low-density paths to penetrate a turbulent ISM and reach high
latitude clouds, such as LDN 1780, the resulting ionization would
be limited to a thin outer shell and the ambient medium of the clouds
\citep{Witt2010}. \citet{Wood2010} estimated that only $\sim$0.05
pc thick skin of LDN 1780 would be photoionized. For such morphology,
the optically thick core regions would appear relatively dark, which
contradicts the observations described in \citet{Mattila07} and \citet{Witt2010}.
However, we note that this does not preclude ionizing photons travelling
to high latitudes. Photoionization would contribute to a portion of
the H$\alpha$ intensity seen in LDN 1780, although dust-scattering
is the dominant process for the H$\alpha$ intensity on the cloud.
Therefore, further investigation is needed to disentangle which process
(dust-scattered H$\alpha$ emission versus in-situ ionized gas) is
more significant in explaining the correlation between the diffuse
FUV and H$\alpha$ emissions.

A weak saturation effect in the correlation between $I_{{\rm FUV}}^{{\rm diffuse}}$
and $I_{{\rm H}\alpha}$, which is similar to but much weaker than
those in the correlation of $I_{{\rm FUV}}^{{\rm diffuse}}$ versus
$N_{{\rm HI}}$ and $I_{{\rm 100\mu m}}$, is also noticeable. The
weak saturation seen in the correlation with $I_{{\rm H}\alpha}$
may be easily understood through the fact that the WIM can also be
approximated with a plane-parallel model but the extinction cross-section
at FUV is higher than that of H$\alpha$.

\subsection{Correlation with Galactic Latitude}

If the diffuse FUV background correlates primarily with Galactic latitude
$|b|$, the FUV background intensity is described by $C_{1}\csc|b|+C_{2}$.
In addition to these, we also noted in the previous section that the
diffuse FUV intensity can be approximated with a simple plane-parallel
model defined by $C_{1}\csc|b|$. We thus fitted the data with two
trial functions, one with two free parameters and the other with only
one free parameter.

The dependence on $\csc|b|$ was also obtained by averaging the diffuse
FUV intensity for a given Galactic latitude. Best regression functions
are then given by

\begin{equation}
I_{{\rm FUV}}^{{\rm diffuse}}=\frac{846.7\pm96.1}{\sin|b|}+(-457.2\pm100.5)\,\,{\rm CU,}\label{eq:fuv_sinb1}
\end{equation}
and
\begin{equation}
I_{{\rm FUV}}^{{\rm diffuse}}=\frac{412.3\pm10.3}{\sin|b|}\,\,{\rm CU}\label{eq:fuv_sinb2}
\end{equation}
for two-parameters and one-parameter models, respectively. The average
intensity of the FUV background versus $|b|$ are also shown in Figure
\ref{correlation_linear}. The regression functions with two and one
parameters are denoted by solid and dashed lines, respectively. The
value of the two-parameters function at $\sin|b|=1$ is a bit lower
than the value obtained by \citet{FCF89}, but still within their
error range. We note that the two-parameters model is better in explaining
the dependence of $I_{{\rm FUV}}^{{\rm diffuse}}$ on Galactic latitude.
However, the physical interpretation of the two-parameters model is
problematic, as discussed in \S\ref{sub:Isotropic-component}.

\subsection{Comparison with the Soft X-ray}

Comparison of the FUV background with the soft X-ray emission has
been mentioned by \citet{Joubert83} and \citet{Zvereva82}. Especially
\citet{Zvereva82} claimed there was a good correlation of the FUV
intensity in high latitude range ($|b|>30^{\circ}$) with soft X-ray
brightness. They interpreted the correlation as an indication of the
significant contribution of the hot ISM to the diffuse FUV background.
However, \citet{Joubert83} argued that hot gas has too small an emission
measure to contribute significantly to the FUV background. Soft X-ray
emission has been observed extensively with ROSAT and is well-known
to be anti-correlated with \ion{H}{1} column density. Indeed, we
found a strong anti-correlation between the diffuse FUV intensity
and the soft X-ray background, as shown in Figure \ref{correlation}(d).
This anti-correlation between the diffuse FUV background and the soft
X-ray also supports the dust-scattered origin of the diffuse FUV background.

\section{Discussion}

\subsection{Log-normal nature of the FUV intensity distribution?\label{sub:discussion-Log-normal}}

In presenting the results, we used logarithmic scales to visualize
large orders of magnitude in observed FUV intensity. Here, we show
that the distribution of the diffuse FUV intensity seems to be well
represented by a log-normal distribution or, equivalently, a Gaussian
distribution in a logarithm scale, although its detailed study is
beyond the scope of the present paper.

\citet{Hill2008} found that the histogram of the H$\alpha$ emission
line observed from the WIM fits well with a log-normal distribution.
They attributed the log-normal nature to the turbulence in the WIM.
A log-normal distribution is the expected probability density function
of density and/or column density for the ISM with a density structure
established by turbulence \citep[e.g., ][]{VP99}. Since the scattered
FUV intensity is directly related to the dust column density, the
log-normal property of the distribution of the diffuse FUV intensity
is a natural consequence of the density structure and the turbulence
property of the ISM. Thus in Figure \ref{log-normal}, we plotted
intensity histograms of $I_{{\rm FUV}}^{{\rm diffuse}}\sin|b|$ for
various Galactic latitude ranges together with best-fit log-normal
functions, and found that the intensity distributions are indeed well
represented with log-normal functions. Here, we removed the Magellanic
Clouds and 19 large-scale regions (the same regions as in \citealp{Hill2008})
with significantly enhanced H$\alpha$ emission to avoid discrete
ionized regions where the contribution of two-photon emission to $I_{{\rm FUV}}^{{\rm diffuse}}$
may be significant. We also removed some high intensity ``streaks''
(most visible in the left panel of Figure \ref{diffuse_map}(b)) that
show up in the exposure maps.

Log-normal distribution shows an extended tail to large intensity
values in a linear scale. Some evidence of the extended tail in an
intensity histogram has already been noticed in the previous studies.
\citet{Joubert83} noted that the distributions of the points in correlations
between UV brightnesses and \ion{H}{1} column density are not symmetrical
with respect to the linear regression lines. There was a high flux
tail in the distribution of the UV intensities at a given value of
$N$(\ion{H}{1}). The intensity histograms in Figure 2 of \citet{PMB1980}
also show such extended tails. These regions with UV intensity excesses
were rejected for the correlation studies in \citet{Joubert83} and
\citet{Perault91} by eliminating the points in the tail with counts
larger than the median by several times the dispersion and iterating
the procedure. \citet{Joubert83} attributed this property to an excess
FUV radiation in certain regions of the sky, perhaps due to two photon
emission by ionized gas. However, \citet{Deharveng82} and \citet{Reynolds92}
showed that the emission from the WIM is unlikely to contribute significantly
to the general diffuse UV background. \citet{Perault91} found that
the excess in the intensity histogram does not obey the Poisson distribution;
they attributed the high-count skewness and tail to the contribution
of the stars brighter than $m({\rm UV})=8$ and unresolved faint stars.
However, we argued that contribution of residual stars to the diffuse
FUV background is not significant. In addition, the faint late-type
stars cannot explain the hardness ratio observed at low intensity
regions, as shown in \S4. \citet{SCH2001} also noted a large variance
in the FUV intensity and attributed the observed variance to the ``cosmic
variance'' predicted by the multi-cloud model combined with multiple
scattering and three-dimensional radiation field variance. We thus
believe that the anomalous high intensity tails shown in previous
studies are in fact due to an intrinsic property of the log-normal
nature of the diffuse FUV background. A detailed statistical property
of the dust density distribution can be investigated through a radiative
transfer modeling of the starlight dust-scattered off in turbulent
media.

It may also be possible to isolate high-intensity tails in the histograms
and identify extended local features in the FUV maps. Some locally
enhanced features were found, but bright regions generally had a patchy
appearance. We also found that some of the enhanced features in the
FUV sky coincided with the H$\alpha$ sky, but not always. We note
here that the observed intensity toward a cloud could be higher, lower,
or equal to its surroundings (as demonstrated in \citealp{Mattila07}),
depending on the scattered intensity, incident intensity from behind
the cloud, and cloud optical depth. Therefore, the high-intensity
tails in histograms do not necessarily correspond to local dust clouds.
In addition, because of the difference in dust-extinction cross-sections
at the FUV wavelengths and H$\alpha$ line, there would not be a strict
correlation between diffuse FUV and H$\alpha$ emissions, even when
all the FUV and H$\alpha$ background photons originate from dust-scattering.
The origin of both the high-intensity tails in the histograms and
the diffuse H$\alpha$ line emission may be elucidated by more detailed
analyses of the dust clouds observed in both the FUV wavelengths and
H$\alpha$ line, along with self-consistent radiative transfer models.
But, such analyses are beyond the scope of the present paper.

\subsection{Relative softness of the diffuse radiation}

We found a weak reddening of the diffuse radiation relative to the
direct starlight in \S4. In order to understand the trend, we calculated
the scattered to direct intensity ratio for a simplified case where
a point source is embedded in a homogeneous spherical dust cloud.
As noted in \citet{Witt1982}, the analytical expression derived by
\citet{Code1973} approximates the scattered to the unattenuated luminosity
ratio very well. In the approximation, the scattered intensity is
given by

\[
\frac{L_{{\rm scatt}}}{L_{0}}=\frac{2}{\left(1+\zeta\right)e^{\xi\tau}+\left(1-\zeta\right)e^{-\xi\tau}}-e^{-\tau}
\]
where $\zeta=\sqrt{\left(1-a\right)/\left(1-ag\right)}$ and $\xi=\sqrt{\left(1-a\right)\left(1-ag\right)}$,
and the attenuated direct starlight by $L_{{\rm direc}}=L_{0}e^{-\tau}$.
Here, the quantities $L_{0}$, $a$, $g$, and $\tau$ refer to the
unattenuated stellar luminosity, grain albedo, phase function asymmetry
factor, and radial optical depth of the spherical cloud measured from
the center to the outer edge.

We calculated ratios $L_{{\rm direc}}/L_{0}$, $L_{{\rm scatt}}/L_{0}$,
and $L_{{\rm scatt}}/L_{{\rm direc}}$ as a function of wavelength
using the extinction cross-section $\sigma(\lambda)$, $a(\lambda)$,
and $g(\lambda)$ from \citet{DRA2003} for an illustrative case of
$\tau(1550$ \AA$)=2$ in Figure \ref{aeffect}. Here, $L_{{\rm 0}}$,
$L_{{\rm direc}}$, and $L_{{\rm scatt}}$ are the input, direct,
and scattered luminosities. Note that the albedo $a(\lambda)$ is
an increasing function with the wavelength. For comparison, we also
plotted the results calculated for three hypothetical cases: (1) the
case of a constant albedo of $a(1550$ \AA) and a variable asymmetry
factor $g(\lambda)$, (2) the case where both $a$ and $g$ are constants
obtained at 1550 \AA, and (3) the case in which the wavelength-dependence
of $a(\lambda)$ is reversed so that the albedo decreases with the
wavelength. The figure shows that assuming the asymmetry factor to
be a constant does not significantly affect the result while the albedo
does. As is evident, the directly-escaped spectrum is redder or softer
than the input spectrum (Figure \ref{aeffect}a). When a constant
albedo is assumed, we obtain a scattered spectrum reddened compared
to the unattenuated input spectrum, but bluer than the (attenuated)
directly-escaped spectrum because of the rapid increase of the absorption
with the wavelength. However, as the albedo increases with the wavelength
in the model of \citet{WD2001} and \citet{DRA2003}, the scattered
spectra rise much more rapidly with the wavelength than the attenuation,
thereby giving a relatively-red scattered spectrum compared to the
directly escaped spectrum. In the case in which the wavelength-dependence
of $a(\lambda)$ is reversed, the scattered spectrum become bluer
relative to both the unattenuated input and the reddened directly-escaped
spectra, which is contrary to the other cases. The same trend is observed
when the optical depth is varied. We therefore conclude that the relative
softness of the diffuse background radiation compared to the stellar
radiation field is attributed to the increase of dust albedo with
the wavelength.

In \S4, we noticed a weak rise in the diffuse spectra longward of
1550 \AA. The rise itself tells nothing about the wavelength-dependence
of dust albedo unless the incident unattenuated spectrum $I_{0}(\lambda)$
is known. Instead, the relative softness or hardness of the scattered
radiation compared to the directly observed radiation can give direct
information about the wavelength-dependence of dust albedo. In this
regard, we note that the dust albedo provided by \citet{Bruzual1988}
is constant or decreases with the wavelength in FUV, and thus is not
consistent with our results. A more detailed analysis on the spectral
shape can be obtained by radiative transfer models including wavelength
dependence in calculating stellar radiation fields rather than that
performed for a single wavelength band as have been done by \citet{WIT1997}
and \citet{SCH2001}.

\subsection{Contribution of two-photon continuum emission\label{sub:two-photon}}

The WIM, which is expected to have a temperature of $\sim8000$ K,
must be a source of two-photon continuum in the FUV wavelengths, potentially
impacting the interpretation of the FUV continuum background \citep{Reynolds92}.
The two-photon continuum intensity at $\sim$ 1600 \AA\ due to the
WIM with a temperature of 8000 K is estimated to be $I_{2\gamma}=57.4(I_{{\rm H}\alpha}/{\rm R})$
CU (\citealp{Reynolds1990,Reynolds92}; see also Figure \ref{two_photon}).
Using the average H$\alpha$ intensity distribution $I_{{\rm H}\alpha}\approx1.2\csc|b|$
R, which was obtained from rather limited observations, Reynolds also
obtained $I_{2\gamma}\approx70\csc|b|$ CU. However, using the result
of \citet{Hill2008} obtained from the full WIM observations, the
distribution of H$\alpha$ for $|b|\ge10^{\circ}$ is fitted on average
by $I_{{\rm H}\alpha}\approx0.625\csc|b|$ R (see also \citealt{Dong2011}).
This leads to a lower two-photon intensity of $I_{2\gamma}\approx36\csc|b|$
CU, which accounts for $\sim9$\% of the FUV background (Eq. \ref{eq:fuv_sinb2}).

The two-photon continuum's contribution to the diffuse FUV intensity
is reduced even further when the model of WIM developed by \citet{Dong2011}
is adopted. Recent observations using the \emph{Wilkinson Microwave
Anisotropy Probe (WMAP)} have found that the ratio of the free-free
radio continuum to H$\alpha$ is surprisingly low in the WIM \citep{Davies2006,Dobler2008,Dobler2009}.
\citet{Dong2011} considered a three-component model consisting of
a mix of (1) photoionized hot gas, (2) gas that is recombining and
cooling after removal of a photoionizing source, and (3) cool \ion{H}{1}
gas. In their standard model for explaining the observed intensity
ratios of the free-free radio continuum to H$\alpha$ and {[}\ion{N}{2}{]}
$\lambda$6583 to H$\alpha$, the authors assumed that the scattered
fraction of the H$\alpha$ originating from the hot gas is $f_{{\rm H}\alpha}^{{\rm (refl)}}=0.2$,
while the fractions of the hot, cooling, and \ion{H}{1} gases are
$f_{{\rm H}\alpha}^{{\rm (hot)}}=0.22$, $f_{{\rm H}\alpha}^{{\rm (cooling)}}=0.56$,
and $f_{{\rm H}\alpha}^{{\rm (HI)}}=0.02$, respectively. The temperature
of the hot gas was found to be $T^{{\rm (hot)}}=9100$ K. The cooling
gas cooled down rapidly from its initial temperature $T^{{\rm (cooling)}}\sim$
10,000 K to $\sim400$ K (Fig. 3 of \citealt{Dong2011}).

Using the two-photon spectral profile \citep{Kwok2007}, the hydrogen
recombination coefficient $\alpha_{{\rm 2S}}$ for case B (Table 3
of \citealt{MartinPG1998}), and the H$\alpha$ emission rate (Eq.\
10 of \citealt{Dong2011}), we estimated the two-photon intensity
averaged over 1370--1710 \AA\ from hot and cooling ionized gas (Figure
\ref{two_photon}): $I_{2\gamma}^{{\rm (hot)}}=59.1(I_{{\rm H}\alpha}/{\rm R})$
CU, and $I_{2\gamma}^{{\rm (cooling)}}=26.0(I_{{\rm H}\alpha}/{\rm R})$
CU. The two-photon intensity of the cooling gas is mostly determined
by the gas with a temperature of 400 K, because the cooling gas is
400 K or lower for more than 98\% of its lifetime (Fig. 3 of \citealt{Dong2011}).
The total two-photon intensity originating from the ionized gas is
then given by $I_{2\gamma}=f_{{\rm H}\alpha}^{{\rm (cooling)}}I_{2\gamma}^{{\rm (cooling)}}+f_{{\rm H}\alpha}^{{\rm (hot)}}I_{2\gamma}^{{\rm (hot)}}=28.4(I_{{\rm H}\alpha}/{\rm R})$
CU. The distribution of two-photon emission is then $I_{2\gamma}\approx18\csc|b|$
CU, corresponding to $\sim4$\% of the FUV background intensity.

In summary, two-photon emission may account for $\sim4-9$\% of the
diffuse FUV background. Although the two-photon emission may be significant
in discrete bright \ion{H}{2} regions, the contribution of two-photon
emission to the diffuse FUV background is much smaller than has previously
been suggested. The large variation of $I_{{\rm FUV}}^{{\rm diffuse}}$
(from a factor of 2--3 up to an order of magnitude) is also unlikely
to be the result of two-photon emission.

\subsection{Isotropic component\label{sub:Isotropic-component}}

The phrase ``isotropic component'' has been used to refer to the diffuse
intensity extrapolated to $N_{{\rm HI}}=0$, or, alternatively, to
$|b|=90^{\circ}$, and is frequently assumed to be extragalactic background.
\citet{Wright92} claimed that the isotropic extragalactic component
of the diffuse FUV background should be obtained by extrapolating
to $\csc(|b|)=0$, instead of to $\csc|b|=1$ as practiced by \citet{FCF89}.
\citet{WP94} discussed systematic errors arising in the method of
extrapolating high latitude measurements of the diffuse FUV background
to $N_{{\rm HI}}=0$ or to $\csc(|b|)=0$ to derive the level of the
cosmic extragalactic background. They argue that the derivation of
the cosmic background from the extrapolation of the diffuse FUV background
to $N_{{\rm HI}}=0$ is more likely to yield a reliable result. They
found, in fact, that extrapolations of the predicted FUV background
intensities through a Monte-Carlo radiative transfer model to $\csc|b|=0$
produce negative values when it should produce a zero intercept. We
also found from the SPEAR data that the extrapolation to $\csc|b|=0$
indeed produces a negative value of $\sim-400$ CU.

A discussion of the origin of the isotropic component is beyond the
scope of the present paper. \citet{Onaka91} observed intensities
as low as 300 CU at high Galactic latitudes. \citet{SCH2001} placed
constraints on the extragalactic background $200\pm100$ CU. We also
found the isotropic component consistent with previous authors, but
a bit higher. It should be noted that a similar level of the isotropic
component was found from the relationship between the diffuse FUV
and H$\alpha$ emissions (Eq. \ref{eq:fuv_halpha}). However, it is
not clear at this time that all of the $\sim300$ CU, the extrapolation
value of the average FUV-$N_{{\rm HI}}$ correlation to $N_{{\rm HI}}=0$,
corresponds to the extragalactic background. In fact, we observed
the FUV intensities lower than 300 CU as is obvious in Figure \ref{histograms}.
\citet{WIT1997} analyzed the FUV data observed in the FAUST experiment
\citep{BOW1993} and estimated a contribution of about 700$\pm$200
CU that is uncorrelated with Galactic parameters. They argued that
the extragalactic component of more than 300 CU is excluded by the
lower limit directly measured by \citet{Onaka91} and concluded that
most likely value of isotropic extragalactic background radiation
would be 160$\pm$50 CU, after correction for Galactic extinction.

\section{Summary}

In this paper, we reported some general results on the diffuse FUV
background and the total FUV radiation observed with SPEAR/FIMS. Our
main conclusions are as follows:
\begin{enumerate}
\item Strong anisotropies are found in not only the diffuse background intensity
but also in the hardness ratio (1370--1520 \AA\ to 1560--1710 \AA).
\item We found a good correlation between the diffuse FUV background and
other Galactic quantities, such as \ion{H}{1} column density, 100
$\mu$m emission, and H$\alpha$ emission. The correlation of the
FUV continuum background with the diffuse H$\alpha$ emission seems
to be better than with the other quantities. Anti-correlation with
the soft X-ray was also found. Correlation of the FUV background with
the direct stellar intensity is rather weaker than the correlation
with other waveband observations, such as 100 $\mu$m and H$\alpha$
emissions.
\item The fact that a linear correlation of the FUV background with the
diffuse H$\alpha$ emission is found at low intensities seems to indicate
that the dust-scattered component of H$\alpha$ becomes increasingly
important at higher latitudes.
\item The spatially-averaged total FUV intensity, including the direct stellar
and diffuse FUV intensities, observed with SPEAR/FIMS is weaker than
the well-known ISRF models, such as the Habing, Mathis, and Draine
models.
\item The spectrum of the diffuse FUV background is in general flat and
a bit softer than the directly-escaped stellar spectrum. The relative
softening can be attributed to the rise of dust albedo as the wavelength
increases.
\item The hardness ratio seems to follow in general the longitudinal distribution
of OB-type stars in the Galactic plane.
\item The ``isotropic component'' obtained by comparing the FUV background
with other ISM tracers, $N_{{\rm HI}}$, 100 $\mu$m, and H$\alpha$
emissions is consistent with the previous results.
\item Evidence that the intensity histogram of the diffuse background is
well represented by log-normal distribution is found, as for the diffuse
H$\alpha$ emission.
\end{enumerate}
The above results well accord with the fact that most of the diffuse
FUV background is scattered light of the FUV stellar radiation by
the interstellar dust. \citet{WIT1997} and \citet{SCH2001} developed
radiative transfer models to interpret the diffuse FUV background.
Our future work will deal with a Monte-Carlo simulation model incorporating
three-dimensional stellar distribution obtained from the \emph{Hipparcos}
catalog and three-dimensional dust distributions modeled by several
authors such as \citet{DRI2001} and \citet{DRI2003}. The model compared
with the \emph{SPEAR/FIMS} data should provide a better understanding
to the observational results.

\acknowledgements{}

The SPEAR/FIMS was supported by NASA grant NAG5-5355 and flew on the
STSAT-1 Mission, supported of the Korea Ministry of Science and Technology.
We acknowledge the use of the Legacy Archive for Microwave Background
Data Analysis (LAMBDA). Support for LAMBDA is provided by the NASA
Office of Space Science. K.-I. S. was supported by a National Research
Foundation of Korea grant funded by the Korean government (grant no.
313-2008-2-C00377). We thank the anonymous referee for comments that
led to improvements in the manuscript.

\clearpage{}

\setcounter{figure}{4}

\begin{figure*}
\begin{centering}
\includegraphics[scale=0.5,angle=90]{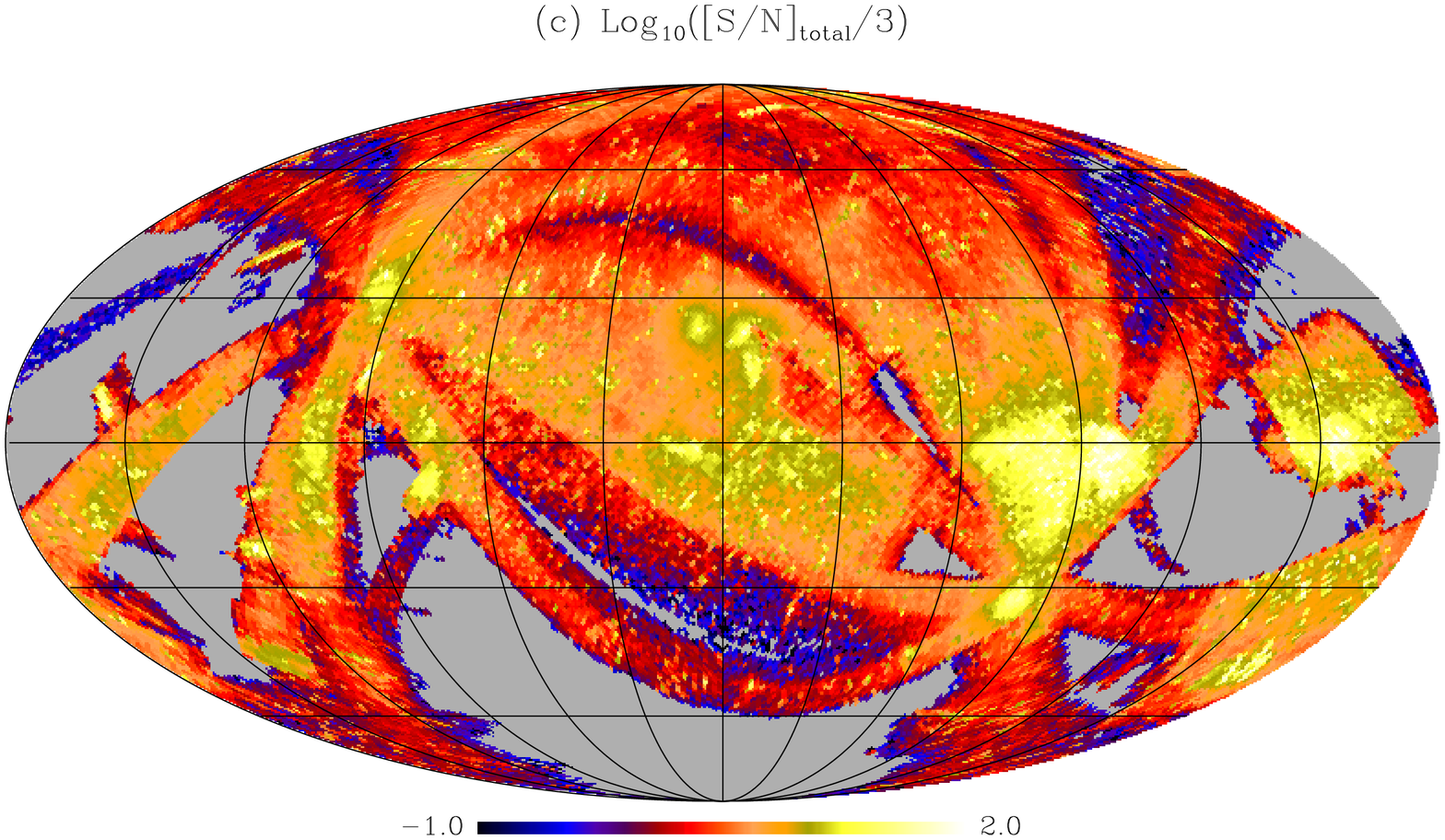}
\par\end{centering}

\bigskip{}

\begin{centering}
\includegraphics[scale=0.5,angle=90]{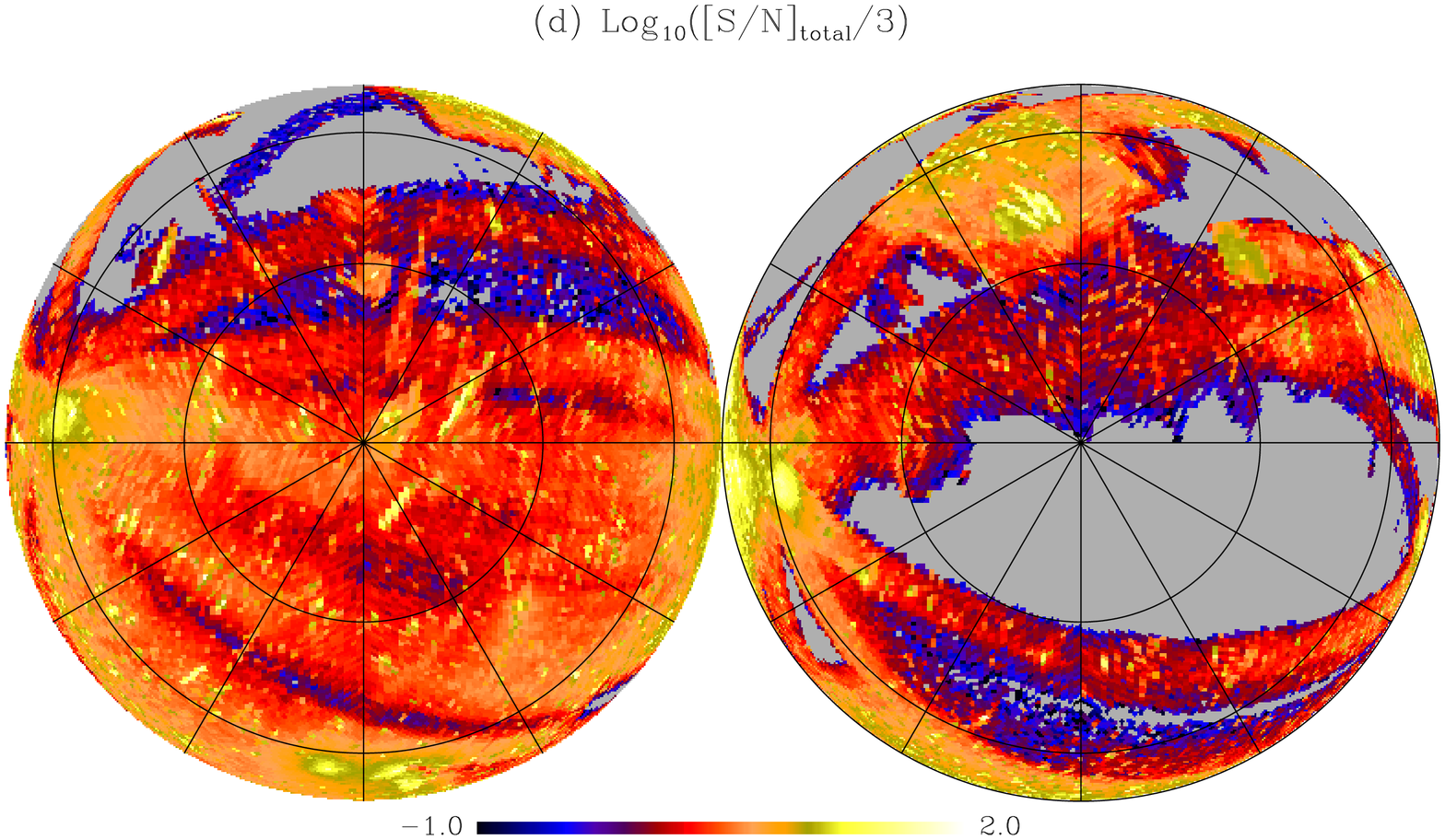}\medskip{}

\par\end{centering}

\caption{\label{isrf_map-1}(c) mollweide and (d) orthographic projections
of signal-to-noise ratio of the total FUV map. The left and right
sides of orthographic projections are centered at the northern and
southern Galactic polar caps, respectively, and their longitude increases
clockwise and counter-clockwise, respectively. In both projections,
$l=0^{\circ}$ is at the six o'clock position. The intensity scales
are logarithmic across the color bars.}
\end{figure*}

\setcounter{figure}{5}

\begin{figure*}[t]
\begin{centering}
\includegraphics[scale=0.5,angle=90]{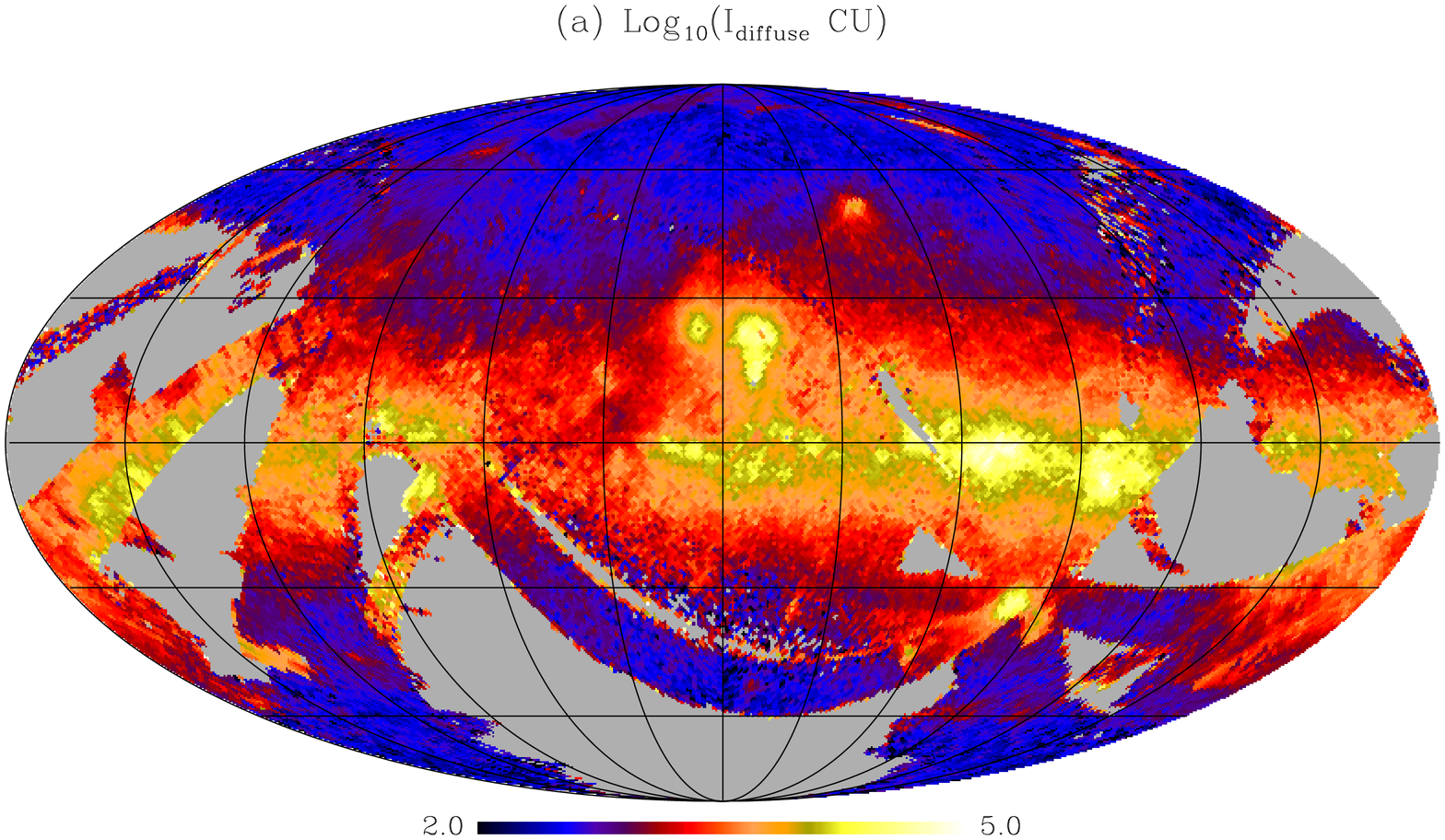}\bigskip{}

\par\end{centering}

\begin{centering}
\includegraphics[scale=0.5,angle=90]{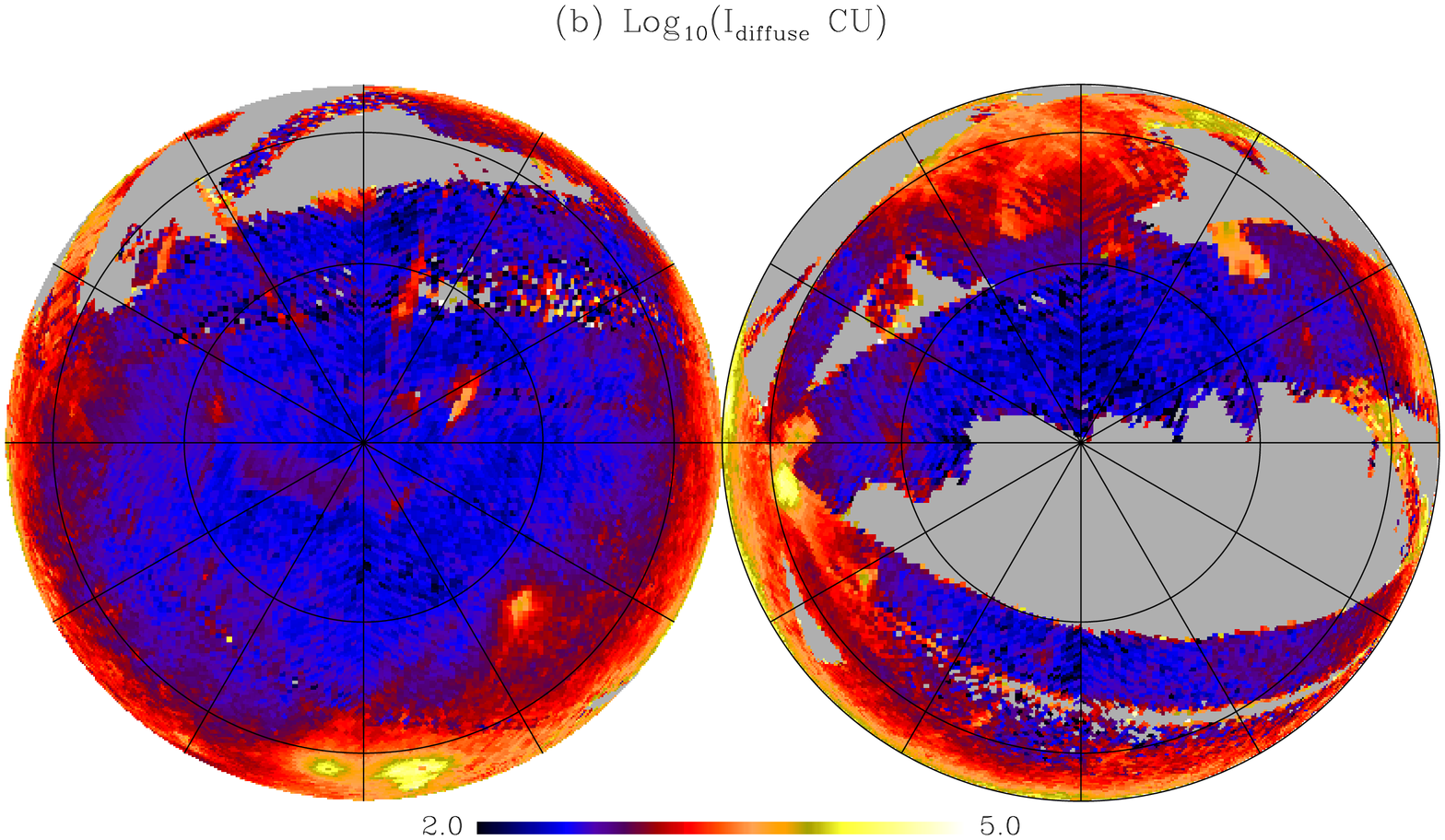}\medskip{}

\par\end{centering}

\caption{\label{diffuse_map}(a) Mollweide and (b) orthographic projections
of the diffuse FUV background map observed with SPEAR/FIMS, after
removal of locally intense pixels (stars) and the instrumental background.
The left and right sides of orthographic projections are centered
at the northern and southern Galactic polar caps, respectively, and
their longitude increases clockwise and counter-clockwise, respectively.
In both projections, $l=0^{\circ}$ is at the six o'clock position.
The intensity scales are logarithmic across the color bars.}
\end{figure*}

\setcounter{figure}{5}

\begin{figure*}[t]
\begin{centering}
\includegraphics[scale=0.5,angle=90]{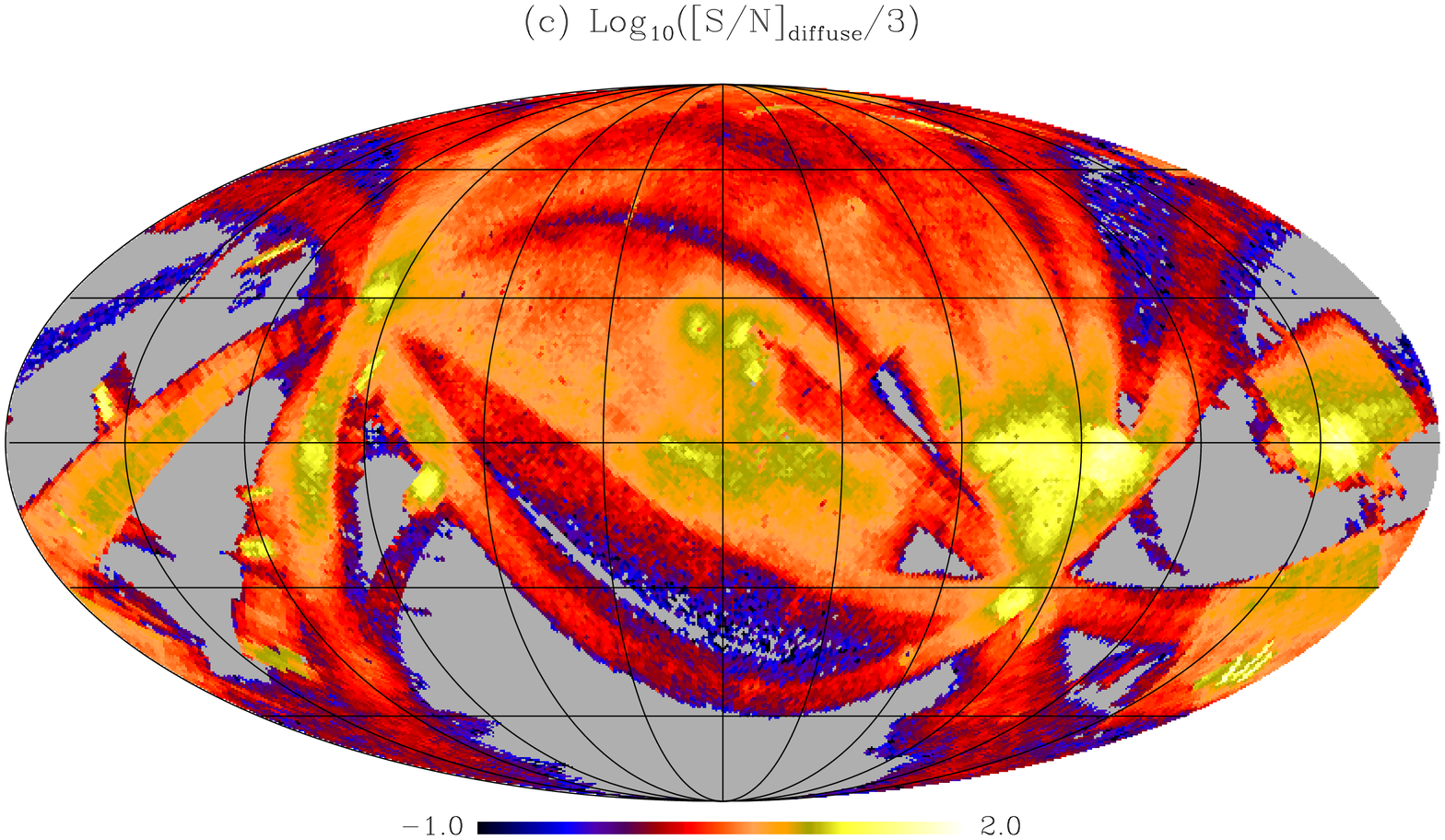}\bigskip{}

\par\end{centering}

\begin{centering}
\includegraphics[scale=0.5,angle=90]{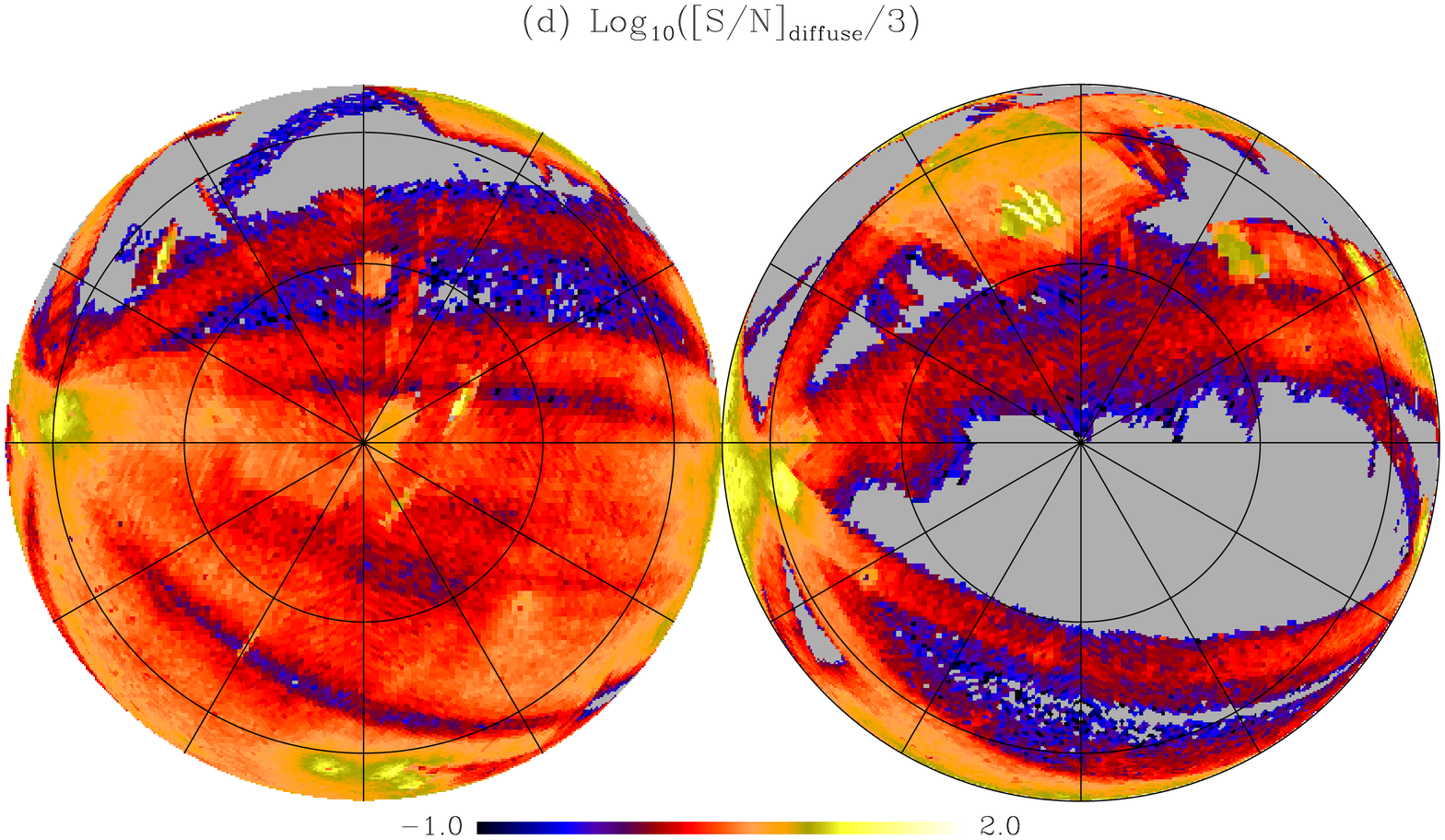}\medskip{}

\par\end{centering}

\caption{\label{diffuse_map-1}(c) mollweide and (d) orthographic projections
of signal-to-noise ratio of the diffuse FUV map. The left and right
sides of orthographic projections are centered at the northern and
southern Galactic polar caps, respectively, and their longitude increases
clockwise and counter-clockwise, respectively. In both projections,
$l=0^{\circ}$ is at the six o'clock position. The intensity scales
are logarithmic across the color bars.}
\end{figure*}

\begin{figure}[t]
\begin{centering}
\includegraphics[scale=0.55]{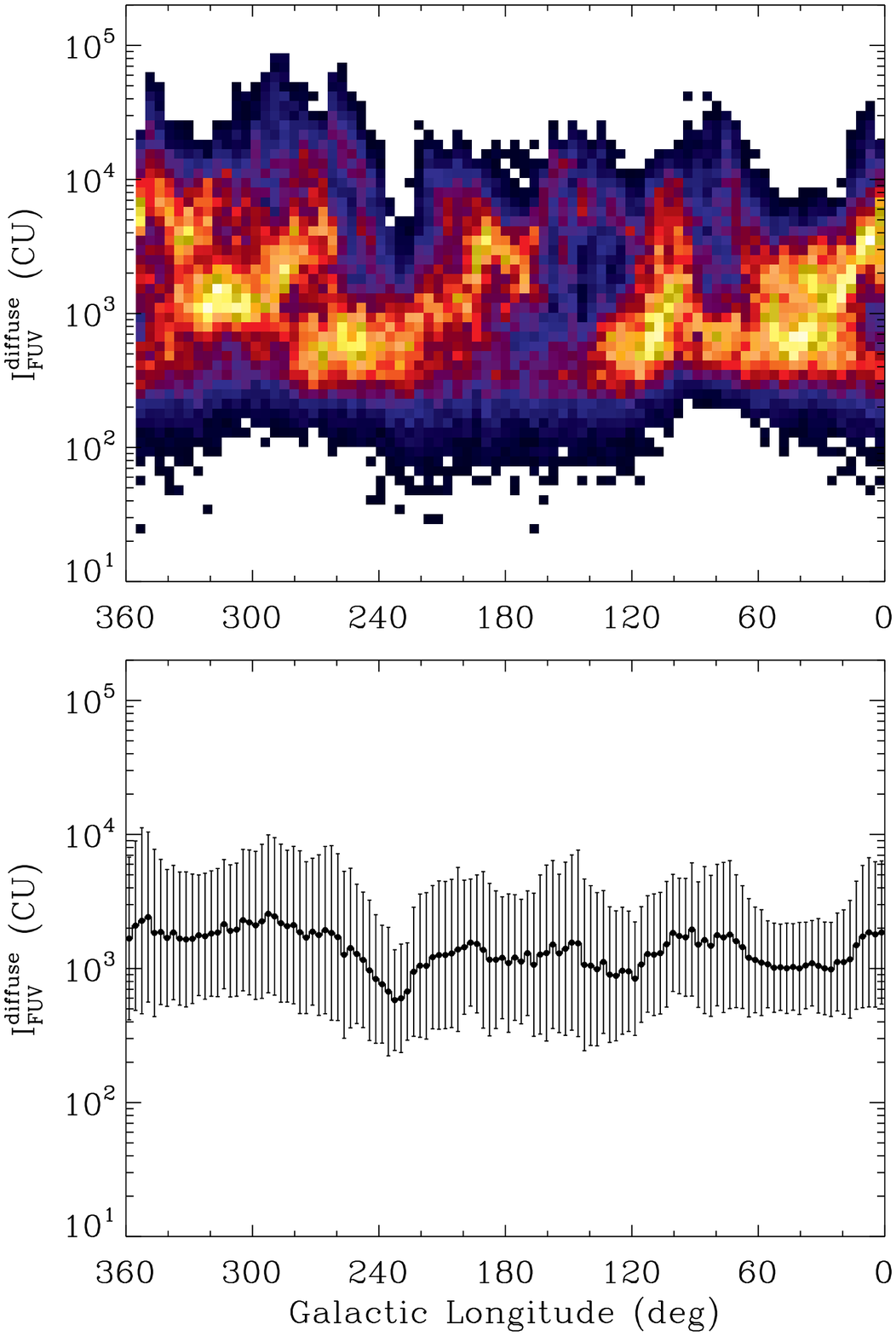}
\par\end{centering}

\caption{\label{intensity_vs_lon}The diffuse FUV intensity versus Galactic
longitude. The top panel plots the two-dimensional histogram of the
diffuse FUV intensity as a function of Galactic longitude. The bottom
panel plots the median value within each of the $3^{\circ}$ longitude
strips from the data. The vertical extent of each plotted bar is determined
from the average deviation about the median within that bin.}
\end{figure}

\begin{figure}[t]
\begin{centering}
\includegraphics[scale=0.55]{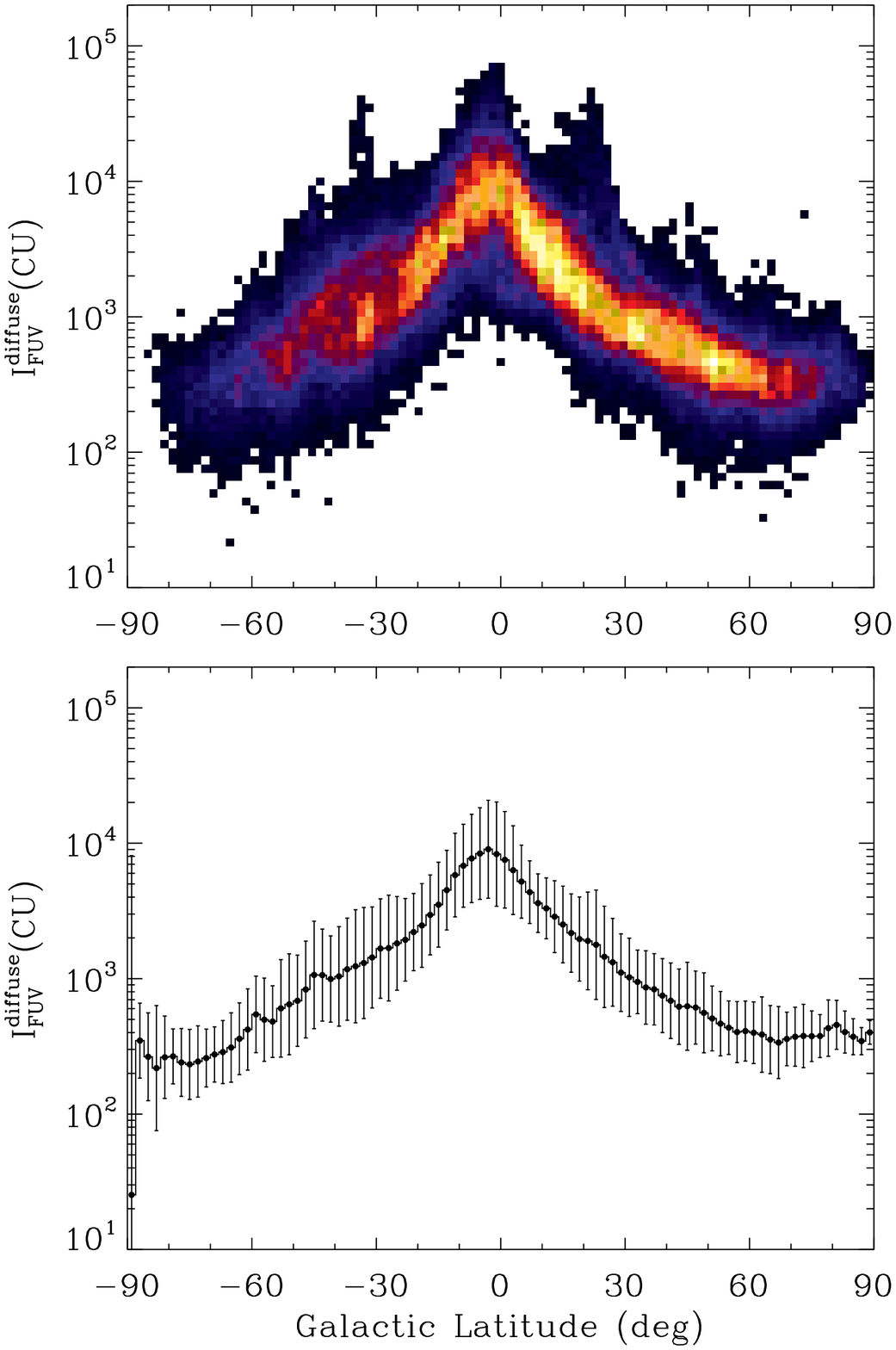}
\par\end{centering}

\caption{\label{intensity_vs_lat}The diffuse FUV intensity versus Galactic
latitude. The top panel plots the two-dimensional histogram of the
diffuse FUV intensity as a function of Galactic latitude. The bottom
panel plots the median value within each of the $2^{\circ}$ latitude
strips from the data. The vertical extent of each plotted bar is determined
from the average deviation about the median within that bin.}
\end{figure}

\clearpage{}

\begin{figure}[tb]
\begin{centering}
\includegraphics[scale=0.55]{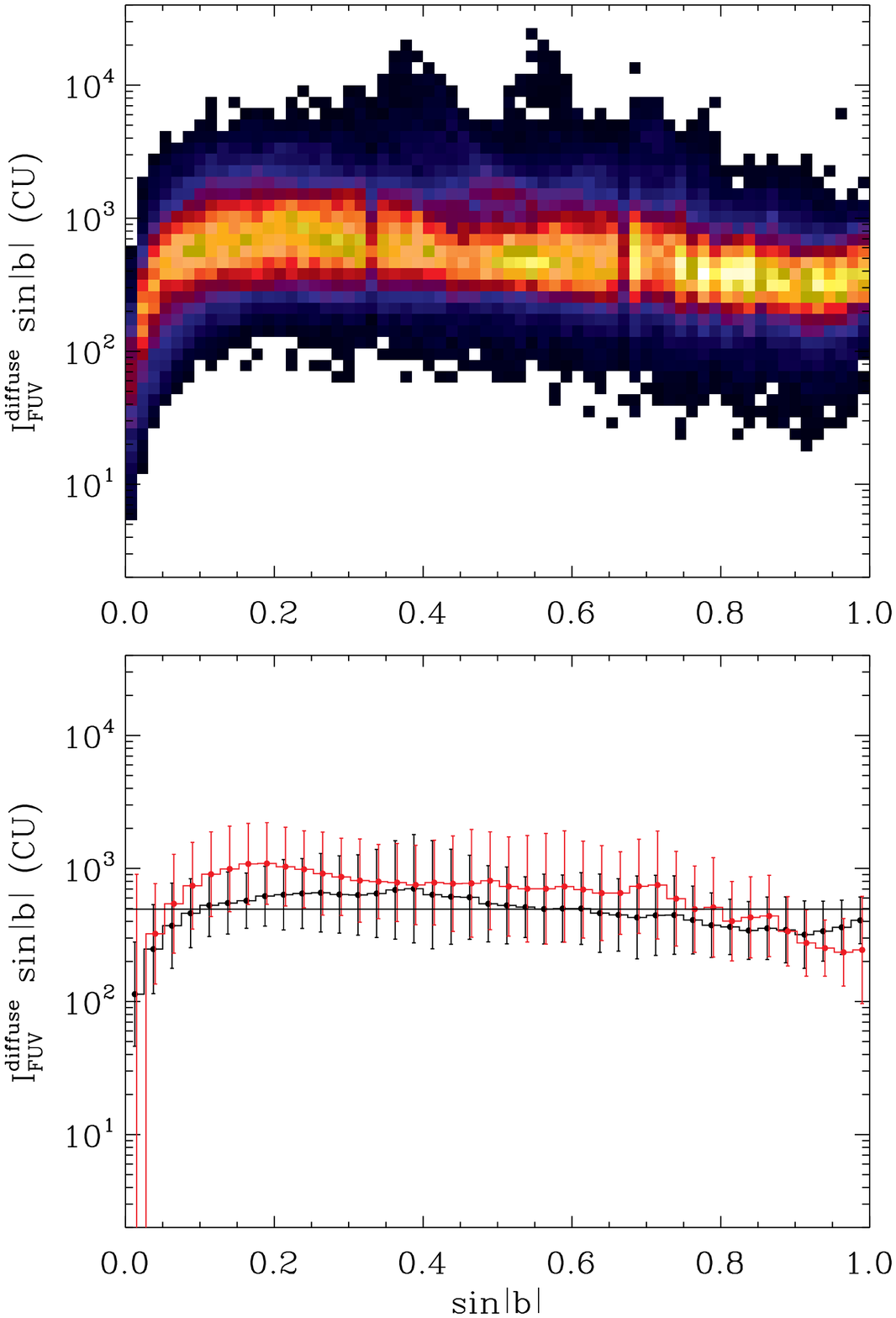}
\par\end{centering}

\caption{\label{isinb_vs_sinb}$I_{{\rm FUV}}^{{\rm diffuse}}\sin|b|$ versus
$\sin|b|$. The top panel plots the two-dimensional histogram as a
function of Galactic longitude. The bottom panel plots the median
value within each of the $\Delta\sin|b|=0.025$ latitude strips from
the data. The vertical extent of each plotted bar is determined from
the average deviation about the median within that bin. The black
and red points represent the data observed toward northern and southern
Galactic latitude, respectively. The horizontal line denotes the median
$I_{{\rm FUV}}\sin|b|$ value.}
\end{figure}

\clearpage{}

\begin{figure*}[t]
\begin{centering}
\includegraphics[scale=0.65]{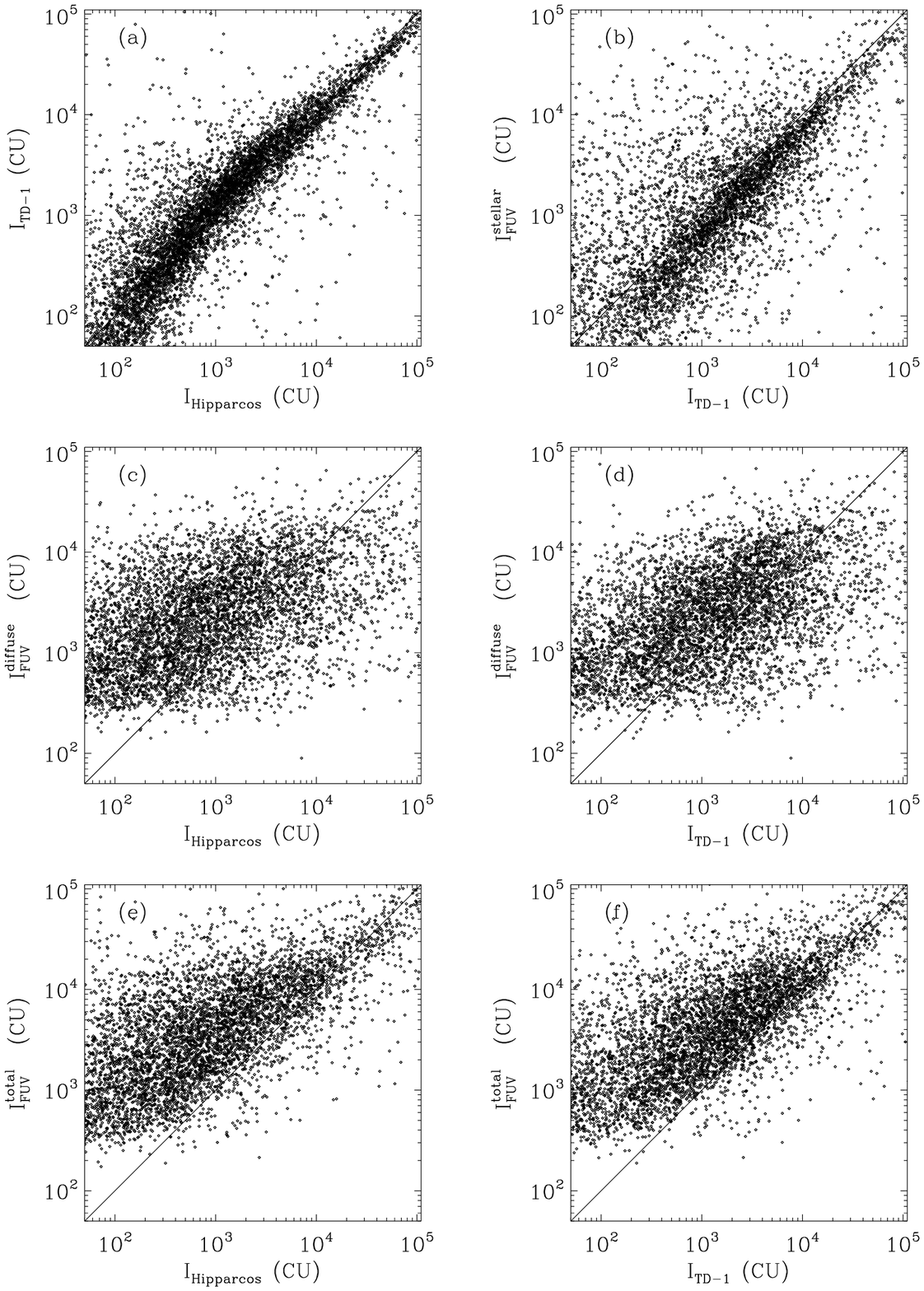}
\par\end{centering}

\caption{\label{compare_stellar}Comparison between the TD-1 stellar equivalent
diffuse intensity (SEDI), the \emph{Hipparcos} SEDI, and the SPEAR/FIMS
intensities. (a) The TD-1 SEDI versus the \emph{Hipparcos} SEDI, (b)
the direct stellar intensity observed with SPEAR/FIMS versus the TD-1
SEDI, (c) the diffuse FUV background versus the \emph{Hipparcos} SEDI,
(d) the diffuse FUV background versus the TD-1 SEDI, (e) the total
(direct stellar + diffuse) FUV intensity versus the \emph{Hipparcos}
SEDI, and (f) the total FUV intensity versus the TD-1 SEDI. Here,
the resolution parameter $N_{{\rm side}}=32$ corresponding to angular
resolution of $\sim1.8^{\circ}$ is used. The TD-1 SEDI was estimated
using the 1565 \AA\ band, and the others were calculated over the
same wavelength bands 1370--1520, 1560--1660, and 1680--1720 \AA.}
\end{figure*}

\begin{figure*}[tp]
\begin{centering}
\includegraphics[scale=0.7]{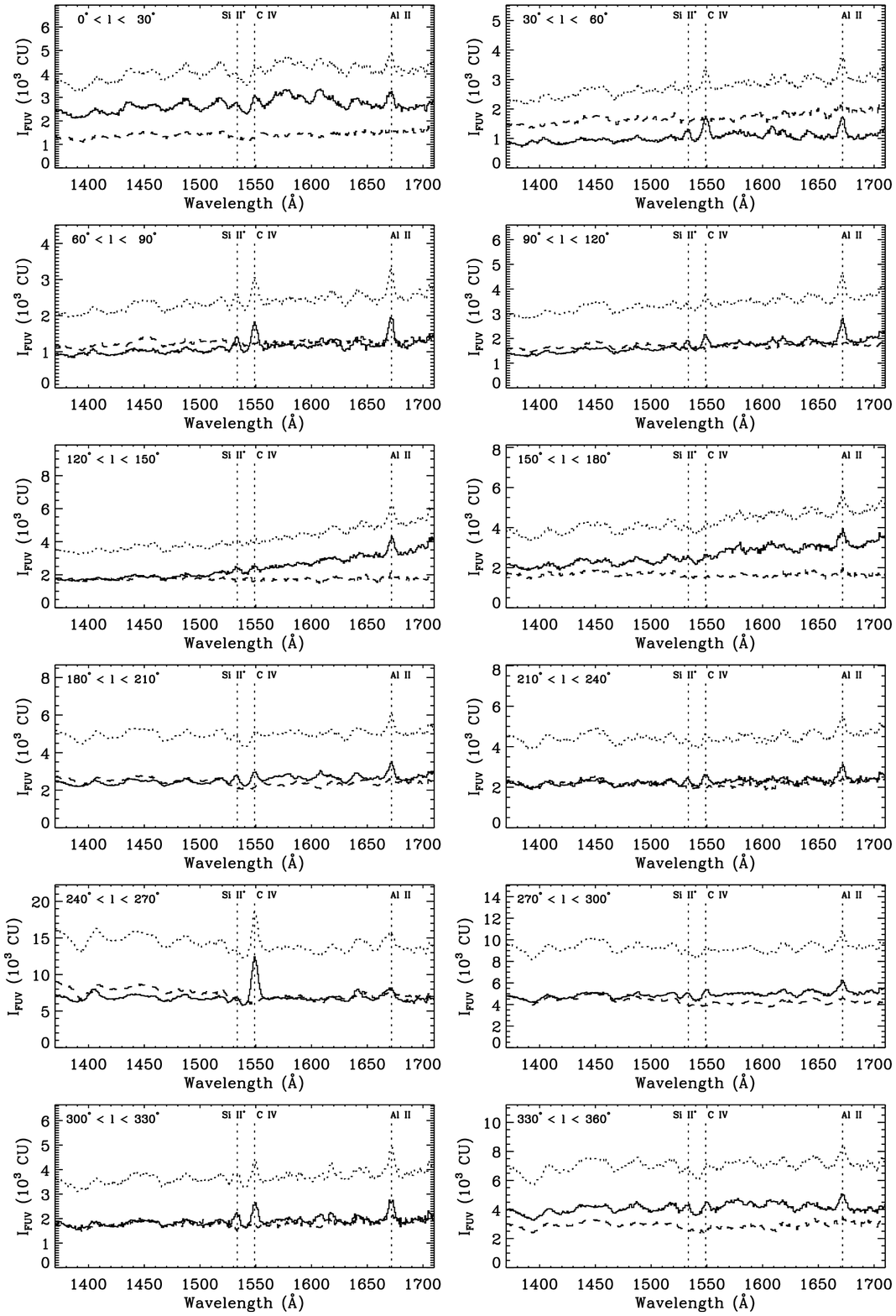}
\par\end{centering}

\caption{\label{spec_vs_lon}Average FUV spectra in various Galactic longitude
ranges. Solid curve shows the diffuse FUV background spectrum. Dotted
and dashed curves represent the total (direct stellar + diffuse) and
the direct stellar spectra, respectively. Some important ionic lines
are also indicated.}
\end{figure*}

\begin{figure*}[tp]
\begin{centering}
\includegraphics[scale=0.7]{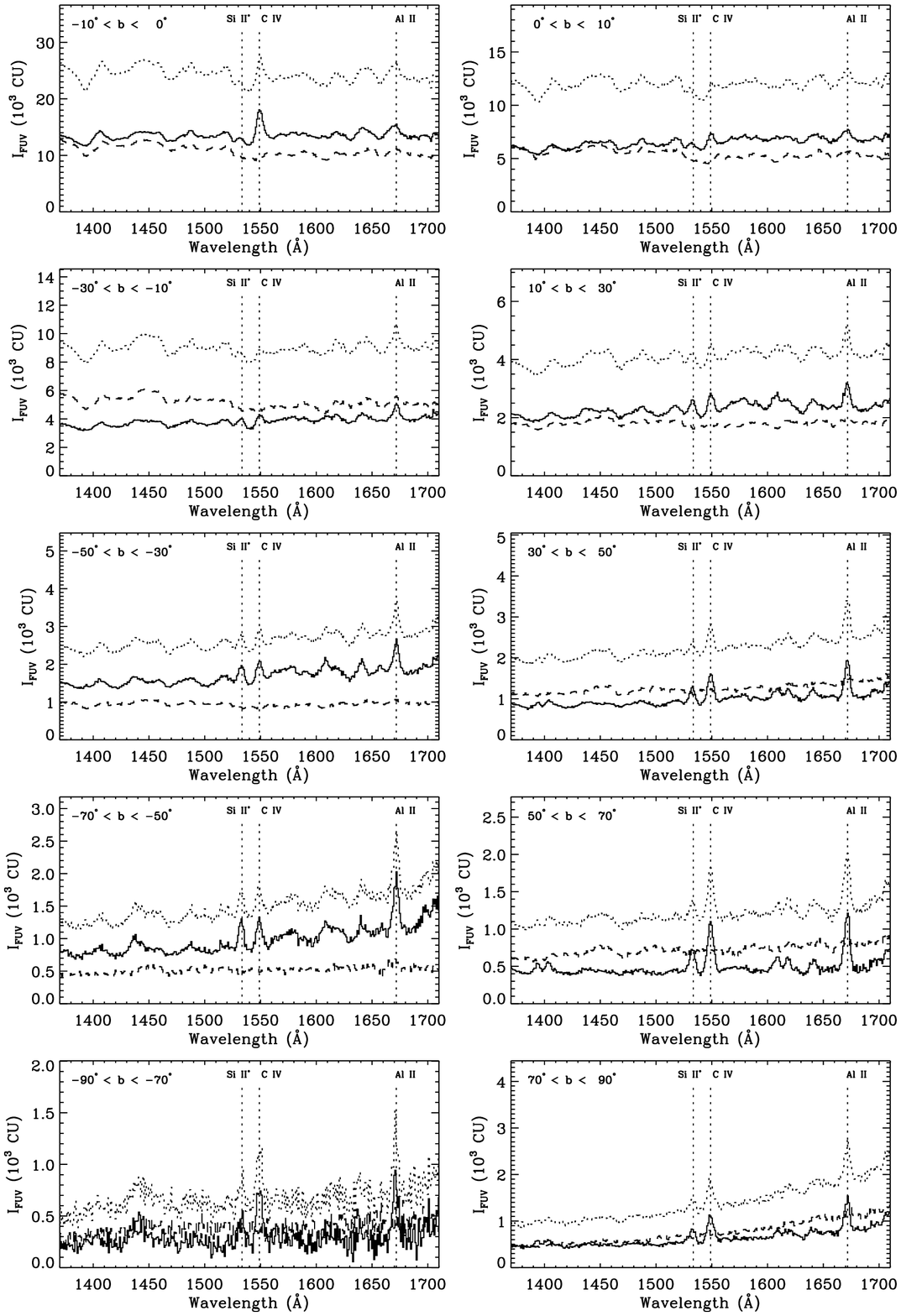}
\par\end{centering}

\caption{\label{spec_vs_lat}Average FUV spectra in various Galactic latitude
ranges. Solid curve shows the diffuse FUV background spectrum. Dotted
and dashed curves represent the total (direct stellar + diffuse) and
the direct stellar spectra, respectively. Some important ionic lines
are also indicated.}
\end{figure*}

\begin{figure}[t]
\begin{centering}
\includegraphics[scale=0.48]{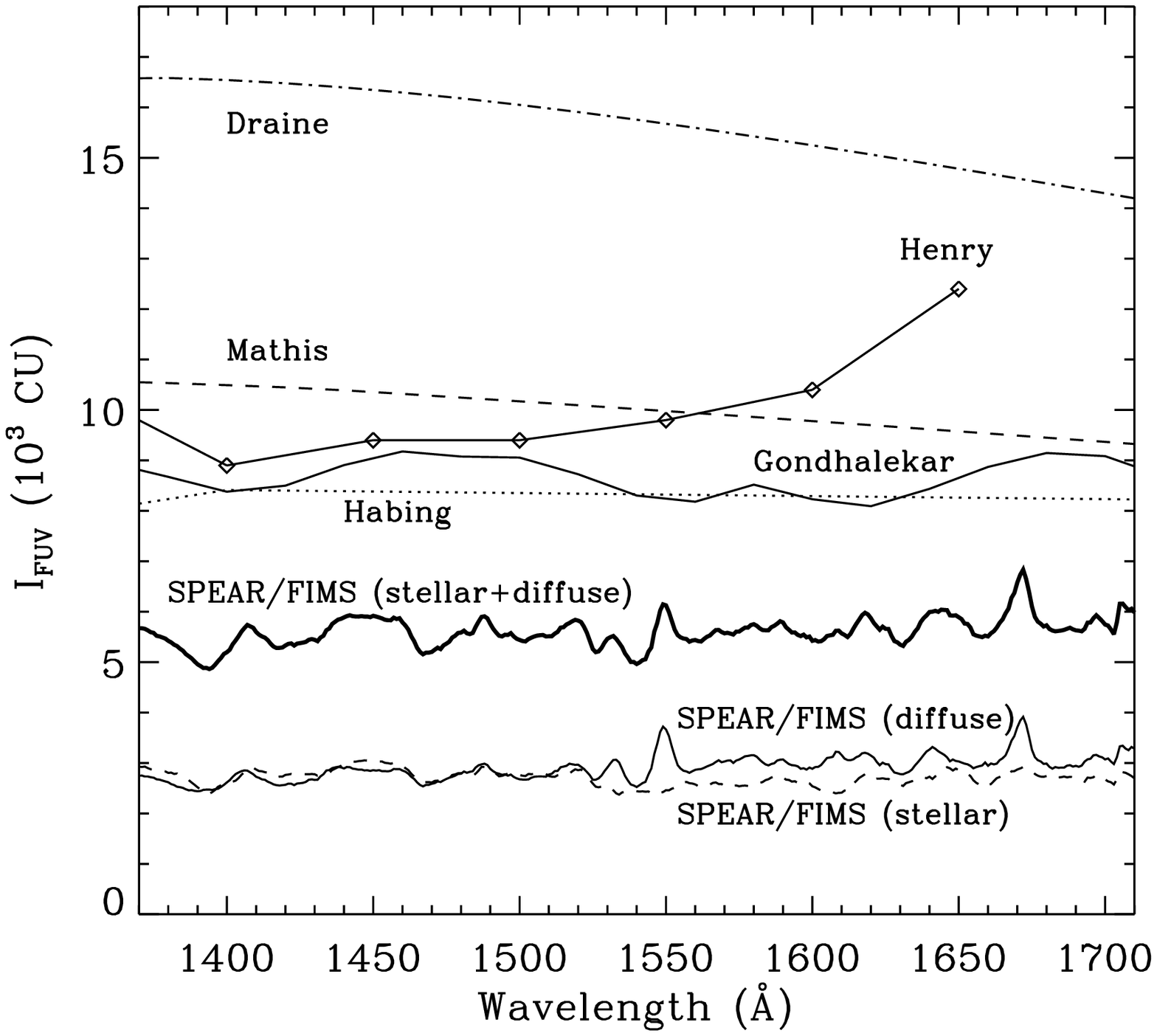}
\par\end{centering}

\caption{\label{isrf_compare}SPEAR/FIMS spectra averaged over all the observations
are compared with the three models of the ISRF. The four models are
from \citet{Harbing68}, \citet{GON80}, \citet{Draine78}, and \citet{Mathis83}.
The spectrum measured by \citet{Henry80} is also compared.}
\end{figure}

\begin{figure*}[t]
\begin{centering}
\includegraphics[scale=0.5,angle=90]{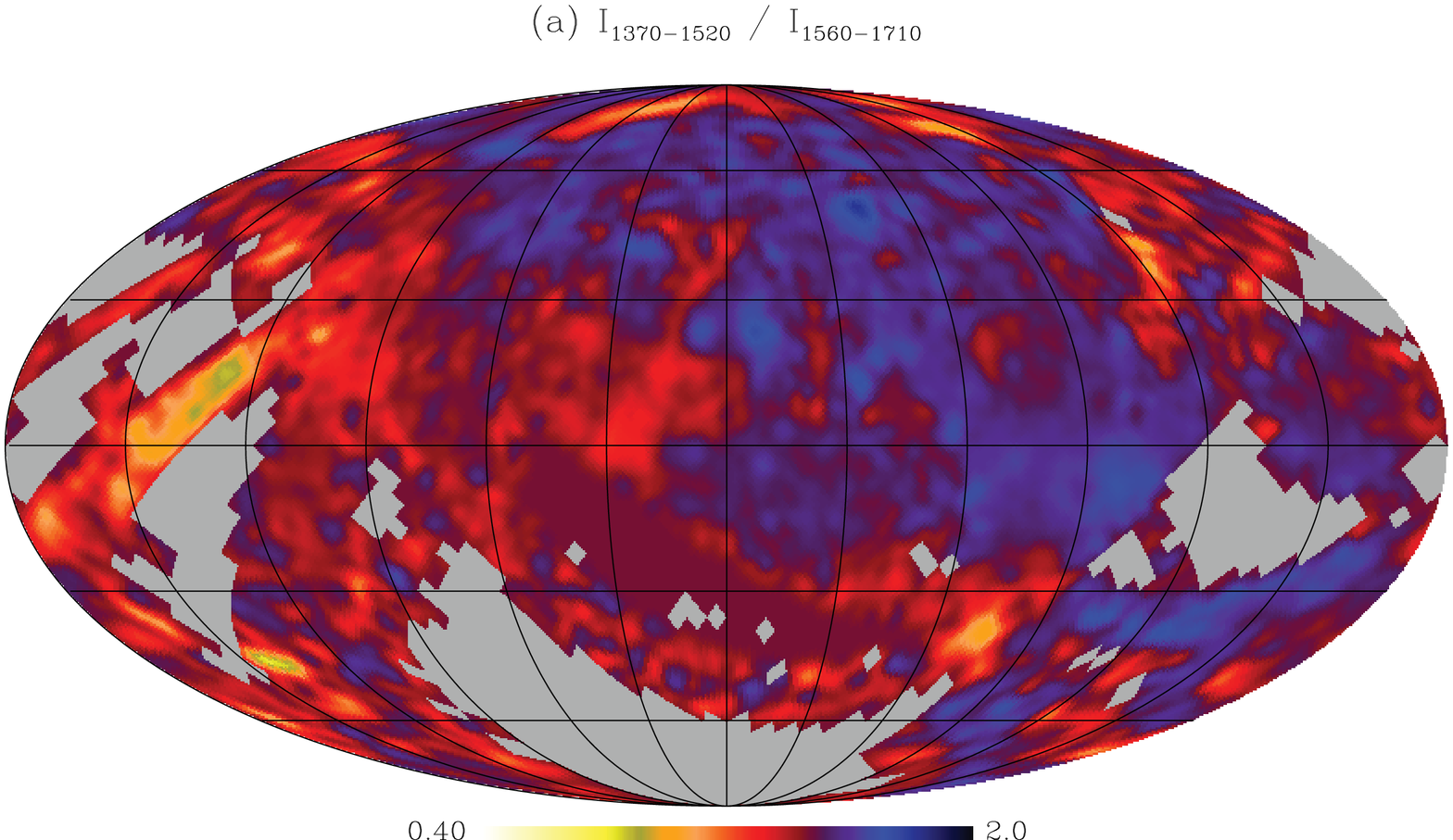}\medskip{}
\includegraphics[scale=0.5,angle=90]{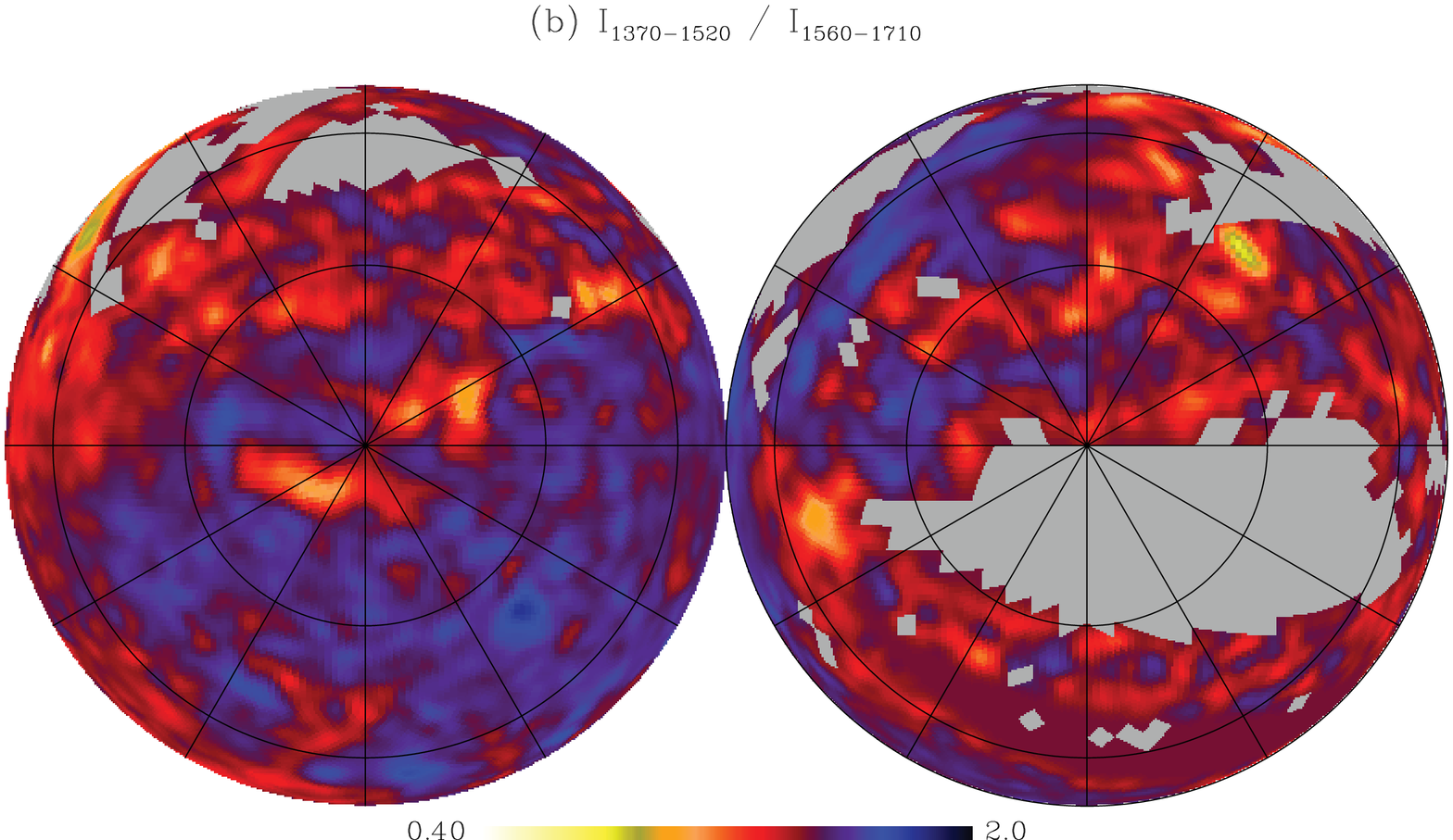}\medskip{}

\par\end{centering}

\caption{\label{hardness_map} (a) Mollweide and (b) orthographic projections
of the hardness ratio (1370--1520 \AA\ to 1560--1710 \AA) map. The
\ion{Al}{2} emission line was excluded in estimating the hardness
ratio. The left and right sides of orthographic projections are centered
at the northern and southern Galactic polar caps, respectively, and
their longitude increases clockwise and counter-clockwise, respectively.
In both projections, $l=0^{\circ}$ is at the six o'clock position.
The scales are linear across the color bars.}
\end{figure*}

\begin{figure*}[t]
\begin{centering}
\includegraphics[scale=0.5,angle=90]{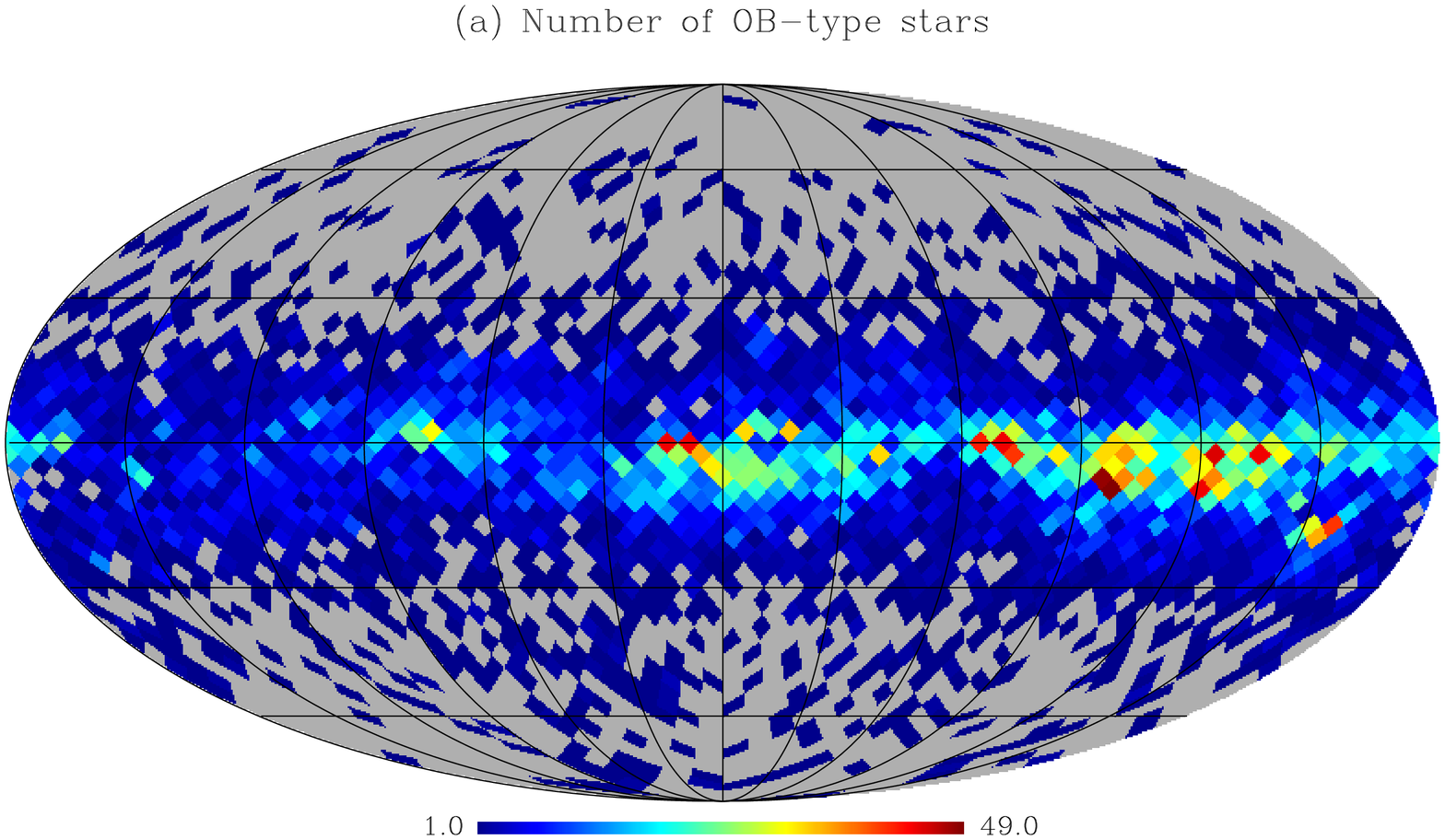}\medskip{}
\includegraphics[scale=0.5,angle=90]{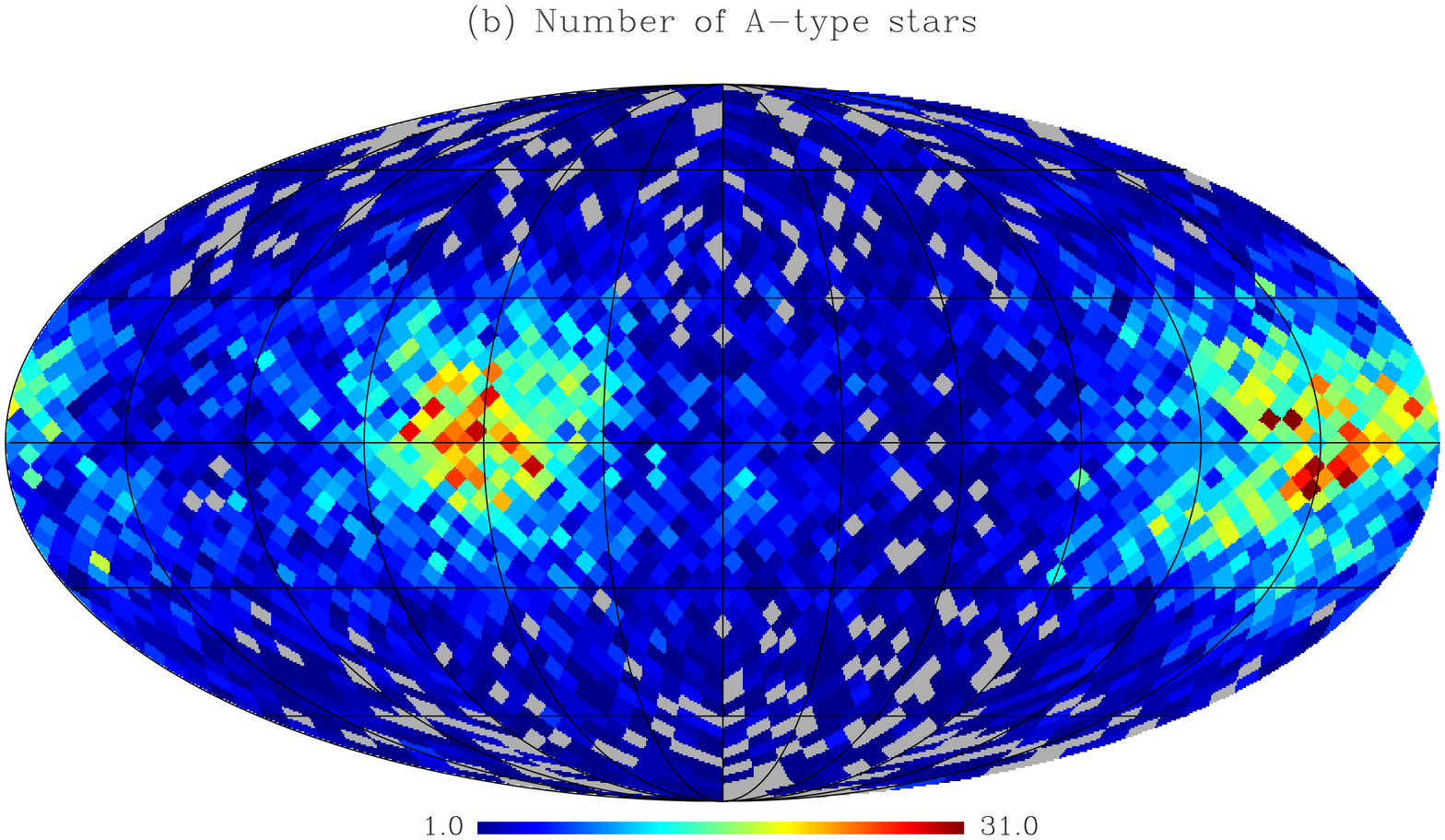}\medskip{}

\par\end{centering}

\caption{\label{td1_stars} (a) Number of OB-type stars and (b) number of A-type
stars from the TD-1 stellar catalog. The maps were made using the
resolution parameter $N_{{\rm side}}=16$ (with $\sim3.7^{\circ}$
pixels).}
\end{figure*}

\begin{figure}[t]
\begin{centering}
\includegraphics[scale=0.55]{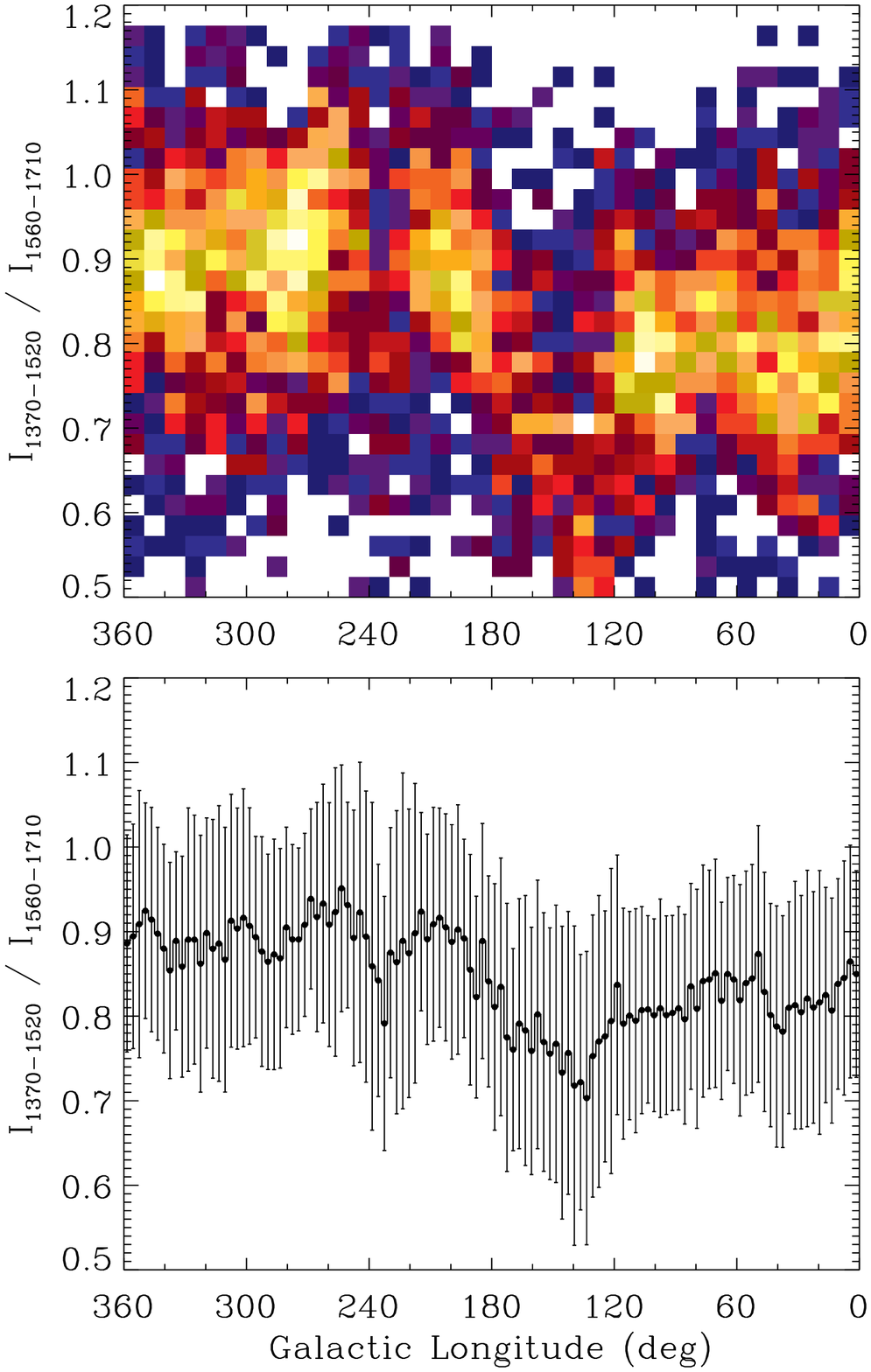}
\par\end{centering}

\caption{\label{hardness_vs_lon}Hardness ratio (1370--1520 \AA\ to 1560--1710
\AA) versus Galactic Longitude. The \ion{Al}{2} emission line was
excluded in estimating the hardness ratio.}
\end{figure}

\begin{figure}[t]
\begin{centering}
\includegraphics[scale=0.55]{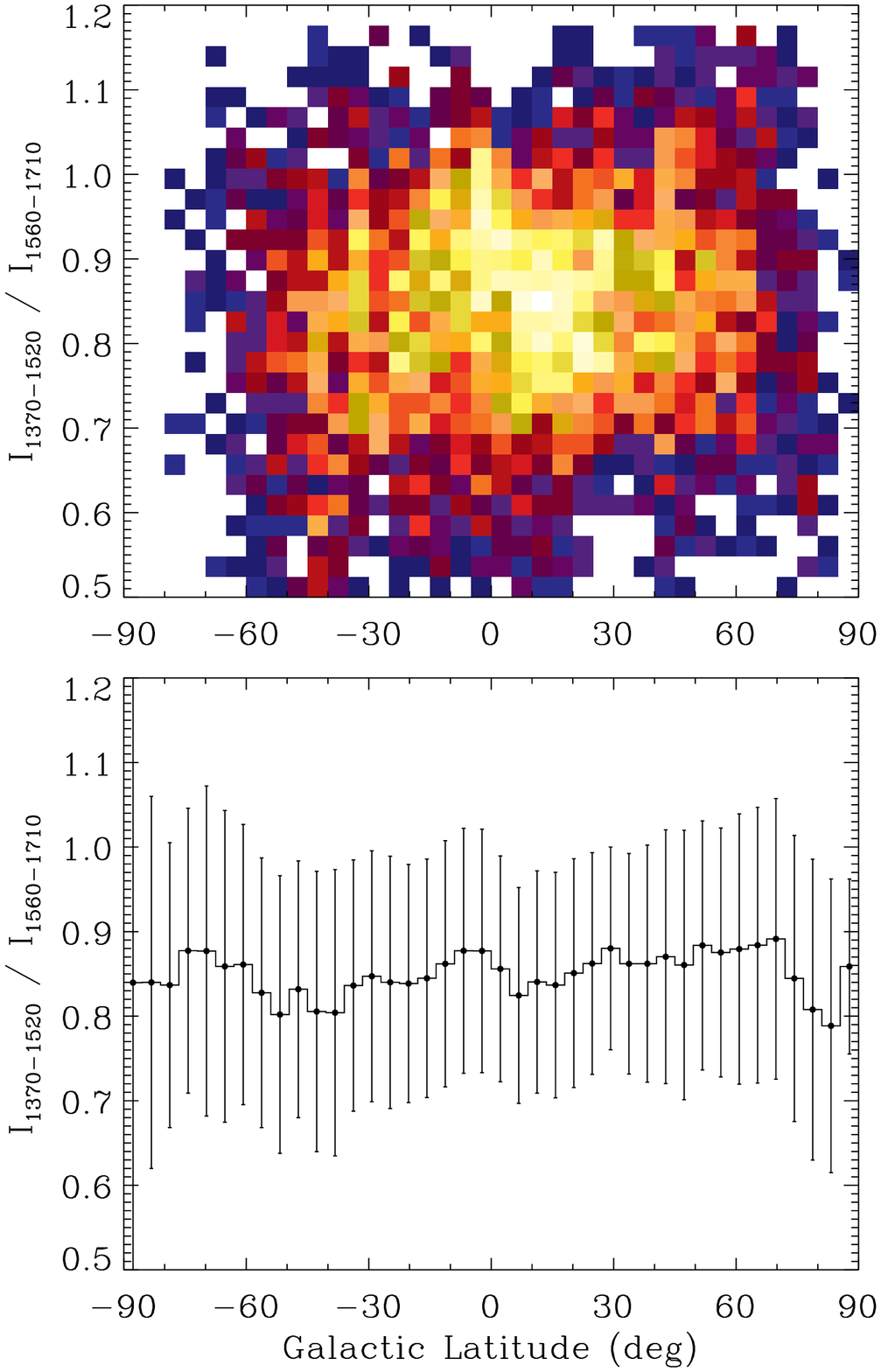}
\par\end{centering}

\caption{\label{hardness_vs_lat}Hardness ratio (1370--1520 \AA\ to 1560--1710
\AA) versus Galactic Latitude. The \ion{Al}{2} emission line was
excluded in estimating the hardness ratio.}
\end{figure}

\begin{figure}[t]
\begin{centering}
\includegraphics[scale=0.55]{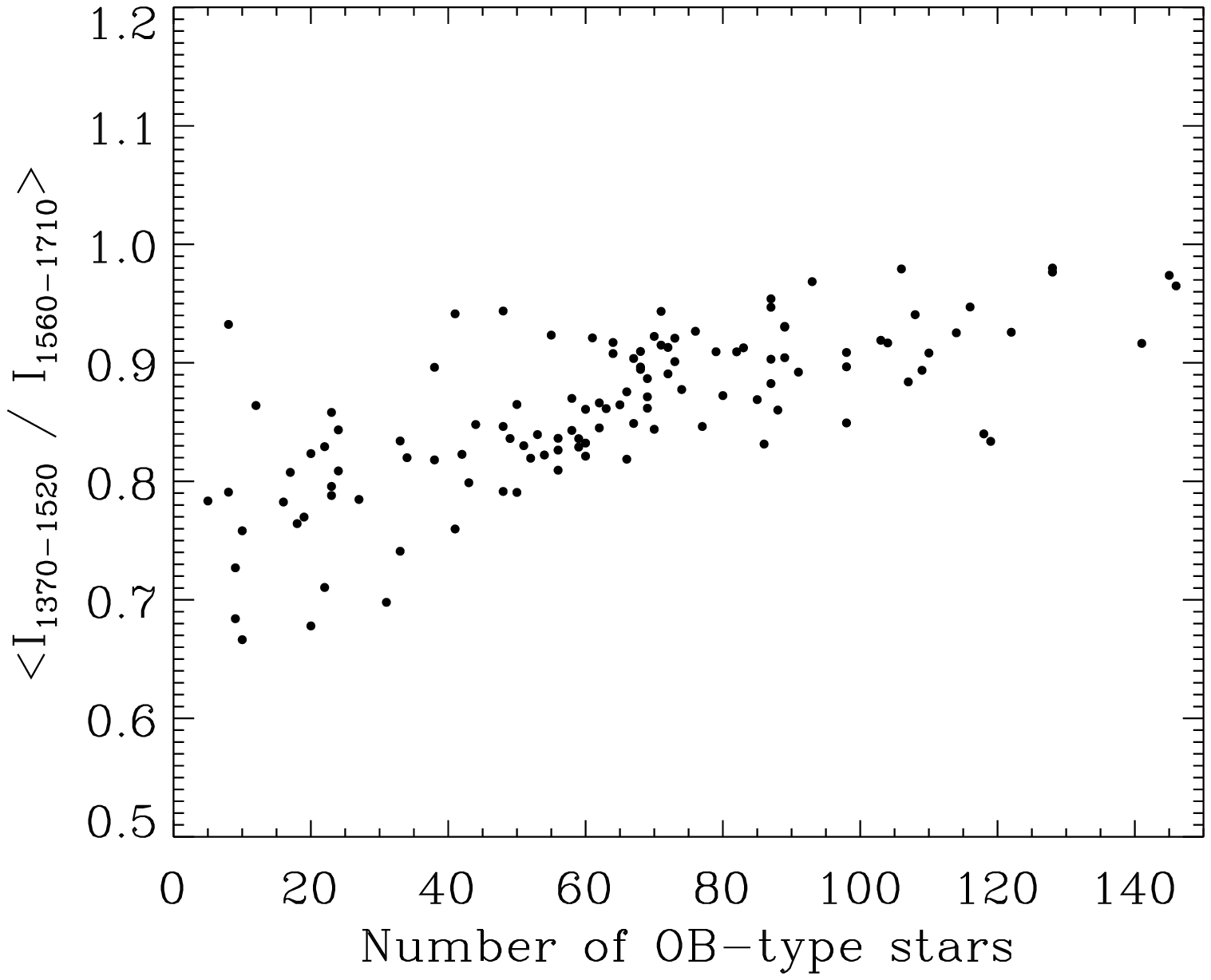}
\par\end{centering}

\caption{\label{hardness_obtype}Correlation between the average hardness ratio
(1370--1520 \AA\ to 1560--1710 \AA) and the number of OB-type stars.
The average hardness ratio and the number of OB-type stars were calculated
within each of the $3^{\circ}$ longitude strips, and then the average
hardness ratio versus the number of OB-type stars was plotted. The
\ion{Al}{2} emission line was excluded in estimating the hardness
ratio.}
\end{figure}

\begin{figure}[t]
\begin{centering}
\includegraphics[scale=0.55]{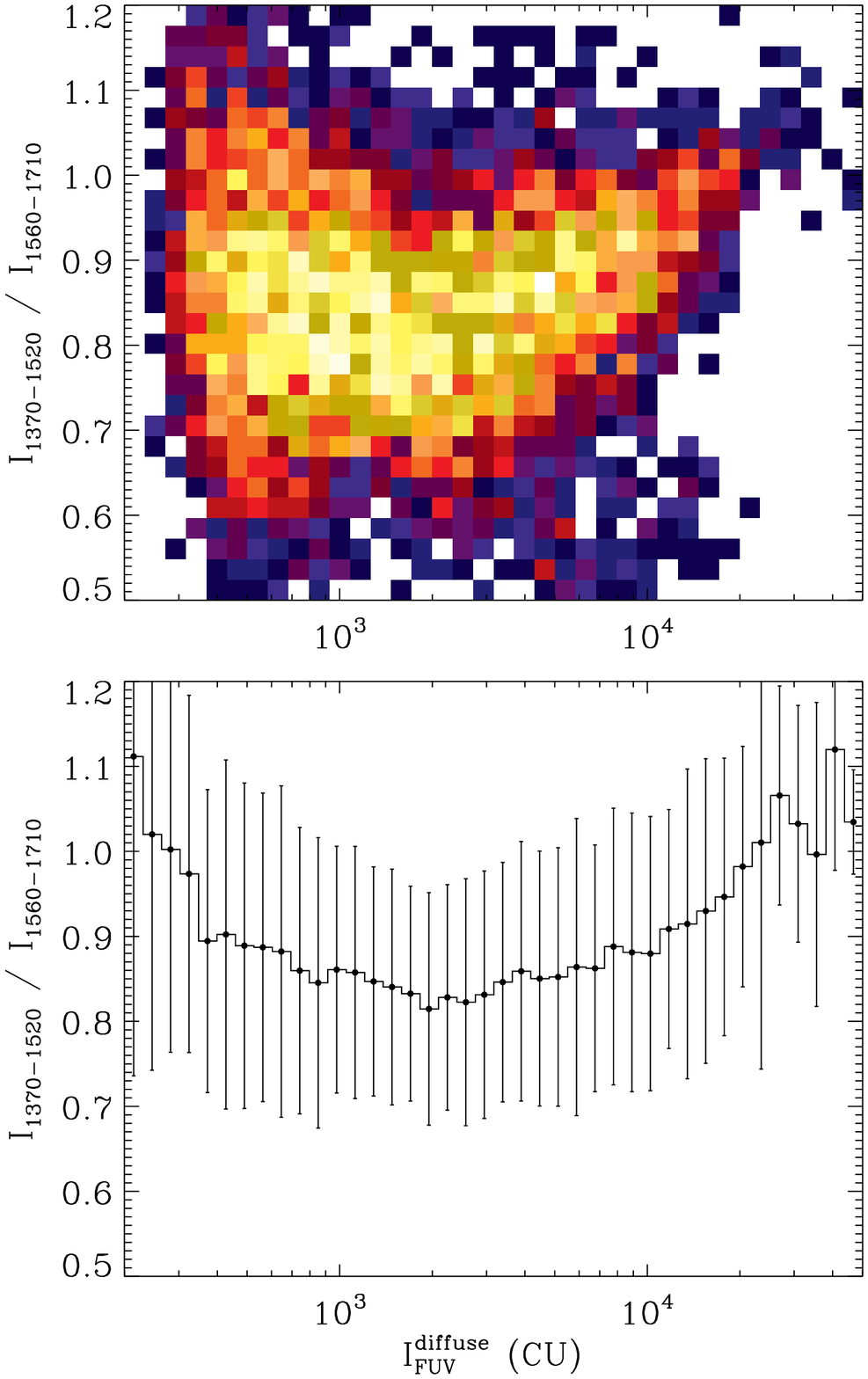}
\par\end{centering}

\caption{\label{hardness_vs_intensity}Hardness ratio (1370--1520 \AA\ to
1560--1710 \AA) versus intensity (1370--1520, 1560--1660, and 1680--1710
\AA). The \ion{Al}{2} emission line was excluded in estimating the
hardness ratio.}
\end{figure}

\begin{figure}[t]
\begin{centering}
\includegraphics[scale=0.55]{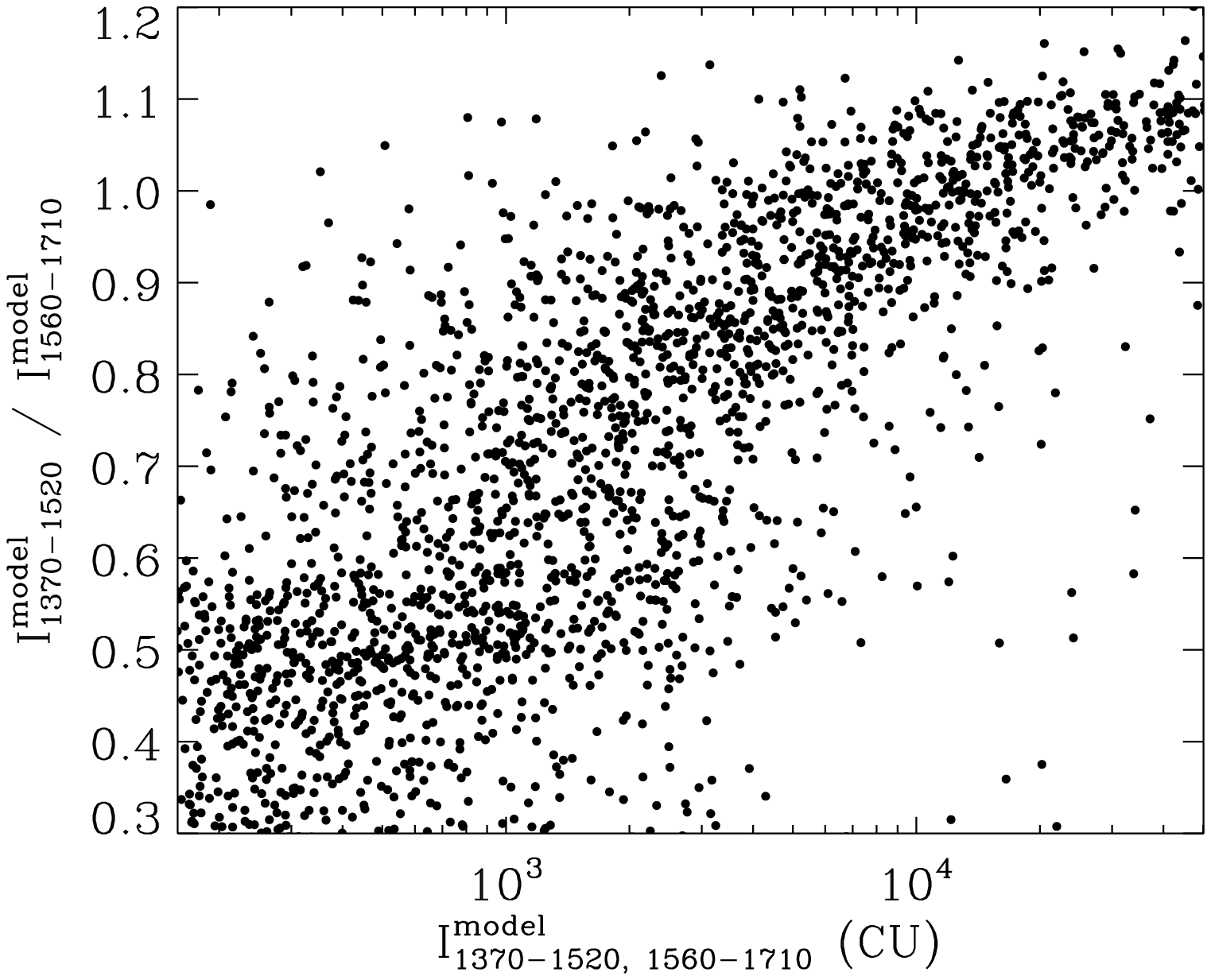}
\par\end{centering}

\caption{\label{hardness_model}Stellar model prediction of hardness ratio
(1370--1520 \AA\ to 1560--1710 \AA) versus intensity (1370--1520,
1560--1660, and 1680--1710 \AA). The \ion{Al}{2} emission line was
excluded in estimating the hardness ratio.}
\end{figure}

\begin{figure*}[t]
\begin{centering}
\includegraphics[scale=0.75]{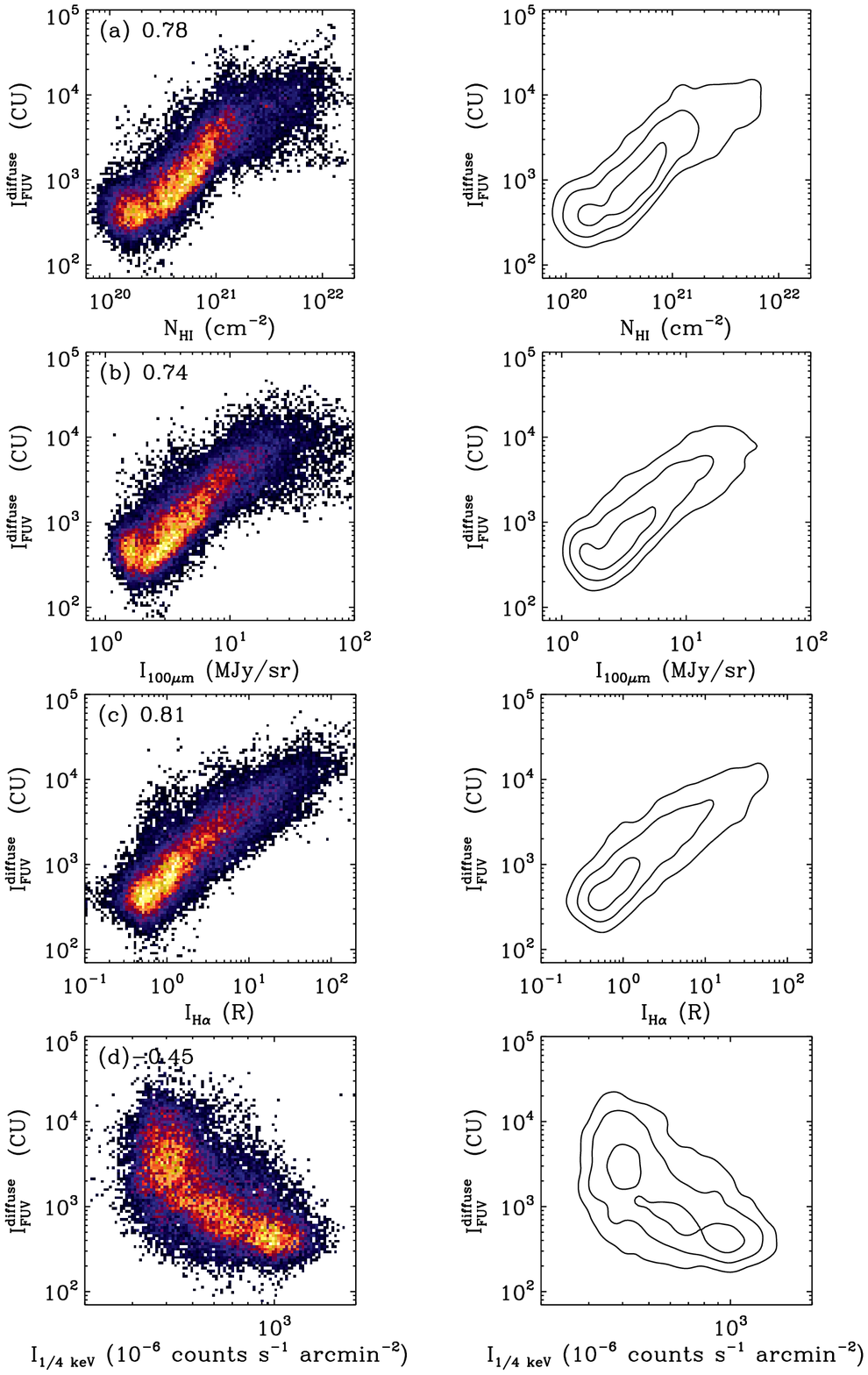}
\par\end{centering}

\caption{\label{correlation}Correlation of the diffuse FUV background with
(a) neutral hydrogen column density, (b) 100 $\mu$m emission, (c)
the diffuse H$\alpha$ emission, and (d) soft X-ray (1/4 keV band)
background. The left panels show two-dimensional histogram and the
right panels contours of the histograms. Numbers in the left panels
are the correlation coefficients estimated in logarithmic scale. The
contours correspond to 0.7, 0.3, and 0.1 of the maximum values of
the histograms.}
\end{figure*}

\begin{figure*}[t]
\begin{centering}
\includegraphics[scale=0.9]{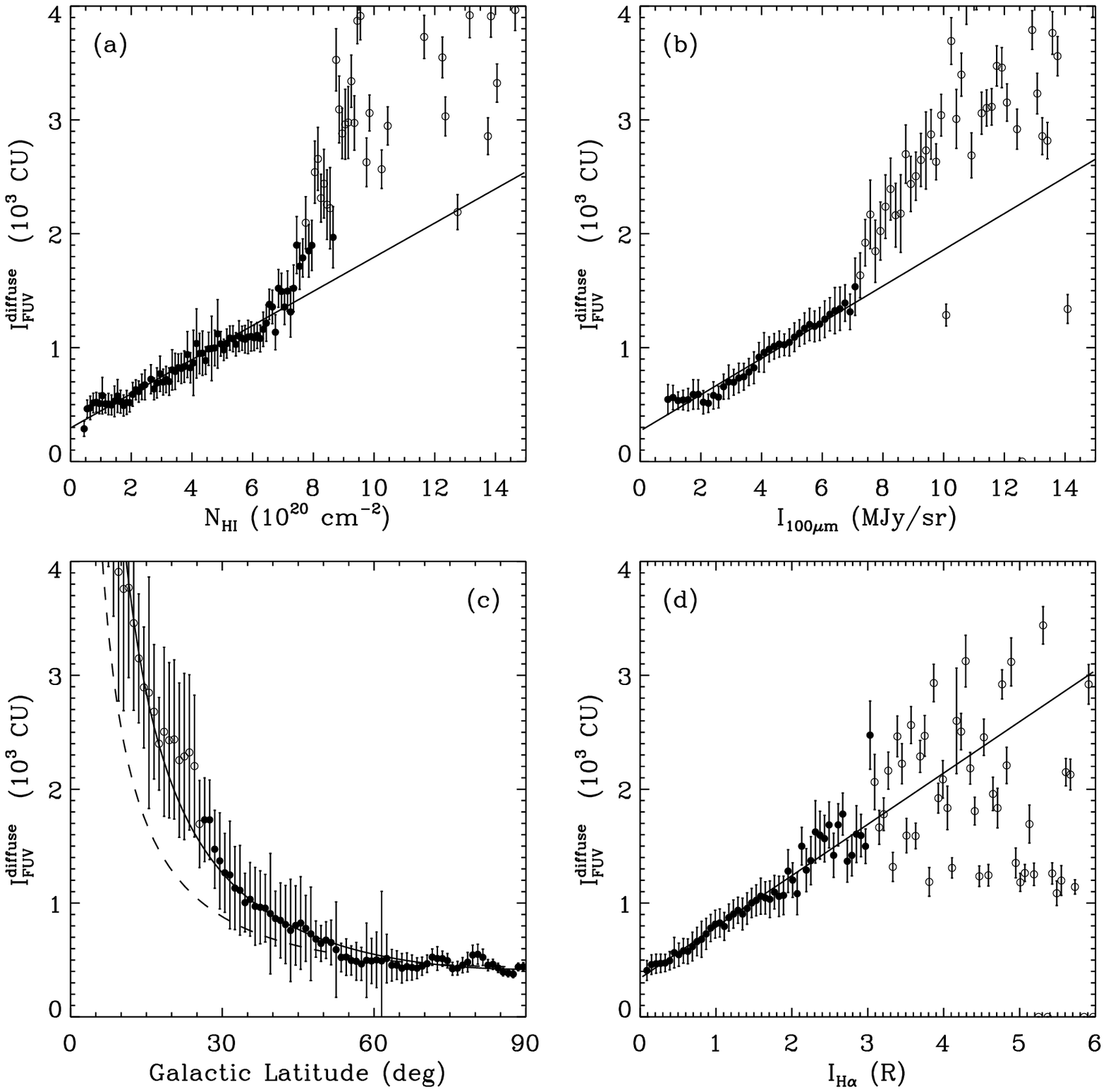}
\par\end{centering}

\caption{\label{correlation_linear}The FUV intensity versus (a) neutral hydrogen
column density, (b) 100 $\mu$m emission, (c) Galactic latitude, and
(d) the H$\alpha$ intensity. Best fits in (a), (b), and (d) are lines
with (a) slope 1.49 CU/(10$^{18}$ cm$^{-2}$), offset 271 CU, (b)
slope 158 CU/(MJy/sr), offset 243 CU, and (d) slope 456 CU/R, offset
309 CU. Best fit curves in (c) are $847\csc|b|-457$ CU for solid
curve, and $412\csc|b|$ CU for dashed curve. Filled and hollow circles
denote the data points, respectively, used and not used for the fit.}
\end{figure*}

\begin{figure*}[t]
\begin{centering}
\includegraphics[scale=0.9]{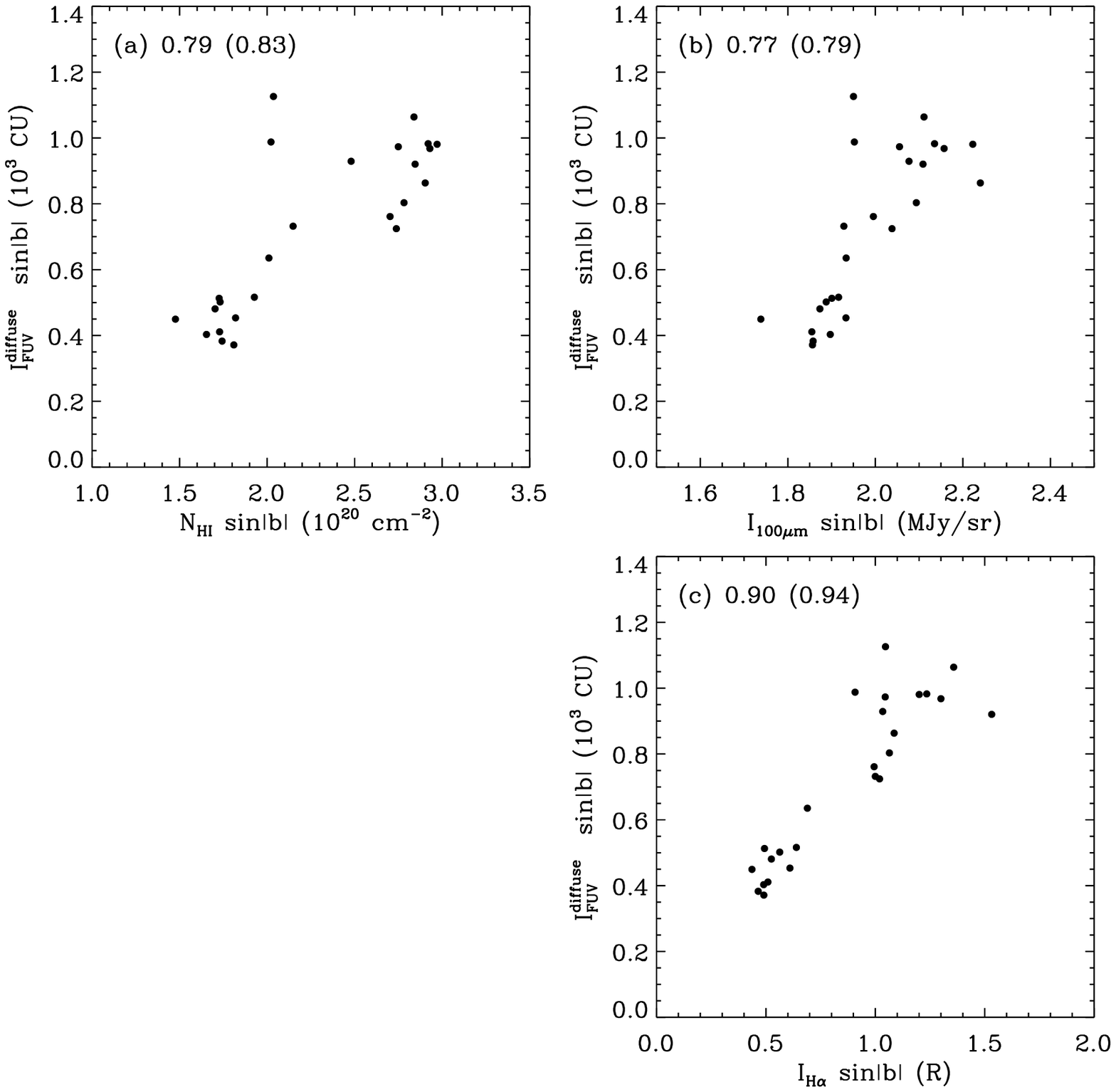}
\par\end{centering}

\caption{\label{correlation_sinb}Correlation of the diffuse FUV continuum
intensity with (a) neutral hydrogen column density, (b) 100 $\mu$m
emission, and (c) the diffuse H$\alpha$ intensity after the removal
of inverse-$\sin|b|$ dependence due to the plane-parallel medium.
Data points were obtained from latitude ranges of $\sin|b|\ge0.5$
($|b|\ge30^{\circ}$). Numbers outside and inside the parentheses
are correlation coefficients estimated in linear and logarithmic scales,
respectively.}
\end{figure*}

\begin{figure*}[tp]
\begin{centering}
\includegraphics[clip,scale=0.88]{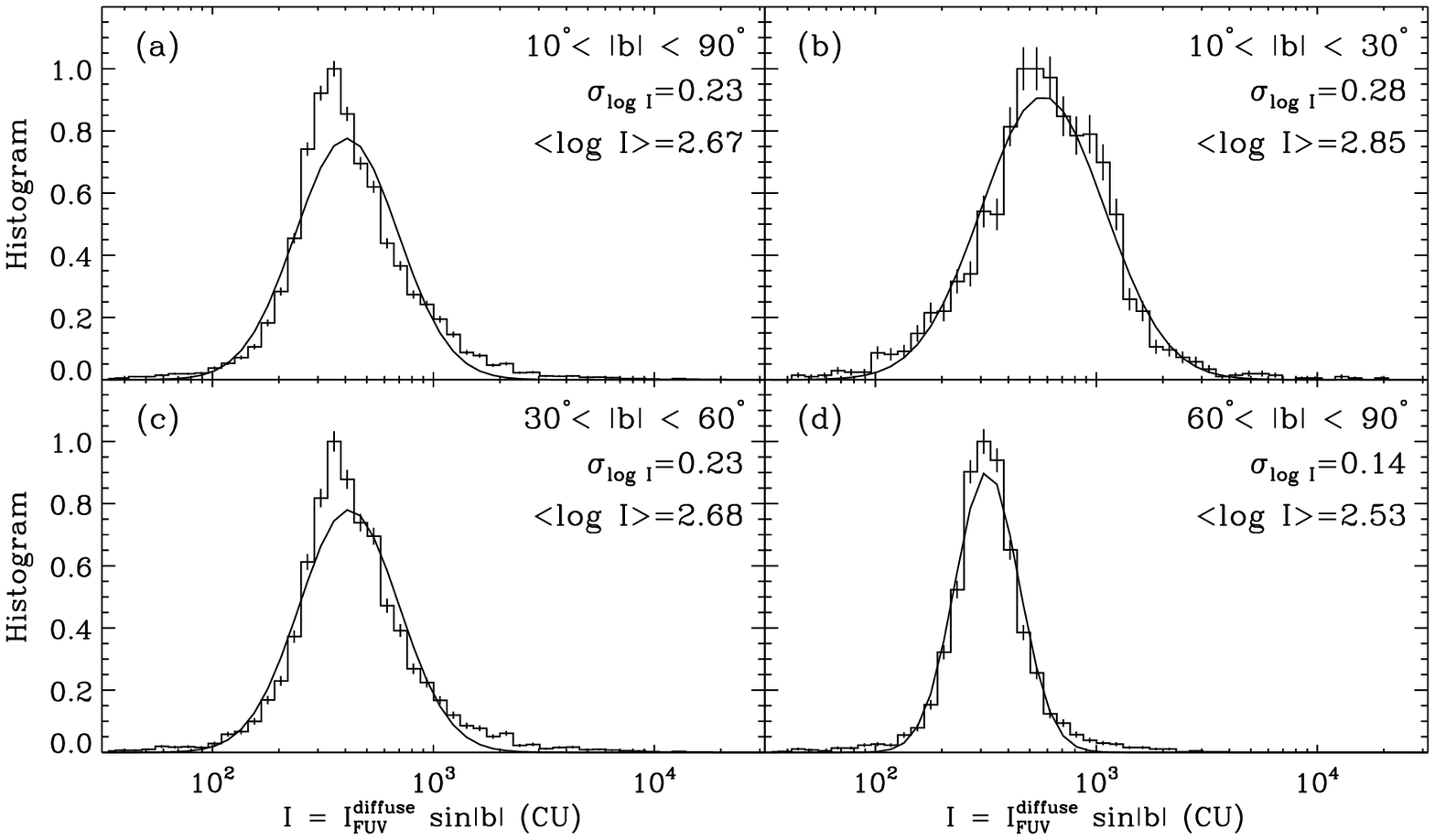}
\par\end{centering}

\caption{\label{log-normal}Histograms of $I_{{\rm FUV}}^{{\rm diffuse}}\sin|b|$
in logarithmic scale. Log-normal fits are also shown. The standard
deviations and the mean values of the distribution in logarithmic
scale are also shown.}
\end{figure*}

\begin{figure}[t]
\begin{centering}
\includegraphics[scale=0.47]{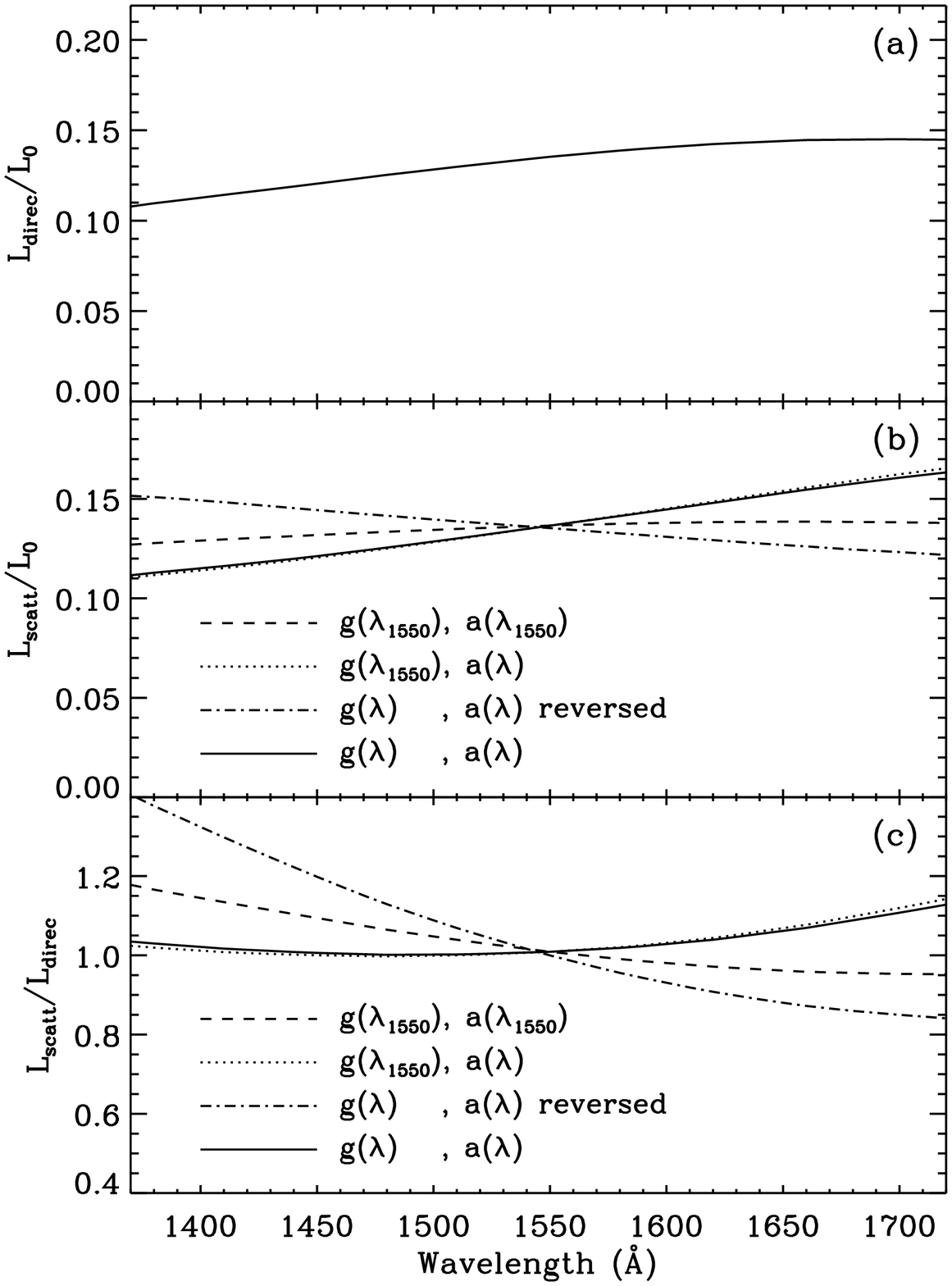}
\par\end{centering}

\caption{\label{aeffect}Scattering and absorption in a homogeneous dust medium.
(a) The directly-escaped and (b) scattered spectra relative to the
input luminosity of the point source. (c) The scattered spectrum relative
to the attenuated direct stellar spectrum. Here, various combinations
of the wavelength-dependences of the dust albedo and asymmetry factor
are examined.}
\end{figure}

\begin{figure}[t]
\begin{centering}
\includegraphics[scale=0.52]{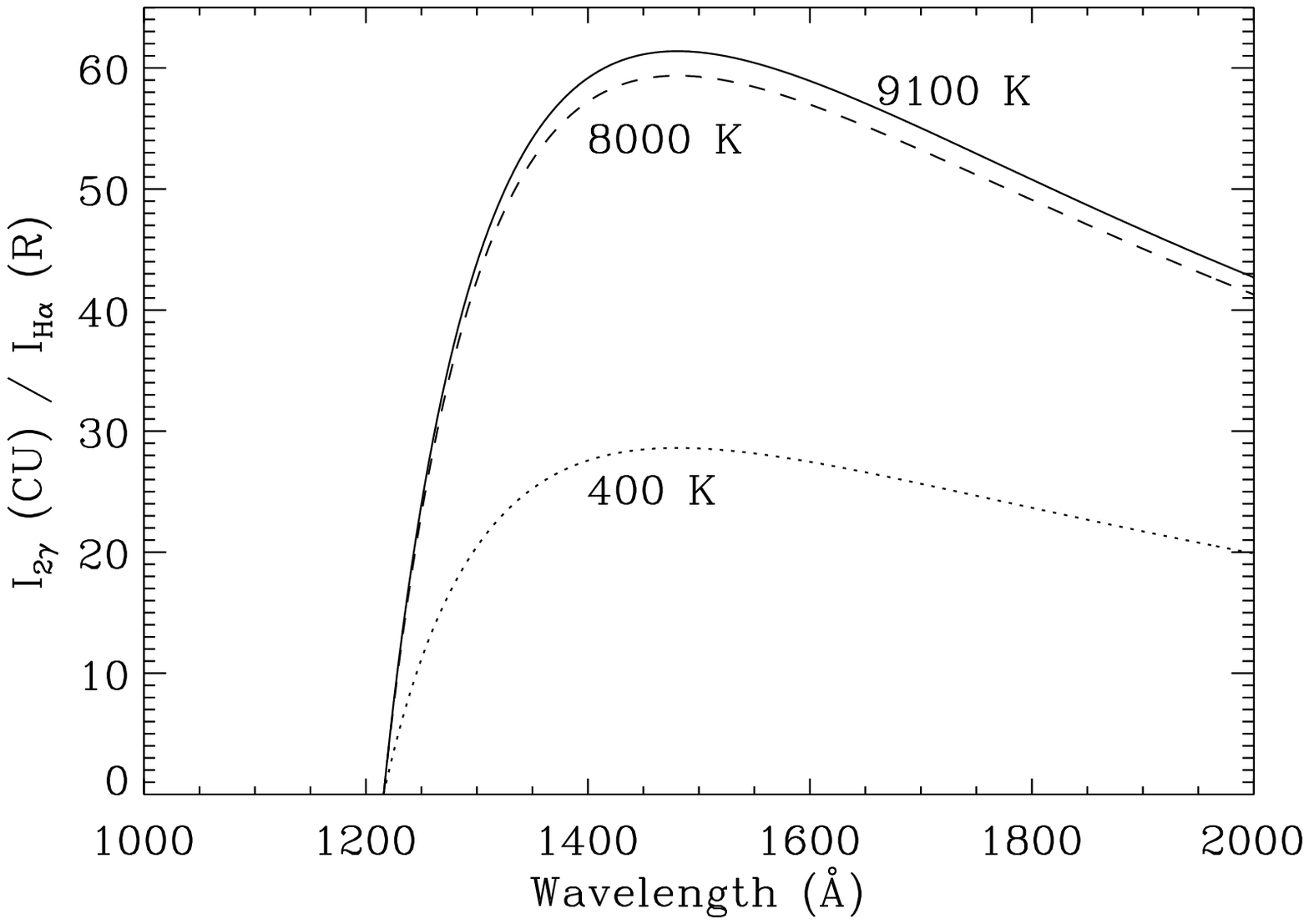}
\par\end{centering}

\caption{\label{two_photon}Two-photon continuum intensity that accompany 1
R of H$\alpha$ emission from ionized interstellar hydrogen. The continuum
intensity is shown for three temperatures: 9100 K, solid line; 8000
K, dashed line; 400 K, dotted line.}
\end{figure}

\end{document}